\documentclass[11pt,a4paper]{article}
\pdfoutput=1
\usepackage{jheppub}

\usepackage{amsmath}
\usepackage{verbatim}
\usepackage{amssymb}
\usepackage{inputenc}
\usepackage{textcomp}
\usepackage{appendix}
\usepackage{mathtools}
\usepackage{multirow}
\usepackage{amsmath,latexsym,amssymb}
\usepackage[makeroom]{cancel}
\usepackage[english]{babel}
 \usepackage{hyperref}
\hypersetup{
	colorlinks=true, 	
	linkcolor=blue,		
    citecolor=cyan,		
}

\newcommand{\e}{\eta}
\newcommand{\be}[1]{\begin{equation}\label{#1} }
\newcommand{\ee}{\end{equation}}
\newcommand{\bea}[1]{\begin{eqnarray}\label{#1} }
\newcommand{\eea}{\end{eqnarray}}
\newcommand{\bes}{\begin{subequations}}
\newcommand{\ees}{\end{subequations}}
\newcommand{\sss}{\scriptscriptstyle}
\newcommand{\cd}{\cdot}
\newcommand*\xbar[1]{%
   \hbox{%
     \vbox{%
       \hrule height 0.5pt 
       \kern0.5ex
       \hbox{%
         \kern+0.1em
         \ensuremath{#1}%
         \kern+0.1em
       }%
     }%
   }%
}

\newcommand{\p}{\partial}
\newcommand{\n}{\nabla}
\newcommand{\nn}{\nonumber}

\newcommand{\tf}{\tilde{f}}
\newcommand{\tdf}{\tilde{f}}
\newcommand{\ga}{\Gamma}

\newcommand{\La}{\Lambda}
\renewcommand{\(}{\left(}
\renewcommand{\)}{\right)}

\newcommand{\lb}{\Big[}
\newcommand{\rb}{\Big]}

\newcommand*\mtK{\mathcal{K}}
\newcommand*\mtO{\mathcal{O}}

\newcommand*\dl{\nabla}

\title{Large $D$ gravity and charged membrane dynamics with nonzero cosmological constant}

\author[a]{Suman Kundu,} \author[b]{Poulami Nandi,} 
\affiliation[a]{Tata Institute of Fundamental Research, Mumbai, 400005, INDIA.\\} 
\affiliation[b]{Indian Institute of Technology Kanpur, Kanpur 208016, INDIA.}

\emailAdd{suman.kundu 290@tifr.res.in, poulamin@iitk.ac.in}

\abstract{In this paper, we have found a class of dynamical charged `black-hole' solutions to Einstein-Maxwell system with a non-zero cosmological constant in a large number of spacetime dimensions. We have solved up to the first sub-leading order using large D scheme where the inverse of the number of dimensions serves as the perturbation parameter. The system is dual to a dynamical membrane with a charge and a velocity field, living on it.  The dual membrane has to be embedded in a background geometry that itself, satisfies the pure gravity equation in presence of a cosmological constant. Pure AdS / dS are particular examples of such background. We have also obtained the membrane equations governing the dynamics of charged membrane. The consistency of our membrane equations is checked by calculating the quasi-normal modes with different Einstein-Maxwell System in AdS/dS.}

\begin{document}
\maketitle

\section{Introduction} 
It is now well-known that black hole solutions simplify a lot in a large number of space-time dimensions (denoted as $D$), the key reason being that the `blackening factor' of the black hole/ brane metric reduces to its asymptotic form exponentially fast in space as we take $D$ to infinity. The sole effect of  the  black hole,  then,  confines within an infinitesimally thin region (referred to as `membrane region') around its event horizon. Also, it turns out that there is a large ${\cal O}(D)$ gap in the spectrum of  linearized fluctuations around these large-$D$ black holes. The slowly varying modes, which are finite in number,  are decoupled from  the fast modes in the sense, that they also decay exponentially outside the same membrane region \cite{EmparanCoupling}. Considering all these facts together, it is natural to expect a non-linear completion for these decoupled Quasi-Normal Modes, leading to new dynamical black hole solutions of Einstein equations.  \\
 The initial development of the subject is found in \cite{Emparan:2013moa, Emparan:2013xia, Emparan:2013oza, Emparan:2014cia,
Emparan:2014jca, Prester:2013gxa,  membrane, Chmembrane,yogesh1, Emparan:2015hwa,Suzuki:2015iha,Emparan:2015gva,Tanabe:2015hda,Tanabe:2015isb,Suzuki:2015axa,Emparan:2015rva,EmparanHydro,Romero-Bermudez:2015bma,Tanabe:2016opw}. More works related to large D are in  \cite{Sadhu:2016ynd,Herzog:2016hob,Rozali:2016yhw,
Chen:2015fuf,Chen:2016fuy,Chen:2017hwm,Chen:2017wpf,Konoplya:2013sba,Guo:2015swu,Chen:2017rxa,Herzog:2017qwp,radiation,yogesh2,Dandekar:2017aiv,SB,SB2,Saha:2018elg}. New dynamical black hole / brane metrics have been constructed for both asymptotically flat and dS/ AdS backgrounds \cite{membrane,Chmembrane,SB,yogesh1,Suzuki:2015iha,Tanabe:2015hda,Tanabe:2016opw,Chen:2017rxa,Saha:2018elg}.  In \cite{Chmembrane} The technique has been extended for Einstein-Maxwell system in asymptotically flat space-times. In \cite{SB} the authors have generalized the technique to any asymptotic geometry as long as it separately satisfies the relevant Einstein equations (with or without cosmological constant) upto the first subleading order. The second subleading order analysis has been carried out in  \cite{SB2}.

These solutions are always perturbative and could be constructed only in a large number of dimensions as an expansion in $\left(\frac{1}{D}\right)$. Nevertheless, they are useful for several purposes. Firstly, they generate a new class of dynamical black hole solutions of Einstein equations which are very difficult to solve otherwise (even numerically).\\
 Secondly, through these constructions, we could see a duality between the dynamical horizons of the black hole/brane metric and a co-dimension one dynamical membrane, embedded in the asymptotic background geometry.  This membrane-gravity duality allows us to analyse the complicated dynamics of the black holes from a different angle, which might turn out to be useful in future for some realistic calculation.

In this paper, our goal is to extend the `background covariant'  technique of \cite{SB} to Einstein-Maxwell system in presence of cosmological constant. For this case, the dual system would be a codimension-one dynamical charged membrane, embedded in the asymptotic dS / AdS metric.\\
 The motivation for our work is two-fold. The first is, of course, to see how the whole technique of background-covariantization works for Einstein-Maxwell system, which,  in terms of complexity,  is just at the next level, compared to the pure gravity system. Indeed we have noticed that unlike the uncharged case, only naive covariantization of the flat space-equations of  the charged membrane (as derived in \cite{Chmembrane}) will not give the correct duality and we need to add a  term proportional to the background curvature even at the first subleading order in ${\cal O}\left(\frac{1}{D}\right)$ expansion. This is indeed one of the interesting observations in our paper. \\
 The second piece of motivation is as follows.  We know that in asymptotically AdS geometry there exist another set of perturbative solutions to Einstein-Maxwell system. These are black holes/branes constructed in a derivative expansion and are dual to dynamical charged fluid living at the boundary of AdS. Recent works can be found in \cite{EmparanHydro,Dandekar:2017aiv,Chen:2018nbh}. At this point, it is natural to ask whether there exists any overlap regime for these two types of perturbations, and if so,  whether the two metrics agree.  In the best possible scenario, the outcome of this comparison could be a duality between the dynamics of a charged fluid and charged membrane in a large number of dimensions, where gravity does not have much role to play. Our construction in this paper is one necessary step towards such duality. \\
 
 \vspace{0.3cm}
 The outline of this paper is described as following. We start with the Einstein-Maxwell-Hilbert action in section 2. We write the Einstein-Maxwell equations in a simplified form and sketch the general solutions of metric and gauge field using the large D perturbative technique. The next section is devoted mainly to guess the initial ansatz for the metric and gauge field. We consider some educated guess to reach the starting point of the perturbation. Section 4 backs up the choice of our ansatz and describe the conditions imposed in leading order for the ansatz to satisfy the Einstein-Maxwell equations. We state the scaling laws of different tensors in powers of D. In section 5, we have covered, in details, the strategy to solve the equations. We have mentioned the subsidiary conditions on the auxiliary functions and the gauge choice to construct the most general structure of metric and gauge field corrections. Our analysis does not require any coordinate dependency of the background. Instead, the sub-leading order corrections are parametrized by some smooth function, a one form and charge field. The subsidiary conditions help to fix these functions in the background. One of the advantages of the large D technique is that it reduces the dimensionality of the problem by one, by integrating the radial direction. In few specific cases, like ours, it reduces the coupled non-linear PDEs (involving two variables) to ODEs. In section 5, this simplification is discussed. To solve the ODE s we need the boundary conditions. We have devoted a subsection in Section 5 for this purpose.  Section 6 deals with the strategy we mentioned in Section 5 to solve the first subleading corrections.  The next section summarise the results of this paper. It include the solutions of the Einstein-Maxwell equations along with the membrane equations governing the dynamics of  membrane. In section 8, we prove the consistency of our membrane equations. We calculate the light quasi-normal modes of few known solutions of Einstein-Maxwell system using our membrane equations. Section 9 wraps up this paper with a conclusion. The explicit details of the calculations are provided in the appendices. 


\section{Set up}

In this section, we shall describe our basic set-up for this problem.

\paragraph{}  We start by looking at the Einstein- Hilbert-Maxwell action in presence of cosmological constant $\Lambda$.
\bea{ehmac}
\mathcal{S}=\int (R-\frac{1}{4}F_{AB}F^{AB}-2\La) \sqrt{-G} \: d^DX
\eea

Here $G$ is the determinant of the space-time metric $G_{AB}$,  the corresponding Ricci scalar is denoted as $R$  and $F_{AB}$ is the field strength tensor for the $U(1)$ gauge field $A_B$. Following \cite{SB}   we scale  to define the cosmological constant $\Lambda$ as:
\be{}
\La=\frac{(D-1)(D-2)}{2}\lambda
\ee
\paragraph{} Varying the Einstein-Maxwell action, we get the following Einstein and Maxwell equation.
\bes \label{eomEM}
\bea{}
\mathcal{E}_{AB}&\equiv&R_{AB}-\frac{1}{2}R \:G_{AB}-\frac{1}{2}\lb F_{AC}F_B^{\:\:C}-\frac{1}{4}F^2G_{AB}\rb+\La G_{AB}=0\\
\mathcal{E}^{N}&\equiv&\xbar{\nabla}_M F^{MN}=0
\eea
\ees
In the Maxwell equation $\xbar{\nabla}$ is the covariant derivative with respect to $G_{AB}$.\\
The first equation \eqref{eomEM} could be simplified a bit. Note that, by contracting this equation with $G^{AB}$, we get a relation among  $R$, $F_{AB} F^{AB}$ and $\Lambda$. Substituting this relation back into the expression of ${\cal E}_{AB}$ we find the simplified version of the Einstein equation, which we shall use for our further computation. So the  final set of differential equations that we are going to solve in this paper using  the `large $D$ perturbation technique', are the following. 
\bes \label{fineom}
\bea{}
\mathcal{E}^{(1)}_{AB}&=& R_{AB}-\frac{1}{2}F_{AC}F^{\:\:C}_B+\frac{1}{4D}F^2G_{AB}-(D-1) \lambda G_{AB}=0\\
\mathcal{E}_{(2)}^{N}&=&\xbar{\nabla}_M (F_{AB}\:G^{MA}\:G^{NB})=0
\eea
\ees

As mentioned before, our objective is to solve the equations perturbatively where the perturbation parameter is $\left(\frac{1}{D}\right)$. Schematically the solution will take the following form.

\bes \label{schemsol}
\bea{}
G_{AB}&=&\bar G_{AB}+ \sum_0^{\infty} \Big(\frac{1}{D}\Big)^k G_{AB}^{(k)}\\
A_M &=& \sum_0^{\infty} \left(\frac{1}{D}\right)^kA^{(k)}_M
\eea
\ees
where $\bar G_{AB}$  is some exact solution of pure Einstein equation in presence of cosmological constant (i.e., the first equation of \eqref{fineom} with gauge field set to zero). 
In this paper, the precise goal of our computation would be to determine $G^{(1)}_{AB}$ and $A_M^{(1)}$ given some arbitrary $\bar G_{AB}$, satisfying the above constraint.

\section{The leading ansatz}
Our goal is to determine the first subleading correction to the metric and the gauge field. But, as it is true in any perturbative calculation, we must know the leading solution before we could determine any subleading term.  On the other hand, the leading ansatz could never be derived, since typically there is no unique answer to this. In some sense, we have to start with an educated  guess for $G^{(0)}_{AB}$ and $ A_M^{(0)}$ so that the metric $\bigg[G_{AB}|_\text{leading} = \bar G_{AB} + G^{(0)}_{AB}\bigg]$ and the gauge field $\bigg[A_M|_\text{leading} = A_M^{(0)}\bigg]$ satisfy the equations \eqref{fineom} at leading order (which turns out to be order ${\cal O}(D^2)$ in this case).\\
Now from \cite{SB}, we know the form of the ansatz for arbitrary $\bar G_{AB}$ but without the gauge field. It is given by almost the same ansatz one has in asymptotically flat space, except that the explicit appearance of the flat space Minkowski metric $\eta_{AB}$ has been changed to $\bar G_{AB}$.
Also from \cite{Chmembrane},  we know the form of the leading ansatz in asymptotically flat space  in presence of gauge field. 
\begin{equation}\label{flatspace}
\begin{split}
\left[G_{AB}\right]_\text{leading}^\text{flat} &= \eta_{AB} + \left[(1 + Q^2)\psi^{-D} - Q^2\psi^{-2D}\right]O_A O_B\\
\left[A_M\right]_\text{leading}^\text{flat} &= \sqrt{2} Q ~\psi^{-D}O_M
\end{split}
\end{equation}
Here $\psi$ and $Q$ are any smooth functions and $O_A~dX^A$  is a one-form which is null with respect to $\eta_{AB}$.
Taking the cue from the uncharged case, one very natural guess for  the leading ansatz in Einstein-Maxwell system in the arbitrary asymptotic background would be to simply replace the explicit appearance of $\eta_{AB}$ to $\bar G_{AB}$, i.e.,
\begin{equation}\label{curvespace}
\begin{split}
\left[G_{AB}\right]_\text{leading} &= \bar G_{AB} + \left[(1 + Q^2)\psi^{-D} - Q^2\psi^{-2D}\right]O_A O_B\\
\left[A_M\right]_\text{leading} &= \sqrt{2} Q ~\psi^{-D}O_M
\end{split}
\end{equation}
and now $O_A$ is null with respect to $\bar G_{AB}$.\footnote{We could intuitively understand why such simple replacement works for the leading ansatz. As described in the introduction, all these geometries will necessarily possess an event horizon and the non-trivial gravity effects of the black holes will be confined within a thin region of the thickness of order ${\cal O}\left(\frac{1}{D}\right)$ around this event horizon, which we shall refer to as `membrane region'. Because of this infinitesimal thickness, inside the membrane region, the details of the asymptotic geometry become irrelevant at the very leading order, and the same ansatz works as long as we replace asymptotic background as required. See \cite{Chmembrane} for a more detailed discussion on this point. }.

Note that the $\psi=1$ hypersurface is null with respect to  the metric $\left[G_{AB}\right]_\text{leading}$. Also at large $D$, as  one goes finitely away from this hypersurface, the metric either blows up (if $\psi <1$ )  or reduces to its asymptotic form $\bar G_{AB}$. However, we could easily see that the metric is non-trivial and finite only within a region of thickness of order ${\cal O}\left(\frac{1}{D}\right)$ in the following way. \\
Suppose we are infinitesimally away from the hypersurface such that  $(\psi = 1 +\frac{R}{D})$ where $R\sim {\cal O}(1)$. This implies $\psi^{-D}$ is non trivial even when $D\rightarrow\infty$.
\begin{equation*}
\begin{split}
&\lim_{D\rightarrow\infty}\psi^{-D}=\lim_{D\rightarrow\infty}\left(1 +\frac{R}{D}\right)^{-D} = e^{-R}\\
\Rightarrow& \lim_{D\rightarrow\infty} \left[(1 + Q^2)\psi^{-D} - Q^2\psi^{-2D}\right] = \left[(1 + Q^2)e^{-R} - Q^2e^{-2R}\right]
\end{split}
\end{equation*}
Clearly, our ansatz does have the required form with a non-trivial membrane region. Also, the singular part of the space-time is shielded by the membrane at $\psi=1$, which we shall identify with the event horizon of the space-time.  Now we have to check whether this ansatz satisfy the relevant equations at leading order.
\section{How  the ansatz solves \eqref{fineom} at leading order}
In this section, we shall see that the above ansatz indeed solves \eqref{fineom} at leading order provided $O_A$ and $\psi$ satisfy certain conditions on the $(\psi=1)$ hypersurface. But before getting into the details of the equations, we shall first describe our $\left(\frac{1}{D}\right)$ expansion in a little more details.
 \subsection{Scaling with D}
 Note that the first equation in \eqref{fineom} is a set of $\left(\frac{D(D+1)}{2}\right)$ equations  and the second equation is a set of $D$ equations for total $\left(\frac{D(D+1)}{2}\right)$ metric components and $ D$ gauge field components. Hence, as we increase $D$ (or decrease our perturbation parameter $\left(\frac{1}{D}\right)$), both  the number of equations  and the number of unknowns increase and no perturbation technique can work in such a situation.
This  has been discussed in detail in \cite{membrane}, \cite{Chmembrane},\cite{yogesh1} and \cite{SB}.  We shall follow their strategy to have a meaningful $\left(\frac{1}{D}\right)$ expansion. We shall assume  that the metric $G_{AB}$ and the gauge field $A_M$ are dynamical only along some fixed finite number of dimensions $p+1$ and the rest of the  $(D-p-1)$ dimensions  are protected by some symmetry. In terms of equation, we mean the following.
 \be{decommet}
 ds^2= G_{AB}dX^A dX^B=\tilde{G}_{ab}({x^a})dx^a dx^b +f({x^a})d\Omega^2
 \ee
where,
$\{a,b\}=\{0,1,\hdots p\}$. $\tilde{G}_{ab}({x^a})$ is a dynamical and finite $p+1$ dimensional metric and $d\Omega^2$ is the line element which takes care of the infinite $(D-p-1)$ symmetric space. $f({x^a})$ is an arbitrary constants of the $(p+1)$ dynamical coordinates.

However, as in \cite{yogesh1,SB}, for our computation, we do not need to use any  detail of this decomposition. The sole effect of this symmetry would be to impose some scaling rules (with $D$) on different derivative structures. Below we are simply stating these rules and would request the reader to go through the discussion in \cite{yogesh1,SB} for their justification.

\begin{itemize}
\item For a generic tensor $T^{A_1 A_2\cdots A_n}$ of order ${\cal O}\left(\frac{1}{D}\right)^k$, its divergence, $\nabla_{A_j} T^{A_1 A_2\cdots A_j\cdots A_n}$, would be of order ${\cal O}\left(\frac{1}{D}\right)^{k-1}$\\
where $\nabla_A$ denotes the covariant derivative with respect to $\bar G_{AB}$.
\item $\bar G_{AB}$ is such that all components of Riemann tensor, evaluated on $\bar G_{AB}$ is of order ${\cal O}(1)$, which further implies that
$$R_{AB}|_\text{on $\bar G_{AB}$}\sim {\cal O}(D),~~R|_\text{on $\bar G_{AB}$}\sim {\cal O}(D^2)$$
\end{itemize}

\subsection{Conditions imposed due to leading ansatz} 
Now we shall simply substitute our leading ansatz \eqref{curvespace} in \eqref{fineom} and shall compute the leading piece, keeping in mind the scaling rules we have mentioned above.  The details of the computation are all presented in appendix \eqref{appndx:source}. Here we shall only present the final result of this computation.
As mentioned before, the leading pieces in both ${\cal E}_{AB}$ and ${\cal E}_A$  turn out to be of order ${\cal O}(D^2)$. Up to corrections of order ${\cal O}(D)$, they have the following form
\begin{equation}\label{leadingeom}
\begin{split}
{\cal E}_{AB}& = \frac{1}{2}\left(O_B\nabla_A + O_A\nabla_B\right)\bigg[f\left(\nabla\cdot O - {DN\over\psi}\right)\bigg] +{f\over 2}\left[(O\cdot\nabla)f\right]\left[\nabla\cdot O- {DN\over\psi}\right]O_A O_B\\
&~~~- {1\over 2}\left(1 + Q^2 - 2 Q^2 \psi^{-D}\right) \left({DN\over\psi} -K\right)O_A O_B+ {\cal O}(D)\\
{\cal E}_M &=\left(\frac{DN}{\psi}\right)\tf\lb \( \frac{DN}{\psi}-K\) O^N+\(\n \cd O-\frac{DN}{\psi}\) n^N\rb\\
\text{where}&~\nabla \equiv\text{Covariant derivative w.r.t }~\bar G_{AB}\\
f &= (1+Q^2)\psi^{-D} -Q^2\psi^{-2D},~~~~\tilde f = Q\psi^{-D}\\
N &=\sqrt{(\partial_A\psi )(\partial_B\psi )\bar G^{AB}},~~~~n_A = \frac{\partial_A\psi}{N},~~~~K = \nabla\cdot n
\end{split}
\end{equation}
In equation \eqref{leadingeom} (and from now on throughout the paper) all raising, lowering and contraction of indices have been done using $\bar G_{AB}$. Note according to our scaling rules, both $(\nabla\cdot O)$ and $(K=\nabla\cdot n)$ are of order ${\cal O}(D)$ since they are divergences of order ${\cal O}(1)$ vectors $n_A$ and $O_A$.

From \eqref{leadingeom} we could easily see that the leading order piece will vanish, or our ansatz will solve the equation at the very leading order ($\mtO(D^2)$) provided
\begin{equation}\label{cond}
\begin{split}
 (K-DN)|_{\psi=1}&=  \mathcal{O}(1)\\ (\nabla.O-DN)|_{\psi=1}&=\mathcal{O}(1)
 \end{split} \end{equation}
Note that we have imposed the conditions only at $(\psi = 1)$. If we are finitely away from $\psi = 1$, the two equations will anyway vanish, because of the $\psi^{-D}$ factor in $f$ and $\tilde f$. To have a non-trivial effect on the equations, we can deviate from this hypersurface only in a power series in $\left(\frac{1}{D}\right)$ and therefore the effect  of such deviation will always be suppressed. We do have to take care of it in our subleading calculation, but at this order, it does not matter.

For convenience, we shall define a unit time-like vector field $u^A $ which is orthogonal to $n^A$ and defined as 
$u^A = n^A - O_A$. In terms of $u^A$ the second equation of \eqref{cond}  reads as
 \be{deldu}
 \n \cd u=\mtO(1) \footnotemark
\ee
 \footnotetext{As \eqref{deldu} comes as a direct consequence of \eqref{cond}, \eqref{deldu} also matches with the conditions imposed on $u^A$ in \cite{SB}}
 Note that in the language of our `$D$ scaling rules' , the  ${\cal O}(1)$ vector $u^A$ is a special case since its divergence is also of ${\cal O}(1)$ instead of being ${\cal O}(D)$.
 
 In summary, the final form of our leading ansatz is given by \eqref{curvespace}, where $\psi$ and $O_A$  satisfy the conditions given in \eqref{cond}.

\section{Subleading corrections:  The strategy}
Once we have our leading ansatz, we are ready to go for the subleading corrections. In this section, first, we shall describe the strategy very briefly, along with the conventions and the choices we shall be using for the solution. See \cite{SB}  for elaborate discussions on these points. Then we shall implement this strategy for our particular case to get the final set of coupled ODEs.

\subsection{Brief description of the algorithm}
In a nutshell,  the algorithm to determine the first subleading correction is as follows.

We have constructed an initial metric and gauge field ansatz $G^{(0)}_{AB}$, $A_M^{(0)}$ and have seen that $G^{[0]}_{AB}=\bar G_{AB}+G^{(0)}_{AB}$ and $A_M^{(0)}$  solve \eqref{fineom} at leading order (which is of $\mtO(D^2)$) provided the null one-form field $O_A$,  the scalar function $\psi$ satisfy \eqref{cond}  on $(\psi =1)$ hypersurface.  Hence, after we impose \eqref{cond}, both ${\cal E}_{AB}$ and ${\cal E}_M$, evaluated on $G^{[0]}_{AB}$ and $A_M^{(0)}$ would be of  order  $\mtO(D)$.  Let us denote these ${\cal O}(D)$ pieces as $S_{AB}$ and $S_M$ respectively. 

Now we add the first subleading corrections, namely  $\frac{1}{D}G^{(1)}_{AB}$  to the metric and $\frac{1}{D}A^{(1)}_M$ to the gauge field. A simple order counting suggests that the leading contribution of these corrections to \eqref{fineom} would also be of order $\mtO(D)$. Let us denote these ${\cal O}(D)$ pieces as $H_{AB}$ and $H_M$ respectively.  Clearly, $H_{AB}$ and $H_M$ would be some differential operators, acting on the unknown functions appearing in  $\frac{1}{D}G^{(1)}_{AB}$ and $\frac{1}{D}A^{(1)}_M$.  Their form would be universal in any order and could easily be computed by treating the subleading corrections as linear perturbations on $G^{[0]}_{AB}$ and $ A_M^{(0)}$.

Since  $H_{AB}$ and $H_M$ are of the same order as $S_{AB}$ and $S_M$, they could potentially cancel each other. This cancellation  gives the final differential equations for the first subleading correction, the schematic form of which
\begin{equation}\label{schemeeqn}
H_{AB}+S_{AB} = {\cal O}(1),~~~~H_M + S_M = {\cal O}(1)
\end{equation}

  Now we shall describe the conventions and gauge choices we have used to determine $H_{AB}$, $H_M$ and $S_{AB}$, $S_M$.

\subsection{Subsidiary conditions}
Following \cite{SB}  we shall write the final answer for $G^{(1)}_{AB}$ in terms of $\psi$, $Q$,  $O_A$ and their derivatives. The advantage of presenting the answer this way is that we never need to choose any specific coordinate system for the background, and therefore our solution will have explicit background covariance.\\
However, such a final answer does not make sense unless $\psi$, $Q$ and  $O_A$ are some known functions of the background.
Note that the conditions \eqref{cond} (which has to be satisfied only at $\psi =1$ hypersurface) are not enough to determine these two functions and the one-form everywhere in the space-time.  Therefore there is a huge ambiguity in fixing these functions everywhere. For convenience, we shall fix this ambiguity, by imposing some conditions (which, following \cite{SB,membrane,Chmembrane}, we shall refer to as `subsidiary conditions') on $\psi$ , $Q$ and $O_A$ externally. Our choice of subsidiary conditions  (these have to be satisfied everywhere in the background space-time) would be the following.
\begin{equation}\label{subcond}
\begin{split}
&\nabla^2\psi^{-D} =0,~~~(n\cdot\nabla)Q=0\\
&O\cdot O = 0,~~~O\cdot n =1,~~~(O\cdot\nabla) O_A\propto O_A
\end{split}
\end{equation}
Note that this choice of subsidiary conditions are consistent with \eqref{cond}, and also maintain `the background covariance', in the sense that to specify them we do not need to choose any coordinate system. \\
Being differential equations, the subsidiary conditions will fix these functions up to some boundary conditions. We shall specify the boundary conditions on $\psi=1$ hypersurface. In other words, given the shape of the $\psi=1$ hypersurface, and $Q$ and $O_A$  on this hypersurface, \eqref{subcond} will determine them everywhere in space, (i.e. at all $\psi\neq 1$)\footnote{ See \cite{radiation} for an explicit construction of $\psi$ in terms of the extrinsic curvature of the $\psi=1$ hypersurface in large-$D$ approximation.} . It will turn out that these boundary values  ( i.e., the $Q$ and projected $O_A$  fields on the hypersurface and its extrinsic curvature) are not completely free and the Einstein-Maxwell system could be solved only if they together satisfy some integrability conditions. These are the equations that govern the dynamics of the dual charged membrane and one of the key results of this paper.

\subsection{Gauge choice}
 In this section, we shall specify a choice of gauge\footnote{It is important to distinguish the subsidiary conditions from coordinate and $U(1)$ gauge covariance of the Einstein-Maxwell system. Here we have chosen to express our final answer in terms of some auxiliary functions $\psi$, $Q$ and $O_A$. The purpose of the subsidiary conditions is to fix or define these functions.  We have defined these auxiliary functions in a way so that we do not need to fix any particular coordinate system for our analysis. On the other hand, our gauge choice does fix the coordinate system upto some possible residual gauge invariance.} and shall parametrize the metric and gauge field correction accordingly. To fix the general coordinate invariance we need $D$ conditions on the metric and the $U(1)$ gauge fixing will require one condition on the gauge field. Our choice would be as follows
\begin{equation}\label{gauge}
\begin{split}
O^A G^{(1)}_{AB} =0,~~~O^M A_M^{(1)}=0
\end{split}
\end{equation}
The most general form of the metric and the gauge field correction consistent with \eqref{gauge} is
\begin{equation}\label{metgaugeform}
\begin{split}
G^{(1)}_{AB} &= {\cal G}^{(s_1)}~ O_A O_B + {\cal G}^{(s_2)}~ \left({P_{AB}\over D}\right) + \left( {\cal G}^{(v)}_A O_B + {\cal G}^{(v)}_B O_A\right) + {\cal G}^{(T)}_{AB}\\
A^{(1)}_M &= {\cal A}^{(s)}~O_M + {\cal A}^{(v)}_M
\end{split}
\end{equation}
where $P_{AB}$ is the projector perpendicular $n_A$ and $u_A$
$$P_{AB} = \bar G_{AB} -n_A n_B + u_A u_B$$ 
$ {\cal G}^{(v)}_A$,  ${\cal A}^{(v)}_M$ and ${\cal G}^{(T)}_{AB}$ satisfy the following conditions
$$n^A {\cal G}^{(v)}_A= u^A {\cal G}^{(v)}_A =n^M{\cal A}^{(v)}_M=u^M{\cal A}^{(v)}_M =0$$
$$n^A{\cal G}^{(T)}_{AB} = u^A{\cal G}^{(T)}_{AB} =0$$

\subsection{Reducing the PDE to ODE}
Naively if we compute the leading contribution of \eqref{metgaugeform} to \eqref{fineom}, we shall get linear \textit{partial} differential operators on the unknown functions (${\cal G}^{(s_1)}$,  ${\cal A}^{(s)}$ etc.) appearing in $G^{(1)}_{AB}$ and $A^{(1)}_M$.
However, the key simplification arises as follows. \\
 Suppose we choose the $\psi=\text{constant}$ surfaces to foliate the space-time. It turns out that at large $D$,  part of the metric and gauge field vary parametrically fast in the direction of increasing $\psi$, compared to the directions along the constant $\psi$ hypersurfaces. 
In other words, each component of the metric and the gauge field correction could be decomposed as a product of `fast-varying' and `slowly-varying' pieces. The derivatives of the `slowly-varying' pieces are parametrically suppressed. As a consequence, if we were to evaluate \eqref{fineom} on  the metric and gauge field correction at a given order of ${\cal O}\left(\frac{1}{D^k}\right)$, (assuming the system of equations have been solved up to ${\cal O}\left(\frac{1}{D^{k-1}}\right)$) `slowly-varying' pieces should be treated as constants. This reduces the complicated PDE of Einstein-Maxwell system to a set of in-homogeneous ODEs along the fast varying direction (namely the direction of increasing $\psi$) and the $\left(1\over D\right)$ expansion takes the form of an effective derivative expansion along these $\left[\psi = \text{constant}\right]$ hypersurfaces.

From the above discussion it is clear that  the first step  in determining the subleading corrections would be to decompose the metric and  gauge field  functions as  products of slowly and fast varying pieces as we have done below.
\begin{equation}\label{split}
\begin{split}
{\cal G}^{(s_1)}&=\sum_{i=1}^{N_S} S^{(i)}_1(\zeta)~ {\mathcal S}^{(i)},~~~~~{\cal G}^{(s_2)}=\sum_{i=1}^{N_S}  S^{(i)}_2(\zeta) ~{\mathcal S}^{(i)},~~~~~{\cal A}^{(s)}=\sum_{i=1}^{N_S}  a^{(i)}_s(\zeta)~ {\mathcal S}^{(i)}\\
{\cal G}^{(v)}_A &=\sum_{i=1} ^{N_V} \mathcal{V}^{(i)}(\zeta)~V^{(i)}_A,~~~~~{\cal A}^{(v)}_A = \sum_{i=1} ^{N_V}a^{(i)}_v(\zeta)~V^{(i)}_A,~~~~~{\cal G}^{(T)}_{AB} =\sum_{i=1}^{N_T} \mathcal{T}^{(i)}(\zeta)~t^{(i)}_{AB}
\end{split}
\end{equation}
Here $\zeta = D(\psi -1)$. Clearly $\zeta$ dependent parts are the fast varying pieces whose derivatives will have explicit factors of $D$. Each of these fast varying functions are multiplied by `slowly-varying'  scalar, vector and tensor structures. It turns out that at a given order there are only a finite number of slowly varying structures that could appear. Scalar structures are denoted by ${\mathcal S}^{(i)}$; vector structures, perpendicular to both $u_A$ and $n_A$ are denoted by $V^{(i)}_A$ and $t^{(i)}_{AB}$ denotes the traceless tensor structure, perpendicular to both $n_A$ and $u_A$. The total number of such  `slowly varying'  scalar, vector and tensor structures at order ${\cal O}(D)$ are denoted by $N_S$, $N_V$ and $N_T$ respectively.  A similar decomposition in terms of these scalar, vector and tensor structures is also possible for the sources $S_{AB}$ and $S_M$ (see the next section for more details). \\
 Once we substitute these decomposed sources, i.e the metric and the gauge field in the schematic equation \eqref{schemeeqn}, its different component reduce to second order  inhomogeneous ODEs for the unknown functions $S^{(i)}_1(\zeta)$,  $S^{(i)}_2(\zeta) $,  $a^{(i)}_s(\zeta)$ , $\mathcal{V}^{(i)}(\zeta)$, $a^{(i)}_v(\zeta)$ and $\mathcal{T}^{(i)}(\zeta)$.  As explained before, these are ODE s (as opposed to PDE s) simply because the derivatives of the slowly varying structures do not contribute at this order. Each structure could be treated as independent constant, thus decoupling the equations in different  superscript `$(i)$' sectors. Of course, along with this,  due to the symmetry of the equations, the scalar, vector and the tensor sector will decouple  as well in the usual way. These two types of decoupling  of the resultant ODEs lead to a vast simplification.
 With an appropriate choice of boundary conditions and a set of constraints on our scalar and vector data (which turns out to be the equation that governs the dynamics of the dual charged membrane - one of the main results of our paper), we could integrate them. 
 
 In a nutshell, this is how we determine the subleading corrections to the metric and the gauge field.
 
 \subsection{Boundary conditions}
As explained above, the relevant equations in the end would be a set of second order ODEs for each of the functions $S^{(i)}_1(\zeta)$,  $S^{(i)}_2(\zeta) $,  $a^{(i)}_s(\zeta)$ , $\mathcal{V}^{(i)}(\zeta)$,    $a^{(i)}_v(\zeta)$ and $\mathcal{T}^{(i)}(\zeta)$. Generically we need two boundary conditions for each of them to fix the integration constants. We shall impose them at  $\zeta\rightarrow \infty$ and $\zeta\rightarrow 0$ .

The fact that asymptotically the full space-time metric $G_{AB}$ should reduce to the background $\bar G_{AB}$ fixes the boundary condition at $\zeta\rightarrow\infty$ end. It simply says all the functions should vanish as $\zeta$ goes to $\infty$. It turns out that  $S^{(i)}_1(\zeta)$ could be unambiguously determined using this single condition. For the rest of the functions, we need another condition at  $\zeta\rightarrow 0$ end of the space-time.

Now, note that our final solution is parametrized by a dual system of a dynamical charged membrane, embedded in the asymptotic geometry, with a velocity field $u^A$ living on it. We need to unambiguously define these parameters in terms of our full space-time geometry and the gauge field containing the black hole. These definitions fix many of the integration constants. 

In the metric sector, we shall follow \citep{arbitrarydim} to fix our choices of parameter. We shall choose the $\psi=1$ hypersurface to be the horizon of the space-time and $u^A$ to be the null generator of the horizon to all orders in $\left(1\over D\right)$ expansion. As explained in \cite{yogesh1} and \cite{arbitrarydim}, this implies the following boundary conditions.
\begin{equation}\label{bc1}
\begin{split}
\lim_{\zeta\rightarrow 0}S^{(i)}_1(\zeta)=0,~~~~~\lim_{\zeta\rightarrow 0}\mathcal{V}^{(i)}(\zeta)=0~~~~\forall~~ i
\end{split}
\end{equation}
Once the membrane is defined, we shall have the all order definition of  the parameter $Q$ through the following equation.
\begin{equation}\label{qdef}
\begin{split}
Q={1\over \sqrt 2}\left(n^M A_M \right)\vert_{\psi=1}
\end{split}
\end{equation}
 
This definition fixes how $a^{(i)}_s(\zeta)$ should behave as $\zeta\rightarrow 0$
 \begin{equation}\label{bc2}
\begin{split}
\lim_{\zeta\rightarrow 0}a^{(i)}_s(\zeta) = 0
\end{split}
\end{equation}

For rest of the two functions $\mathcal{T}^{(i)}(\zeta)$ and $a^{(i)}_v(\zeta) $, the integration constants at $\zeta = 0$ ends are fixed by demanding that the solution has to be finite on the horizon.

 \section{First subleading correction: Explicit solution}
 In this section, we shall implement the strategy outlined in the previous section to determine the first subleading correction to the metric and the gauge field. We shall refer to the appendices \eqref{appndx:eom}, \eqref{appndx:source} and \eqref{appndx:homo} for some of the details of the calculation.\\
 
 \subsection{Classification of structures}
 As explained in the previous section the first step would be to classify the slowly-varying structures that can appear at the first subleading order.
 Below in table \ref{table:list1} we shall give a list of such structures that will appear in the final answer.
\begin{table}[ht]\caption{List of membrane data} 
\vspace{0.5cm}
\centering 
\begin{tabular}{|c| c| c|}
\hline
Scalar&Vector&Tensor\\
\hline
\hline
${\cal S}^{(1)}\equiv{(u\cdot\nabla)K\over K}$&$V^{(1)}_A\equiv P^C_A\left({\nabla_C K\over K}\right)$&$t_{AB}\equiv  P^C_AP^{C'}_B\left[\left(\frac{\nabla_C O_{C'} +\nabla_{C'}O_C}{2}\right) - {P_{CC'}\over D}\left(\nabla\cdot O\right)\right]$\\
\hline
${\cal S}^{(2)}\equiv (u\cdot K\cdot u)$&$V^{(2)}_A\equiv P^C_A(u\cdot\nabla) O_C$&\\
\hline
${\cal S}^{(3)}\equiv (\hat\nabla\cdot u)$&$V^{(3)}_A\equiv P^C_A(u\cdot\nabla) u_C$&\\
\hline
${\cal S}^{(4)}\equiv {\hat\nabla^2K\over K^2}$&$V^{(4)}_A\equiv P^C_A\left({\hat\nabla^2u_C\over K}\right)$&\\
\hline
${\cal S}^{(5)}\equiv \left(u\cdot{\cal Q}\right)$&&\\
\hline
${\cal S}^{(6)}\equiv{ \nabla\cdot{\cal Q}\over K}$&&\\
\hline
${\cal S}^{(7)}\equiv {K\over D}$&&\\
\hline
\end{tabular}\vspace{.5cm}
\label{table:list1} 
\end{table}
\noindent
Here  ${\cal Q}_A \equiv {\nabla_A Q\over Q}$.\\

Apart from the list of scalars mentioned in table (\ref{table:list1}), we could also have some scalar terms proportional to the background curvature. However, we are not giving a separate name to such terms and shall be writing them explicitly whenever they occur.

\subsection{Source}
Now we shall write the explicit expression for the source $S_{AB}$ and $S_M$ at first subleading order, which turns out to be of order ${\cal O}(D)$. We shall decompose the sources into different components.

\begin{equation}\label{splitsource}
 \begin{split}
S_{AB} \equiv&~K\bigg[{\mathfrak S}_0 ~n_A n_B+{\mathfrak S}_1 ~O_A O_B + {\mathfrak S}_2 ~(n_A O_B + n_B O_A) + {\mathfrak S}_3~ P_{AB}\\
&~~~~~~+({\mathfrak V}^{(1)}_A O_B +{\mathfrak V}^{(1)}_B O_A) + ({\mathfrak V}^{(2)}_A n_B +{\mathfrak V}^{(2)}_B n_A)  + {\mathfrak T}_{AB}\bigg]\\
\\
S_M \equiv&~K\bigg[{\mathfrak A}^{(1)} ~n_M + {\mathfrak A}^{(2)} ~O_M + {\mathfrak A}_M\bigg]
 \end{split}
 \end{equation}
 
 Now the different components could be further decomposed into different scalar vector and tensor structures as appeared in the table (\ref{table:list1}).
 
  \begin{equation}\label{splitTerm}
 \begin{split}
{\mathfrak S}_0=~&0,~~~~~{\mathfrak S}_3=0~~~~~{\mathfrak V}_A^{(2)}=0\\
{\mathfrak S}_1=~&\tilde{f}^2 (1-f) \left[ {\cal S}^{(1)}-{\cal S}^{(2)}+{\cal S}^{(7)}\right] + {f\over 2}\left(f -\tilde f^2\right){\cal S}^{(3)} +\left(Q \tf - \tilde f^2\right)\left(f{\cal S}^{(5)}-{\cal S}^{(6)}\right)\\
{\mathfrak S}_2=~&{f\over 2}\left[{\cal S}^{(3)} \right] +{\tilde f^2}\left(-{\cal S}^{(1)} +{\cal S}^{(2)}-{{\cal S}^{(3)}\over 2}-\mathcal{S}^{(7)}\right) + \left(Q \tf - \tilde f^2\right){\cal S}^{(5)}\\
{\mathfrak V}_A^{(1)}=~&{f\over 2} \left[- V^{(1)}_A  + V^{(2)}_A + V^{(4)}_A\right] +{\tilde f^2\over 2}\left[V^{(1)}_A -
V^{(3)}_A\right]\\
{\mathfrak T}_{AB}=~&\tilde f^2~t_{AB}
 \end{split}
 \end{equation}
 \begin{equation}\label{splitTermM}
 \begin{split}
{\mathfrak A}^{(1)}=~&-\tilde f\left[{\cal S}^{(1)} -{\cal S}^{(2)} + {\cal S}^{(3)}  + {\cal S}^{(5)}\right],~~~~~~
{\mathfrak A}^{(2)}=~\tilde f\left[{\cal S}^{(6)} +\frac{\bar R_{uu}}{K}\right]\\
{\mathfrak A}_A=~&\tilde f\left[ V^{(1)}_A  - V^{(2)}_A - V^{(4)}_A\right] \\
 \end{split}
 \end{equation}
Here, $\bar{R}_{uu}=u^A \: \bar{R}_{AB} \: u^B$ and $\bar{R}_{AB}=(D-1)\lambda \bar{G}_{AB}$ is the Ricci tensor w.r.t the background metric $\bar{G}_{AB}$.

\subsection{Homogeneous part }
As mentioned before, the schematic form of the equations at order ${\cal O}(D)$ is
$${\cal E}_{AB} \Rightarrow H_{AB} + S_{AB} \approx 0,~~~~~~{\cal E}_M \Rightarrow H_M + S_M \approx 0$$
where $H_{AB}$ and $H_M$ consist of the  linear differential operators acting on the unknown functions appearing  in equation \eqref{metgaugeform} and $S_{AB}$ and $S_M$ are the sources that do not depend on $G^{(1)}_{AB}$ and $A_M^{(1)}$. \\
Now once we have  decomposed the functions into slow and fast pieces (see  equation \eqref{split}),  the homogeneous part - $H_{AB}$ and $H_M$ reduce to ordinary linear differential operator on the fast varying functions (the functions that depend on $\zeta$).
Below we are presenting the final form of $H_{AB}$ and $H_M$. See appendix \eqref{appndx:homo} for the detailed derivation.

For convenience, we shall first decompose $H_{AB}$ and $H_M$  in the following way.
\begin{equation}\label{decomph}
\begin{split}
H_{AB}&=H_{AB}^{\text{scalar}}+H_{AB}^{\text{vector}}+H_{AB}^{\text{trace}}+H_{AB}^{\text{tensor}}+H_{AB}^{\text{Gauge scalar}}+H_{AB}^{\text{Gauge vector}}\\
H^M&=H^{M}_{\text{scalar}}+H^{M}_{\text{vector}}
+H^{M}_{\text{trace}}+H^{M}_{\text{tensor}}+H^{M}_{\text{Gauge scalar}}+H^{M}_{\text{Gauge vector}}
\end{split}
\end{equation}

where,
\begin{eqnarray}
    \nonumber H_{AB}^{\text{scalar}} &=& \frac{DN^2}{2}(f-1)\sum_i\left( \ddot{S}^{(i)}_1 +\dot{S}^{(i)}_1\right)\mathcal{S}^{(i)} O_A O_B\\
     && + \frac{DN^2}{2}\sum_i\left( \ddot{S}^{(i)}_1 + \dot{S}^{(i)}_1 \right)\mathcal{S}^{(i)} (n_A O_B + n_B O_A)
\end{eqnarray}
\begin{eqnarray}
    \nonumber H_{AB}^{\text{trace}} &=& \sum_i -\left(\frac{DN^2}{4}\right)(f-\tf^2)\dot{S}_2^{(i)} [(f-1) O_A O_B + (n_A O_B + n_B O_A)] -\left(\frac{DN^2}{2}\right)\ddot{S}_2^{(i)} n_A n_B \\
  \nn &&  -\left(\frac{N^2}{2}\right)\lb \( (1-f)\ddot{S}^{(i)}_2+\dot{S}_2^{(i)}\)\mathcal{S}^{(i)} P_{AB}+2 \dot{S}_2^{(i)}\mathcal{S}^{(i)}n_A n_B\rb+\left(\frac{N^2}{2}\right)\tf^2 \dot{S}_2^{(i)}\mathcal{S}^{(i)}P_{AB}+\mathcal{O}(1)\\
\end{eqnarray}
\begin{eqnarray}
    \nonumber H_{AB}^{\text{vector}} &=&\left(\frac{DN^2}{2}\right)\sum_i\left( \ddot{\mathcal{V}}^{(i)} + \dot{\mathcal{V}}^{(i)}\right)
    \lb f (O_A V^{(i)}_B + O_B V^{(i)}_A)+(u_A V^{(i)}_B + u_B V^{(i)}_A)\rb\\
    &&+\left( \frac{N}{2}\right)\sum_i(\nabla.V^{(i)})\lb \dot{\mathcal{V}}^{(i)}(O_A n_B + O_B n_A)-(f-\tf^2)\mathcal{V}^{(i)}O_A O_B \rb + \mathcal{O}(1)~~~~~~~~
\end{eqnarray}
\begin{eqnarray}
    \nonumber H_{AB}^{\text{tensor}} &=& -\frac{DN^2}{2}\sum_i\left[ (1-f)\ddot{\tau}^{(i)} + \dot{\tau}^{(i)}\right] t^{(i)}_{AB} + \frac{N}{2}\sum_i\dot{\tau}^{(i)}[n_B \nabla_C (t^{(i)C}_A)+n_A \nabla_C (t^{(i)C}_B)]\\
     &&  + \frac{DN^2}{2} \tf^2 \sum_i\dot{\tau}^{(i)} t^{(i)}_{AB} 
\end{eqnarray}
\begin{eqnarray}
   H_{AB}^{\text{Gauge scalar}} &=& -DN^2 \sum_i 2 \dot{a}_s^{(i)}  \tilde{f} \mathcal{S}^{(i)}[n_A O_B + n_B O_A + (f-1)O_A O_B]
\end{eqnarray}
\begin{eqnarray}
  H_{AB}^{\text{Gauge vector}} &=& -DN^2\sum_i \dot{a}_v^{(i)} \tilde{f}\lb u_A V_B^{(i)} +u_B V_A^{(i)} + f(O_A V_B^{(i)} +O_B V_A^{(i)} \rb 
\end{eqnarray}

And for the second equation in \eqref{fineom},
$$H^{M}=H^{M}_{\text{scalar}}+H^{M}_{\text{vector}}
+H^{M}_{\text{trace}}+H^{M}_{\text{tensor}}+H^{M}_{\text{Gauge scalar}}+H^{M}_{\text{Gauge vector}}$$
where,
\be{}
H^N_\text{scalar}=0
\ee

\be{}
H^N_\text{trace}=\frac{DN^2\tdf}{2}\sum_{i} \dot{S_2}^{(i)} u^N \dot{S}_2
\ee

\be{}
H^N_\text{vector}=N\tdf\sum_{i} \mathcal{V}^{(i)}(\n.V^{(i)}) O^N-\frac{DN^2\tdf}{\psi}\sum_{i} \dot{\mathcal{V}}^{(i)}V^{(i)N}
\ee
\be{}
H^N_\text{tensor}=0
\ee
\be{}
H^N_\text{gauge scalar}=-\sum_i DN^2(\ddot{a}_s^{(i)}+\dot{a}_s^{(i)})\mathcal{S}^{(i)}u^N
\ee
\be{}
H^N_\text{gauge vector}=\sum_i \lb DN^2\left(\ddot{a}_v^{(i)}(1-f)+\dot{a}_v^{(i)}(1-\tf^2)\right)\rb V^{(i)N}+N(fO^N-n^N)(\n \cd V^{(i)}) \dot{a}_V^{(i)}
\ee

\subsection{Decoupling the ODE s}
The set of coupled ODE s mentioned in the previous subsection could easily be decoupled, by taking appropriate projection. In this subsection we shall present the decoupled ODEs  and their solutions in the form of integrals. 
\subsubsection{Trace-less tensor sector in the metric correction} 
Consider the following combination,
\be{}
P^A_D(H_{AB}+S_{AB})P^B_C=0
\ee
This combination reduces to the decoupled ODE s for ${\cal T}^{(i)}(\zeta)$ s . Now from the table (\ref{table:list1}) we could see that there exists only one tensor structure at this order, or in other words, according to the notation of equation \eqref{split}, $N_T =1$. To unclutter the notation in this case, let us denote ${\cal T}^{(i)}(\zeta)$, simply by ${\cal T}(\zeta)$ without the subscript. The relevant equation turns out to be the following.
\begin{equation}\label{tensor}
\begin{split}
~&(1-f)~\ddot{\cal T}+(1-\tilde f^2)~\dot{\cal T}-\frac{2D}{K}\tilde f^2=0\\
\Rightarrow~&e^{-\zeta}\frac{d}{d\zeta}\left[e^\zeta~(1-f)~ \dot{\cal T}\right] -\frac{2D}{K}\tilde f^2 =0
\end{split}
\end{equation}
After imposing the boundary conditions (finite at $\zeta=0$ hypersurface and vanishing as $\zeta\rightarrow\infty$) this equation could be integrated .
\begin{equation}\label{soltensor}
\begin{split}
{\cal T}(\zeta) =2\left(D\over K\right)\log\left[1-Q^2e^{-\zeta}\right]
\end{split}
\end{equation}
For $Q\to 0$ limit, the traceless tensor sector correction vanishes, which is consistent with  \cite{SB}. 
\subsubsection{Vector sector } 
Consider the following three combinations
\begin{equation}\label{veceq}
\begin{split}
O^A{\cal E}_{AC}P^C_B
\equiv~&\sum_i\bigg[ \ddot{\mathcal{V}}^{(i)}+\dot{\mathcal{V}}^{(i)}-2\dot{a}_v^{(i)}\tf\bigg]V_B^{(i)}
+ \dot{\cal T}  \left(\nabla_C t^C_B\over K\right)=0\\
u^A{\cal E}_{AC}P^C_B\equiv~&(f-1)\sum_i\bigg[\left(\frac{N}{2}\right)\left( \ddot{\mathcal{V}}^{(i)}+\dot{\mathcal{V}}^{(i)}-2\dot{a}_v^{(i)}\tf\right) \bigg]V_B^{(i)}+ {\mathfrak V}_B^{(1)}=0\\
{\cal E}^AP_{AB}\equiv~& -\sum_iN\left[ \tf \dot{\mathcal{V}}^{(i)}- \left( \ddot{a}_v^{(i)}(1-f) + \frac{\dot{a}_v}{\psi}(1-\tf^2)\right)\right] V^{(i)}_B + {\mathfrak A}_B=0
\end{split}
\end{equation}
These three equations give a set of coupled ODE s for the unknown functions ${\cal V}^{(i)}$ and $a_v^{(i)}$. Note that, by construction, these equations are decoupled for each independent vector structure $V_A^{(i)}$ appearing in the source ${\mathfrak V}_A^{(1)}$ and ${\mathfrak A}_A$, i.e., the superscript $i$ s are not mixed. However, for a given $i$,  we have to do some work to decouple ${\cal V}^{(i)}$ and $a_v^{(i)}$. Moreover, we have three equations for two unknown functions, leading to the following consistency constraint.
\begin{equation}\label{consistency}
\begin{split}
 (f-1)\dot{\cal T} \left(\nabla_C t^C_A\over 2K\right)=\left(D\over K\right){\mathfrak V}^{(1)}_A + {\cal O}\left( \frac{1}{D}\right)
\end{split}
\end{equation}\\
From the theory of `constraint equations' in any gauge theory, it follows that if we satisfy such constraints along one constant $\zeta$  hypersurface,  they will be satisfied everywhere, provided we solve the dynamical equations ( in this case the equations involving two  derivatives w.r.t $ \zeta$  and therefore determining  the `$\zeta$ evolutions'  of the unknown functions ) correctly \cite{WaldBook}.
Here we shall impose these constraints on $\zeta=0$ or $\psi=1$ hypersurface, which will lead to the constraint equations on our membrane data.\\

Note that the LHS of equation \eqref{consistency} vanishes at $(f=1)$  (or equivalently $\psi =1$) hypersurface which further implies that the combination of vector structures ${\mathfrak V}^{(1)}_A$, which generically should be of ${\cal O}(1)$ 
\footnote{{The source $S_{AB} \: \text{and}\: S_{M}$} are of $\mtO(D)$ with a clear $\mtO(D)$ factor $K$ in the front. Hence the different decomposed components are of $\mtO(1)$ in general.},
 turns out to be of order ${\cal O}(\frac{1}{D})$ because of equation \eqref{consistency}.
\begin{equation}\label{membranevec}
\begin{split}
&{\mathfrak V}^{(1)}_A\vert_{\psi=1} ={\cal O}\left( \frac{1}{D}\right)\\
\Rightarrow~& \left[- V^{(1)}_A  + V^{(2)}_A + V^{(4)}_A\right] +Q^2\left[V^{(1)}_A -
V^{(3)}_A\right]={\cal O}\left( \frac{1}{D}\right)\\
\text{More explicitly,} ~~ & \Big [ \frac{\hat{\dl} ^2 u_\nu}{\mtK}-(1-Q^2)\frac{\hat{\dl}_\nu \mtK}{\mtK}+u^\alpha K_{\alpha\nu}-(1+Q^2)(u.\hat{\dl} u_\nu)\Big]\mathcal{P}^\nu_\mu=0
\end{split}
\end{equation}
This is clearly one of the `integrability condition' for our set of ODEs. This  membrane equation is one of the key results of our paper. This is one of those equations that govern the coupled dynamics of the membrane's shape, its charge and the velocity field. Here, $\hat{\n}$ is the covariant derivative wrt to the intrinsic membrane coordinates denoted by the Greek indices ${\mu,\nu}$.$$\mathcal{P}^\nu_\mu=\text{projector perpendicular to } u_\mu=\delta^\nu_\mu+u^\nu u_\mu$$
$\mtK_{\mu \nu}$ is the extrinsic curvature of the membrane in terms of the membrane coordinates and $\mtK$ is the trace of the extrinsic curvature.

As we have explained, this membrane equation ensures that \eqref{consistency} is satisfied only at the hypersurface $\psi =1$. However, for the consistency of the set of ODE s we need \eqref{consistency} to be satisfies at every value of $\psi$. This is an internal consistency test of  our systems of equations. It  is a consequence of the theory of `constraint equations' in any theory with gauge invariance. 
For our case we could easily verify it and the proof is given in appendix \eqref{constraints}.
\\
Next, we proceed to decouple and integrate. We shall use the second and the third equation for this purpose. Let us first  introduce some new notation to  denote the decomposition of  the sources in our basis of vector structure as given in table (\ref{table:list1})
\begin{equation}\label{vecs}
\begin{split}
&{\mathfrak V}^{(1)}_A \equiv \sum_i v_\text{metric}^{(i)}(\zeta)~V_A^{(i)},~~~~~{\mathfrak A}_A \equiv \sum_i v_\text{gauge}^{(i)}(\zeta)~V_A^{(i)},\\
&\text{where}\\
&v_\text{metric}^{(1)} = -{f-\tilde f^2\over 2},~~v_\text{metric}^{(2)} =v_\text{metric}^{(4)}={f\over 2},~~v_\text{metric}^{(3)} =-{\tf^2\over 2}\\
&v_\text{gauge}^{(1)}= \tf,~~v_\text{gauge}^{(2)}=-\tf,~~v_\text{gauge}^{(3)}=0,~~v_\text{gauge}^{(4)}=-\tf\\
\end{split}
\end{equation}
Here the $\zeta$ dependence of $f$ and $\tf$ are given as 
\be{}
f=(1+Q^2) e^{-\zeta}-Q^2 e^{-2\zeta},\:\:\: \tf=Q e^{-\zeta}
\ee

Above notation would help us to decouple different `$i$' sectors in \eqref{veceq}. The second equation of \eqref{veceq} will take the following form. \footnotemark 
\footnotetext{For $Q\to0$ limit, the second term of second equation of \eqref{veceq} gives back the vector constraint on the membrane data for the uncharged case and it is also suppressed by $\mathcal{O}(\frac{1}{D})$ outside $\psi=1$. Then the vector sector correction in the metric becomes zero. For more discussion on this, see \cite{SB}.}
\begin{equation}\label{2ndeq}
\begin{split}
&{d\over d\zeta}\left[e^\zeta~\dot {\cal V}^{(i)} - 2Q ~a_v^{(i)}\right] = 2 \left(D\over K\right)\left[e^\zeta ~v^{(i)}_{metric}\over{ 1-f}\right]\\
\Rightarrow~&\dot {\cal V}^{(i)} = 2\tf a_v^{(i)} + 2\left(D\over K\right)e^{-\zeta}\int_0^\zeta d\rho \left[e^\rho ~v^{(i)}_{metric}(\rho)\over{ 1-f(\rho)}\right]
\end{split}
\end{equation}
Substituting \eqref{2ndeq} in the third equation of  \eqref{veceq} we get the decoupled equation for $a_v^{(i)}$
\begin{equation}\label{3rdeq}
\begin{split}
 &2\tf^2 a_v^{(i)} - \left( \ddot{a}_v^{(i)}(1-f) + \frac{\dot{a}^{(i)}_v}{\psi}(1-\tf^2)\right)= \left(D\over K\right)\left(v_{gauge}^{(i)}(\zeta)- 2 e^{-\zeta} \int_0^\zeta d\rho \left[e^\rho ~v^{(i)}_{metric}(\rho)\over{ 1-f(\rho)}\right]\right)\\
 \Rightarrow~&{d\over d\zeta}\bigg[e^{3\zeta}(1-f)(f-\tf^2)^2~{d\over d\zeta}\left(e^{-\zeta}~a_v^{(i)}\over {f-\tf^2}\right)\bigg]= -{\mathfrak v}^{(i)}(\zeta)\\
\text{where}~~&{\mathfrak v}^{(i)}(\zeta)\equiv\left(D\over K\right)e^{2\zeta}(f-\tf^2)\left({ 2 e^{-\zeta}}\int_0^\zeta d\rho \left[e^\rho ~v^{(i)}_{metric}(\rho)\over{ 1-f(\rho)}\right]-v_{gauge}^{(i)}(\zeta)\right)\\
\end{split}
\end{equation}
Equation \eqref{3rdeq} could be easily integrated and therefore \eqref{2ndeq} could be easily integrated. 
\begin{equation}\label{finsolvec}
\begin{split}
&a_v^{(i)}(\zeta)= e^\zeta~ [f -\tf^2]\int_\zeta^\infty\bigg({e^{-3\rho}\over{[1-f]~[f-\tf^2]^2}}\bigg)\int_0^\rho d\rho' ~{\mathfrak v}^{(i)}(\rho')\\
& {\cal V}^{(i)}(\zeta) = -2\int_{\zeta}^\infty d\rho~\tf(\rho)~ a_v^{(i)}(\rho)  - 2\left(D\over K\right)\int_{\zeta}^\infty d\rho ~e^{-\rho}\int_0^\rho d\rho' \left[e^{\rho' }~v^{(i)}_{metric}(\rho')\over{ 1-f(\rho')}\right]
\end{split}
\end{equation}

\subsubsection{Scalar sector}
For every $i$ (i.e., for every independent scalar structure in table \eqref{table:list1}) we have  unknown functions in the scalar sector, namely $S_1^{(i)}$,  $S_2^{(i)}$ and $a_s^{(i)}$. But there are 5 equations in the scalar sector. Therefore naively the consistency demands two constraints.

Below we are first quoting the equations relevant in the scalar sector.
\begin{equation}\label{scaleqmetric}
\begin{split}
O^A{\cal E}_{AB}O^B
\equiv~& - \frac{N}{2}\sum_{i=1}^{N_S}\left(\ddot{S}^{(i)}_2\right)\mathcal{S}^{(i)}+{\mathfrak S}_0 =0\\
u^A{\cal E}_{AB}u^B
\equiv~& \frac{N}{2}(f-1)\sum_i\bigg[\left( \ddot{S}_1^{(i)}+\dot{S}_1^{(i)}\right)-{1\over 2}(f-\tf^2)\dot{S}_2^{(i)}-4\tf~ \dot{a}_s^{(i)}\bigg]\mathcal{S}^{(i)}\\
&-\frac{N}{2}(f-\tf^2)\sum_i^{N_V}\mathcal{V}^{(i)}\left(\n \cd V^{(i)}\over K\right) +{\mathfrak S}_1 =0\\
O^A{\cal E}_{AB}u^B
\equiv~& \frac{N}{2}\sum_{i=1}^{N_S}\bigg[\left( \ddot{S}_1^{(i)}+\dot{S}_1^{(i)}\right)-\frac{1}{2 }(f-\tf^2)\dot{S}_2-4\tf~ \dot{a}_s^{(i)}\bigg]\mathcal{S}^{(i)}\\
&+\frac{N}{2}\sum_{i=1}^{N_V}\dot{\mathcal{V}}^{(i)}\left(\n \cd V^{(i)}\over K\right) + {\mathfrak S}_2=0\\
\end{split}
\end{equation}
\begin{equation}\label{scaleqguage}
\begin{split}
O^M{\mathcal E}_M\equiv~&\frac{N}{2}\sum_{i=1}^{N_S}\bigg[\tf \dot{S}_2-2 \left( 
\ddot{a}_s^{(i)}+\dot{a}_s^{(i)}\right)\bigg]\mathcal{S}^{(i)}-N\sum_{i=1}^{N_V} \dot{a}_v^{(i)}\left(\n \cd V^{(i)}\over K\right) + {\mathfrak A}^{(1)}=0\\
u^M{\mathcal E}_M\equiv~&-\frac{N}{2}\sum_{i=1}^{N_S}\bigg[\tf \dot{S}_2-2 \left( 
\ddot{a}_s^{(i)}+\dot{a}_s^{(i)}\right)\bigg]\mathcal{S}^{(i)}\\
&+N\sum_{i=1}^{N_V}\bigg[\tf \mathcal{V}^{(i)} +f\dot{a}_v^{(i)}\bigg]\left(\n \cd V^{(i)}\over K\right) + {\mathfrak A}^{(2)}=0\\
\end{split}
\end{equation}

The consistency of this set of 5 equations demands
\begin{equation}\label{consisscal}
\begin{split}
{\cal C}_\text{metric}\equiv~&\left[(1-f)O^A +u^A\right]u^B {\cal E}_{AB} \\
\equiv~&{N\over 2}\sum_{i=1}^{N_V}\bigg[(1-f)\dot{\mathcal{V}}^{(i)} - (f-\tf^2){\mathcal{V}}^{(i)}\bigg] \left(\n \cd V^{(i)}\over K\right)+ (1-f)~{\mathfrak S}_2 + {\mathfrak S}_1 ={\cal O}\left( \frac{1}{D}\right)\\
{\cal C}_\text{guage}\equiv~&\left[O^A +u^A\right]{\cal E}_{A}\\
\equiv~&N\sum_{i=1}^{N_V}\bigg[\tf \mathcal{V}^{(i)} -(1-f)\dot{a}_v^{(i)}\bigg]\left(\n \cd V^{(i)}\over K\right) + {\mathfrak A}^{(1)}+ {\mathfrak A}^{(2)}={\cal O}\left( \frac{1}{D}\right)
\end{split}
\end{equation}
Note that on $\psi =1$ hypersurface the above two constraints \eqref{consisscal} \footnote  {Actually there is one more constraint. Note that if we take projected trace of $H_{AB}$, then it is of order ${\cal O}(D)$, whereas the naive $D$ counting suggests that it should be of order ${\cal O}(D^2)$ since $H_{AB}$ itself is of order ${\cal O}(D)$. This imposes a serious constraint on the  source at order ${\cal O}(D)$, namely 
$${P^{AB}\over D}S_{AB}={\mathfrak S}_3= {\cal O}(1)$$
From equation \eqref{splitTerm} we could see that this is indeed true.} reduces to the following
\begin{equation}\label{scalmem}
\begin{split}
&{\mathfrak S}_1\vert_{\psi=1} ={\cal O}\left( \frac{1}{D}\right),~~~~~~\bigg[{\mathfrak A}^{(1)} + {\mathfrak A}^{(2)}\bigg]_{\psi=1} ={\cal O}\left( \frac{1}{D}\right)\\
&\text{More explicitly},~~\hat{\n}\cd u={\cal O}\left( \frac{1}{D}\right)\\
& \frac{\hat{\nabla}^2 Q}{\mtK} - u.\hat{\dl} Q - Q\left[ \frac{u.\hat{\dl}\mtK}{\mtK} - u.\mtK.u  - \frac{\bar{R}_{uu}}{\mtK}\right] ={\cal O}\left( \frac{1}{D}\right)
\end{split}
\end{equation}
These two are the two scalar constraints on the membrane dynamics. We shall refer to them as  scalar and charge membrane equations.

Equation \eqref{membranevec} and the two equations in \eqref{scalmem}  are the key results of our paper.  Note that in \eqref{membranevec} and in the first equation of \eqref{scalmem},  if we just replace the covariant derivatives  $\nabla$ with partial derivatives $\partial$,  they reduce to two of the constraint equations derived in \cite{membrane}. In other words, these two equations are the simplest possible covariantization of their `flat-space' counterpart. However, the same could not be said about the second equation in \eqref{scalmem} as it includes a term proportional to the Ricci tensor with both indices projected along the $u$ direction. The charge conservation equation does depend on the background curvature ( hence, the cosmological constant) in a non-trivial way. Note also, that without the charge there is no such non-trivial curvature dependence in the membrane equations (see \cite{SB}. Our equations, in $Q \rightarrow 0$ limit,  very simply reduce to that of \cite{SB}). \\

We are yet to solve for the scalar sector. However, before going into the solution of the scalar sector, we want to check that the equations \eqref{consisscal} are satisfied for all $\psi$,  which, in abstract terms, is a consequence of gauge and coordinate invariance of our theory. It is a similar kind of consistency check, as we did for the vector sector. For easier understanding, we have explicitly showed the verification of  the equations  \eqref{consisscal} in appendix \eqref{constraints}. \\
Once the consistency is ensured we could proceed to decouple and integrate the equations. We shall start with the solution of $S_2^{(i)}$. The first equation in \eqref{scaleqmetric} gives the decoupled ODE for this unknown structure. Using the fact that ${\mathfrak S}_0$ vanishes in our case, the equation reduces to
\begin{equation}\label{eqs2}
\begin{split}
\ddot S_2^{(i)}=0~\Rightarrow ~S_2^{(i)}= k_1 ~\zeta + k_2,~~~\{k_1,k_2\}=\text{arbitrary constants}
\end{split}
\end{equation}
Now the boundary condition $S_2\rightarrow0$ as $\zeta\rightarrow\infty$ simply set both $k_1$ and $k_2$ to zero. So finally
$$S_2^{(i)}=0 ~~\text{for every}~~i$$

Now we shall solve for $S_1^{(i)}$ and $a_s^{(i)}$. We shall use the third and the second equations of \eqref{scaleqmetric} and \eqref{scaleqguage}. Note that in the second equation of \eqref{scaleqguage}, once we substitute the solution for $S_2^{(i)}$, it gives a decoupled ODE for $a_s^{(i)}$, which we could easily integrate. Substituting the solution of $a_s^{(i)}$ and $S_2^{(i)}$ in the third equation of \eqref{scaleqmetric}, we get the decoupled ODE for $S_1^{(i)}$, which again has a very simple homogeneous piece and therefore could be integrated easily.

Before carrying out these steps let us introduce few notations to unclutter our equations. As in vector sector, we shall decompose the sources in the third and the second equations of \eqref{scaleqmetric} and \eqref{scaleqguage} in terms of different scalar structures listed in table (\ref{table:list1}).

\begin{equation}\label{scaldecomp}
\begin{split}
&\frac{N}{2}\sum_{i=1}^{N_V}\dot{\mathcal{V}}^{(i)}\left(\n \cd V^{(i)}\over K\right) + {\mathfrak S}_2 \equiv\sum_{i=i}^{N_S} s_\text{metric}^{(i)}(\zeta) ~{\cal S}^{(i)}\\
&N\sum_{i=1}^{N_V}\bigg[\tf \mathcal{V}^{(i)} +f\dot{a}_v^{(i)}\bigg]\left(\n \cd V^{(i)}\over K\right)  + {\mathfrak A}^{(2)}\equiv\sum_{i=i}^{N_S} s_\text{gauge}^{(i)}(\zeta) ~{\cal S}^{(i)}\\
\text{where}&\\
&s^{(1)}_\text{metric}=\bigg[\frac{N}{2}\bigg(\dot{\mathcal{V}}^{(2)}+\dot{\mathcal{V}}^{(3)}+\dot{\mathcal{V}}^{(4)}\bigg)-\tf^2\bigg],~~
s^{(2)}_\text{metric}=\bigg[\frac{N}{2}\bigg(-\dot{\mathcal{V}}^{(2)}+\dot{\mathcal{V}}^{(3)}\bigg)+\tf^2\bigg]\\
&s^{(3)}_\text{metric}=\bigg[\frac{f-\tf^2}{2}\bigg],~~ s^{(4)}_\text{metric}=\bigg(\frac{N}{2} \dot{\mathcal{V}}^{(1)}\bigg)\\
&s^{(5)}_\text{metric}=(Q\tf-\tf^2),~~s^{(6)}_\text{metric}=0,~~s^{(7)}_\text{metric}=-\tf^2,~~s^{R_{uu}}_\text{metric}=\bigg[\frac{N}{2\, K}\bigg(-\dot{\mathcal{V}}^{(2)}+\dot{\mathcal{V}}^{(3)}\bigg)\bigg]\\
&s^{(1)}_\text{gauge}=N\lb \tf \mathcal{V}^{(2)}+f \dot{a}_v^{(2)}+ \tf \mathcal{V}^{(4)}+f \dot{a}_v^{(4)}\rb,~~s^{(2)}_\text{gauge}=N\lb -\tf \mathcal{V}^{(2)}-f \dot{a}_v^{(2)}+ \tf \mathcal{V}^{(3)}+f \dot{a}_v^{(3)}\rb\\
&s^{(4)}_\text{gauge}=N\lb \tf \mathcal{V}^{(1)}+f \dot{a}_v^{(1)}\rb, ~~s^{(6)}_\text{gauge}=\tf,~~ s^{(3)}_\text{gauge}=s^{(5)}_\text{gauge}=s^{(7)}_\text{gauge}=0\\
&s^{R_{uu}}_\text{gauge}=\frac{N}{K}\lb -\tf \mathcal{V}^{(2)}-f \dot{a}_v^{(2)}+ \tf \mathcal{V}^{(3)}+f \dot{a}_v^{(3)}\rb+\tf
\end{split}
\end{equation}

In this new notations it is easier to decouple the different `i' sectors, and  the third and the second equation  of \eqref{scaleqmetric} and \eqref{scaleqguage} take the following form.
\begin{equation}\label{finscaleq}
\begin{split}
&{d\over d\zeta}\left(e^\zeta\dot{S}_1^{(i)}-4Q~{a}_s^{(i)}\right)=-\left({1\over N}\right)e^{\zeta}s^{(i)}_\text{metric}(\zeta)\\
&{d\over d\zeta}\left(e^\zeta\dot{a}_s^{(i)}\right) =-\left({1\over N}\right)e^{\zeta}s^{(i)}_\text{gauge}(\zeta)\\
\end{split}
\end{equation}
Integrating the second equation  of \eqref{finscaleq} we get 
\begin{equation}\label{intsec}
\begin{split}
a_s^{(i)} (\zeta) = {1\over N}\int_\zeta^{\infty}d\rho~ e^{-\rho} \int_0^\rho d\rho'~e^{\rho'}~s^{(i)}_\text{gauge}(\rho') 
\end{split}
\end{equation}
Substituting the solution the first equation of \eqref{finscaleq} we get the solution for ${S}_1^{(i)}(\zeta)$. \footnotemark
\footnotetext{In the $Q\to 0$ limit, the $s^{(i)}_\text{metric}$ vanishes (as there is no metric vector sector correction in the uncharged limit) except for $s^{(3)}_\text{metric}$. Using equation \eqref{scalmem} and the same logic in the vector sector, the RHS of the first equation in \eqref{scaldecomp} is suppressed in this order and hence there is no scalar sector correction in the uncharged limit. Thus, this result is consistent with \cite{SB}. }

\begin{equation}\label{intsec}
\begin{split}
{S}_1^{(i)} (\zeta) =- 4Q\int_\zeta^\infty d\rho~a_s^{(i)} (\rho) + {1\over N}\int_\zeta^{\infty}d\rho~ e^{-\rho} \int_0^\rho d\rho'~e^{\rho'}~s^{(i)}_\text{metric}(\rho') 
\end{split}
\end{equation}
  \section{Final results}
In an expansion in the inverse power of dimension, we found a class of dynamical `black hole' solutions  to Einstein-Maxwell equation in presence of  cosmological constant. Our algorithm, in principle, works to all order. We have calculated the explicit solution upto the first subleading order.\\
In this section, we shall summarize our final result. For convenience,  we shall repeat some of the definition and conventions, described in the previous sections, again here.
\begin{equation}\label{res1}
\begin{split}
&G_{AB} = \bar G_{AB} +fO_A O_B + \left({1\over D}\right)G^{(1)}_{AB} + {\cal O} \left({1\over D}\right)^2\\
&A_M = \sqrt{2}\left[\tilde f ~O_M + \left({1\over D}\right)A^{(1)}_M + {\cal O} \left({1\over D}\right)^2\right]\\
\end{split}
\end{equation}
  where
\begin{equation}\label{ftf}  
f = (1+ Q^2)\psi^{-D} - Q^2\psi^{-2D},~~~\tilde f = Q~\psi^{-D}
\end{equation}
\begin{itemize}
\item$\bar G_{AB}$ is any  solution of Einstein equation in presence of cosmological constant and vanishing electromagnetic field. 
\item $\psi$ is a smooth function such that $\psi=1$ is the horizon of the full space-time and \\
$\psi^{-D}$ is a harmonic function with respect to the background metric $\bar G_{AB}$.
\item $O^A$ is a null geodesic vector field in the background satisfying
\begin{equation}\label{defeq}
\begin{split}
&O^A\partial_A\psi = \sqrt{\bar G^{AB}\left( \partial_A\psi\right)\left(\partial_B\psi\right)}\\
&u^A\vert_{\psi=1}=\left[O^A-\left({\bar G^{AB}\partial_B\psi\over \sqrt{\bar G^{AB}\left( \partial_A\psi\right)\left(\partial_B\psi\right)}}\right)\right]_{\psi=1}=\text{Null generator of the horizon}
\end{split}
\end{equation}
\item $Q$ is another smooth function satisfying
\begin{equation}\label{defqq}
\bar G^{AB}(\partial_A\psi)(\partial_B Q)=0,~~~~Q\vert_{\psi=1} =\frac{1}{\sqrt{2}}\left( u^MA_M\right)\vert_{\psi=1}
\end{equation}

\end{itemize}

  The first subleading correction $G^{(1)}_{AB}$ and $A^{(1)}_M$ are parametrized in the following way.
\begin{equation}\label{metgaugeformrep}
\begin{split}
G^{(1)}_{AB} &= {\cal G}^{(s_1)}~ O_A O_B + \left( {\cal G}^{(v)}_A O_B + {\cal G}^{(v)}_B O_A\right) + {\cal G}^{(T)}_{AB}\\
A^{(1)}_M &= {\cal A}^{(s)}~O_M + {\cal A}^{(v)}_M
\end{split}
\end{equation}  

where
\begin{equation}
\begin{split}
{\cal G}^{(s_1)}&=\sum_{i=1}^{N_S} S^{(i)}_1(\zeta)~ {\mathcal S}^{(i)},~~~~~{\cal G}^{(s_2)}=\sum_{i=1}^{N_S}  S^{(i)}_2(\zeta) ~{\mathcal S}^{(i)},~~~~~{\cal A}^{(s)}=\sum_{i=1}^{N_S}  a^{(i)}_s(\zeta)~ {\mathcal S}^{(i)}\\
{\cal G}^{(v)}_A &=\sum_{i=1} ^{N_V} \mathcal{V}^{(i)}(\zeta)~V^{(i)}_A,~~~~~{\cal A}^{(v)}_A = \sum_{i=1} ^{N_V}a^{(i)}_v(\zeta)~V^{(i)}_A,~~~~~{\cal G}^{(T)}_{AB} =\sum_{i=1}^{N_T} \mathcal{T}^{(i)}(\zeta)~t^{(i)}_{AB}
\end{split}
\end{equation}
The scalar, vector and tensor structures $\mathcal{S}^{(i)}$, $V_A^{(i)}$ and $t_{AB}^{(i)}$ respectively are given in  table \eqref{table:list1}.\\

The solution of the first subleading order metric and gauge field corrections are:
\begin{equation}\label{solutions}
\begin{split}
&{\cal T}(\zeta) =2\left(D\over K\right)\log\left[1-Q^2e^{-\zeta}\right]\\
&a_v^{(i)}(\zeta)= e^\zeta~ [f -\tf^2]\int_\zeta^\infty\bigg({e^{-3\rho}\over{[1-f]~[f-\tf^2]^2}}\bigg)\int_0^\rho d\rho' ~{\mathfrak v}^{(i)}(\rho')\\
&\text{where}~~{\mathfrak v}^{(i)}(\zeta)\equiv\left(D\over K\right)e^{2\zeta}(f-\tf^2)\left({2 e^{-\zeta}}\int_0^\zeta d\rho \left[e^\rho ~v^{(i)}_{metric}(\rho)\over{ 1-f(\rho)}\right]-v_{gauge}^{(i)}(\zeta)\right)\\
& {\cal V}^{(i)}(\zeta) = -2\int_{\zeta}^\infty d\rho~\tf(\rho)~ a_v^{(i)}(\rho)  - 2 \left(D\over K\right)\int_{\zeta}^\infty d\rho ~e^{-\rho}\int_0^\rho d\rho' \left[e^{\rho' }~v^{(i)}_{metric}(\rho')\over{ 1-f(\rho')}\right]\\
&a_s^{(i)} (\zeta) = {1\over N}\int_\zeta^{\infty}d\rho~ e^{-\rho} \int_0^\rho d\rho'~e^{\rho'}~s^{(i)}_\text{gauge}(\rho') \\
& {S}_1^{(i)} (\zeta) =- 4Q\int_\zeta^\infty d\rho~a_s^{(i)} (\rho) + {1\over N}\int_\zeta^{\infty}d\rho~ e^{-\rho} \int_0^\rho d\rho'~e^{\rho'}~s^{(i)}_\text{metric}(\rho') 
\end{split}
\end{equation}
where $v^{(i)}_{\text{metric}},\: s^{(i)}_\text{metric},\:s^{(i)}_\text{gauge}$ are given in \eqref{vecs} and \eqref{scaldecomp}. In the $Q\to0$ limit, the tensor, vector and scalar sector corrections in the metric vanish, which is consistent with \cite{SB}.\\

The dual membrane is defined in terms of a smooth function $\psi$, a smooth one form $O$, a velocity field $u_A$ and charge field $Q$ everywhere in the background $\bar{G}_{AB}$.
The "integrable condition"  or the membrane equations governing the dynamics of the membrane are:
\begin{equation}\label{membraneeqnts}
\begin{split}
\hat{\n}\cd u={\cal O}\left( \frac{1}{D}\right)\\
\frac{\hat{\nabla}^2 Q}{\mtK} - u.\hat{\dl} Q - Q\left[ \frac{u.\hat{\dl}\mtK}{\mtK} - u.\mtK.u  - \frac{R_{uu}}{\mtK}\right] ={\cal O}\left( \frac{1}{D}\right)\\
\Big [ \frac{\hat{\dl} ^2 u_\nu}{\mtK}-(1-Q^2)\frac{\hat{\dl}_\nu \mtK}{\mtK}+u^\alpha K_{\alpha\nu}-(1+Q^2)(u.\hat{\dl} u_\nu)\Big]\mathcal{P}^\nu_\mu={\cal O}\left( \frac{1}{D}\right)
\end{split}
\end{equation}
$\hat{\n}$ is the covariant derivative wrt the induced metric on the membrane. All the quantities used in \eqref{membraneeqnts} are w.r.t the coordinates intrinsic to the membrane, denoted by the Greek indices. All the lowering and raising has been done w.r.t the induced metric. The greek indices can take $(D-1)$ values.
The final results of our paper are \eqref{solutions}, and \eqref{membraneeqnts}.

\section{Quasi Normal Mode calculations}
In this section, we will be showing the consistency of our membrane equations. Membrane equations are well-posed initial value problem. In principle, it gives the complete dynamical description of any black-hole system. Hence, we will be deriving the light quasi-normal modes for few known solutions of Einstein-Maxwell equation using our membrane equations and will be comparing them with the pre-obtained results calculated purely from the gravitational analysis.\\


\subsection{QNM for charged black brane}
We will be looking at the light quasi-normal modes of the charged black-brane solution in AdS/dS background. For simplicity, we will be starting with AdS charged black brane in  Poincare Patch. The black brane geometry is given as,
\be{}
ds^2=ds^2_{\text{bgrd}}+\frac{1}{r^2}\bigg[ \frac{2M}{r^{D-3}}-\frac{q^2}{r^{2(D-3)}}\bigg]\bigg(rdt+\frac{dr}{r}\bigg)^2
\ee
The background metric in Poincare patch is,
\begin{eqnarray}
\nonumber ds^2_{bgrd}&=& g_{AB}dx^A dx^B= -r^2 dt^2 + \frac{dr^2}{r^2}+r^2 dx^a dx_a
\end{eqnarray}
The horizon of the charged black brane is at $ r_0=1 $ {\cite{SB}} and on the horizon, the velocity field is $ u_0 = -1$. We  consider a small perturbation on this exact solution as,
\begin{eqnarray}
\nonumber r&=&1+\epsilon \delta r (t,x^a)\\
u&=&u_0 dt + \epsilon(\delta u_t(t,x^b)dt + \delta u_a(t,x^b) dx^a)\\
\nonumber Q&=& Q_0+\epsilon \delta Q(t,x^a)
\end{eqnarray}
where $\epsilon$ is the linearization parameter, $ a $ represents the $(D-2)$  linear $x^a$ coordinates. The coordinates along the membrane ($t,a$) are denoted by the Greek indices $\mu$. The induced metric on the membrane up to linear order in $\epsilon$ is,
\begin{equation}
ds^2_{ind}=g^{ind}_{\mu\nu}dx^{\mu}dx^{\nu}=-(1+2\epsilon \delta r)dt^2 +(1+2 \epsilon \delta r) dx^a dx_a
\end{equation}
We follow a convention that,
\begin{eqnarray}
\nonumber \widehat{\nabla}_{\mu} &\rightarrow& \mbox{ Covariant derivative with respect to induced metric }g^{ind}_{\mu\nu}\\
\nonumber \partial_a &\rightarrow& \mbox{ Covariant derivative with respect to the metric of the coordinates } (a,b) \text{ which is } \delta_{ab} 
\end{eqnarray}
Using normalization condition of the velocity field $g^{ind}_{\mu\nu}u^{\mu}u^{\nu}=-1$, in linear order we get,
\begin{equation}
u(t,a)=-(1+\epsilon \delta r(t,a)) 
\end{equation}
The projector $\mathcal{P}_{\mu\nu}$ on the membrane and perpendicular to $ u_{\mu} $ direction are,
\begin{eqnarray}
\nonumber \mathcal{P}_{\mu\nu}&=& g^{ind}_{\mu\nu}+u_{\mu}u_{\nu}\\
\mathcal{P}^t_t= 0 ,&& \mathcal{P}^a_b= \delta^a_b\\
\nonumber \mathcal{P}^t_a=\epsilon \delta u_a, && \mathcal{P}^a_t=-\epsilon \delta u_a
\end{eqnarray}
The structure of the vector membrane equation is written  similar to \cite{SB},
\begin{eqnarray}\label{qnmveceq}
E^{total}_{\mu}&=&\mathcal{P}^{\nu}_{\mu}E_{\nu}\\
\text{where, }E_{\mu}&=&\frac{\widehat{\nabla}^2 u_{\mu}}{\mathcal{K}}- (1-Q^2)\frac{\widehat{\nabla}_{\mu} \mathcal{K}}{\mathcal{K}}+u^{\nu}K_{\mu\nu}-(1+Q^2)u^{\nu}\widehat{\nabla}_{\nu} u_{\mu}
\end{eqnarray}
where $\mtK_{\mu\nu},\mtK$ are the extrinsic curvature and trace of the extrinsic curvature wrt the membrane coordinates. $\mtK_{\mu\nu}$ is the pull back of  the extrinsic curvature $K_{MN}$ (wrt the full spacetime coordinates)\cite{SB}.
\be{}
\mtK_{\mu\nu}=\left(\frac{\p X^M}{\p y^\mu}\right)\left(\frac{\p X^N}{\p y^\nu}\right)K_{MN}\Big|_{r=1+\epsilon \delta r}
\ee

Now we will consider the scalar membrane equation,
\begin{eqnarray}
\nonumber &&\widehat{\nabla}.u=0\\ 
\nn \text{which implies,} &&~~\partial^a \delta u_a+(D-2)\partial_t \delta r=0\\ \label{delua}
 \text{or, }~~&&\partial^a \delta u_a=-(D-2)\partial_t \delta r
\end{eqnarray}
The relevant quantities that we need to calculate to determine the QNM frequencies are, (For detailed calculation the readers are requested to go through \cite{SB})
\begin{eqnarray}
\nonumber \mathcal{K}_{tt}&=& -(1+\epsilon \delta r)-\epsilon \partial^2_t \delta r\\
\nonumber \mathcal{K}_{ta}&=& \epsilon \partial_t \partial_a \delta r\\
\nonumber \mathcal{K}_{ab}&=& (1+2\epsilon \delta r)\delta_{ab}-\epsilon \partial_a\partial_b\delta r\\
\nonumber \mathcal{K}&=& (D-1)+\epsilon \partial^2_t \delta r -\epsilon \partial^2 \delta r\\
\nonumber u^{\mu}\mtK_{\mu t} &=& -1\\
\nonumber u^{\mu}\mtK_{\mu a} &=& \epsilon \delta u_a -\partial_t \partial_a \delta r\\
\nn \nonumber \frac{\widehat{\nabla}_t \mathcal{K}}{\mathcal{K}}&=& \mtO(\epsilon)\\
\nonumber \frac{\widehat{\nabla}^2 u_a}{\mathcal{K}} &=&\mtO(\epsilon)\\
\nonumber \frac{\widehat{\nabla}_a \mathcal{K}}{\mathcal{K}}&=&\frac{\epsilon}{D}\partial_a(\partial^2_t-\partial^2)\delta r\\
\nonumber \frac{\widehat{\nabla}^2 u_a}{\mathcal{K}} &=& \frac{\epsilon}{D}(\partial^2-\partial^2_t)\delta u_a\\
\nonumber u.\widehat{\nabla} u_a &=& \epsilon(\partial_t\delta u_a+\partial_a\delta r)\\ \label{qnm2}
 u.\widehat{\nabla} u_t &=&\mtO(\epsilon)
\end{eqnarray}
We are summarizing here the algorithm to calculate \eqref{qnm1} in the way similar to \cite{SB}. The vector membrane equation \eqref{qnmveceq} can be written as:
\bes \label{qnm1}
\bea{}
\nn E^{tot}_t&=&E_t \mathcal{P}^t_t+E_b \mathcal{P}^b_t\\
E^{tot}_a&=&E_t \mathcal{P}^t_a+E_b \mathcal{P}^b_a
\eea
\ees
The translation symmetry in the background along the $x^a$ directions are  broken by the small fluctuations that we have considered. So, $E_b \sim \mtO (\epsilon)$. Also, $\mathcal{P}^t_t=0,\: \mathcal{P}^b_t\sim \mtO(\epsilon) $. Hence $$ E^{tot}_t \sim \mtO(\epsilon^2)$$ which is considered to be zero at linear order in $\epsilon$.\\
Again,
for $E^{tot}_a$, we consider only $\mtO(\epsilon^0)$ terms for $E_t$ as $\mathcal{P}^t_a\sim \mtO(\epsilon)$.

Using the above mentioned quantities in \eqref{qnm2} and the decomposition,
\begin{eqnarray}\label{brane_decompos}
\nonumber \delta r&=&  \delta r_{0} e^{-\iota (\omega^s_k t+k_a x^a)}\\
u^a &=& \delta u_a e^{-\iota (\omega^v_k t+k_a x^a)}\\
\nonumber \delta Q &=& \delta Q_0 e^{-\iota (\omega^q_k t+k_a x^a)}
\end{eqnarray}

we get,
\begin{eqnarray}\label{medstep}
\nonumber &&E^{tot}_a=0\\
\nonumber&\Rightarrow&E_t \mathcal{P}^t_a+E_b \mathcal{P}^b_a=0\\
\nn&\Rightarrow&\frac{\epsilon}{D}(\partial^2-\partial^2_t)\delta u_a -(1-Q^2)\frac{\epsilon}{D}\partial_a(\partial^2_t-\partial^2)\delta r-(1+Q^2)\epsilon(\partial_t\delta u_a+\partial_a\delta r)-\epsilon \partial_t \partial_a \delta r=0\\
\end{eqnarray}
where  $\p^2=\p_a \p^a$.
This is a vector equation with $\delta_{ab}$ metric. Taking the divergence $\p^a E_a^{tot}$ with respect to $x^a$ and using the relation derived from scalar equation \eqref{delua}, we get,
\begin{equation}\label{qnm3}
-(1+Q^2_0)\partial^2\delta r - 2\partial_t \partial^2\delta r+(1+Q^2_0)D\partial^2_t+(1-Q^2_0)\frac{{(\partial^2)}^2\delta r}{D}=0
\end{equation}

We use the fact that $\partial^2\delta r$ is of $\mathcal{O}(D)$, and consider only the terms which are of $\mtO(D)$ in \eqref{qnm3}.  Thus we get the scalar frequencies,
\begin{equation}\label{scbrfr}
\omega^s_{\pm}=-i\frac{1}{1+Q^2_0}\frac{k^2}{D}\pm \frac{k}{\sqrt{D}}\sqrt{1-{\left( \frac{Q^2_0}{1+Q^2_0}\right)}^2\frac{k^2}{D} }
\end{equation}

The scalar frequencies in the  uncharged case is 
\begin{equation}\label{qnm4}
\omega^s_{\pm}=-i\frac{k^2}{D}\pm \frac{k}{\sqrt{D}}
\end{equation}
which match with the results in \cite{EmparanCoupling,Emparan:2015gva,Emparan:2015rva,yogesh2,SB}.\\

The general solution for shape function with momentum $k_a$ is,
\begin{eqnarray}
\delta r &=& \left( \delta r_+ e^{-i\omega^s_+ t}+\delta r_- e^{-i\omega^s_- t}\right) e^{i k_a x^a}
\end{eqnarray}

Now let,
\begin{equation}
\delta u_a =  \left( \delta r_+ V_{a+} e^{-i\omega^s_+ t}+\delta r_- V_{a-} e^{-i\omega^s_- t}+V_a e^{-i\omega^v_k t}\right) e^{i k_a x^a}
\end{equation}
where $V_{a+}, V_{a-} $ are vectors in the direction of $k_a$ and  $v_a k^a=0$.\\
Putting these two expansions in \eqref{medstep} and equating different modes, we get,
\begin{eqnarray}\label{vecbrfr}
\text{vector frequency,}~~\omega^v_k&=&-i\frac{k^2}{D}\frac{1}{1+Q^2_0}\\
\text{and the vector modes,}~~V_{a\pm}&=& \left( \frac{i}{1+Q^2_0}\mp\sqrt{\frac{D}{k^2}-{\left( \frac{Q_0^2}{1+Q^2_0}\right)}^2  } \right) k_a
\end{eqnarray}

For Charge quasi-normal mode case, we have calculated the quantities that we need. They are mentioned below,
\begin{eqnarray}
\nonumber \widehat{\nabla}^2 Q &=& \epsilon \partial^2 \delta Q\\
\nonumber u.\widehat{\nabla} Q &=& \epsilon \partial_t \delta Q\\
\mathcal{K} &=& D-1\\
\nonumber u.K.u &=&  -(1+2\epsilon\delta r)-\epsilon\partial^2_t\delta r\\
\nonumber {u.\widehat{\nabla}\mathcal{K}}&=& 0
\end{eqnarray}

We take the expansion of charge fluctuation as,
\begin{equation}
\delta Q =  \left( \delta r_+ Q_{+} e^{-i\omega^s_+ t}+\delta r_- Q_{-} e^{-i\omega^s_- t}+\delta Q_0 e^{-i\omega^q t}\right) e^{i k_a x^a}
\end{equation}
where,  we have used the subsidiary condition $n.\widehat{\nabla}\delta Q=0$.\\
Using these quantities and the expansion in the charge membrane equation (the second equation in \eqref{membraneeqnts}) and equating the coefficients of $  e^{-i \omega^q t} $ we get,
\begin{eqnarray}\label{chbrfr}
\Rightarrow \omega^q &=& -i \frac{k^2}{D}
\end{eqnarray}

Also equating the scalar modes $  e^{-i \omega^s_{\pm} t} $ we get,
\begin{equation}
Q_{\pm}=Q_0\frac{2+\omega^s_{\pm}}{i \omega^s_{\pm}-\frac{k^2}{D}}
\end{equation}

which completes the quasi-normal mode calculation for charged black-brane solution. In section(3.4) of \cite{EmparanHydro} the quasi normal frequencies has been already calculated. Our results \eqref{scbrfr}, \eqref{vecbrfr} and \eqref{chbrfr} match with \cite{EmparanHydro} completely if we use their notations:\footnotemark \footnotetext{The effective equations of charged black brane in \cite{EmparanHydro} involves the scaling in $\mathcal{O}\Big(\frac{1}{\sqrt D}\Big)$. In \cite{yogesh2,SB}, the effective equations of uncharged black brane has been reproduced from covariant uncharged membrane equations. We have also considered a scaled co-ordinate transformation and a particular choice of scaling to reproduce the QNM's of charged black brane which is consistent with our results and also with \cite{EmparanHydro}. The details are in \eqref{QNM for the scaled membrane equation}.  } 
$$\epsilon=-1 (\text{For AdS black brane}), k\to\frac{k}{\sqrt{D}}, a_+=\frac{1}{1+Q_0^2},a_-=\frac{Q_0^2}{1+Q_0^2}$$

\subsection{QNM for Reissner-Nordstorm blackhole in global AdS}
To calculate the light quasi normal modes of the AdS Reissner-Nordstr\"{o}m  solution, we  consider the global AdS/dS as the background metric,
\begin{eqnarray}
\nonumber ds^2_{bgrd}&=& g_{AB}dx^A dx^B
=-\left( 1-\frac{\sigma r^2}{L^2}\right) dt^2 + \frac{dr^2}{\left( 1-\frac{\sigma r^2}{L^2}\right)}+r^2 \Omega_{ab}d\theta^a d\theta^b
\end{eqnarray}
where $ L $ is the AdS or dS radius and the value of $ \sigma $ is $0$ for flat space, $+1$ for AdS, $-1$ for dS.
The horizon of RN black hole is at $ r_0=1 $ and on the horizon, the velocity field is $ u_0 = -\left( 1-\frac{\sigma}{L^2}\right)^{\frac{1}{2}}$. We will consider a small perturbation on this exact solution as follows,
\begin{eqnarray}
\nonumber r&=&1+\epsilon \delta r (t,\theta^a)\\
u&=&u_0 dt + \epsilon(\delta u_t(t,\theta^b)dt + \delta u_a(t,\theta^b) d\theta^a)\\
\nonumber Q&=& Q_0+\epsilon \delta Q(t,\theta^a)
\end{eqnarray}
where $ a $ represents the ($D-2$) angular $\theta^a$ coordinates and $\mu$ denotes coordinates along the membrane $(t,\theta^a)$. The induced metric on the membrane upto linear order in $\epsilon$ is,
\begin{equation}
ds^2_{ind}=\left( 1-\frac{\sigma}{L^2}(1+2\epsilon \delta r)\right)dt^2 +(1+2 \epsilon \delta r)  \Omega_{ab} d\theta^a d\theta^b
\end{equation}
We will follow a convention that,
\begin{eqnarray}
\nonumber \widehat{\nabla}_{\mu} &\rightarrow& \mbox{ Covariant derivative with respect to induced metric,}\\
\nonumber \overline{\nabla}_a &\rightarrow& \mbox{ Covariant derivative with respect to } \Omega_{ab}(\text{ The metric along (D-2) dimensional unit sphere})
\end{eqnarray}
Using the normalization condition $g^{ind}_{\mu\nu}u^{\mu}u^{\nu}=-1$, in linear order we get,
\begin{equation}
\delta u_t(t,a)=\frac{\sigma}{L^2}\left( 1-\frac{\sigma}{L^2}\right) \delta r(t,a) 
\end{equation}
The projector on the membrane perpendicular to $u_{\mu}$ is, (The detailed calculation of the below mentioned quantities are available in \cite{SB})
\begin{eqnarray}
\nonumber \mathcal{P}_{\mu\nu}&=& g^{ind}_{\mu\nu}+u_{\mu}u_{\nu}\\
\mathcal{P}^t_t= 0 && \mathcal{P}^a_b= \delta^a_b\\
\nonumber \mathcal{P}^t_a= \left( 1-\frac{\sigma}{L^2}\right)^{\frac{1}{2}}(\epsilon \delta u_a)&& \mathcal{P}^a_t= \left( 1-\frac{\sigma}{L^2}\right)^{-\frac{1}{2}}(\epsilon \delta u_a)
\end{eqnarray}
Let us consider the structure of the membrane equation in the limit $Q\rightarrow 0$, which is the membrane equation for  the uncharged case in \cite{SB}.
\begin{eqnarray}
E^{total}_{\mu}&=&\mathcal{P}^{\nu}_{\mu}E_{\nu}\\
E_{\mu}&=&\frac{\widehat{\nabla}^2 u_{\mu}}{\mathcal{K}}- \frac{\widehat{\nabla}_{\mu} \mathcal{K}}{\mathcal{K}}+u^{\nu}\mtK_{\mu\nu}-u^{\nu}\widehat{\nabla}_{\nu} u_{\mu}
\end{eqnarray}
where $\mtK_{\mu \nu}$ is the pull back of extrinsic curvature $K_{AB}$ along the membrane and $\mtK$ is the trace of extrinsic curvature $\mtK_{\mu \nu}$. 
In our case (where $Q\neq 0$) the vector membrane equation is :
$$\frac{\widehat{\nabla}^2 u_{\mu}}{\mathcal{K}}-(1-Q^2) \frac{\widehat{\nabla}_{\mu} \mathcal{K}}{\mathcal{K}}+u^{\nu}K_{\mu\nu}-(1+Q^2)u^{\nu}\widehat{\nabla}_{\nu} u_{\mu}$$
Hence, the extra terms appearing in our case proportional to Q are,
\begin{equation*}
\triangle E_{\mu} = Q^2 \frac{\widehat{\nabla}_{\mu}\mathcal{K}}{\mathcal{K}}-Q^2 u^{\nu} \widehat{\nabla}_{\nu} u_{\mu}
\end{equation*}
Hence, our equation can be represented as,
\begin{equation}
E^{tot}_{\mu}=\mathcal{P}^{\nu}_{\mu}(E_{\nu}+\triangle E_{\mu})
\end{equation}

The  components that would be relevant for the linearized membrane equation are  taken from \cite{SB},
\begin{equation}\label{eq:list}
\begin{split}
u^\nu {\cal K}_{\nu t} &= \frac{\sigma}{L^2} + {\mathcal O}(\epsilon) \\
u^\nu {\cal K}_{\nu a} &= \left( 1-\frac{\sigma}{L^2} \right)^{-1} (-\epsilon \partial_t \bar\nabla_a \delta r) + \left( 1-\frac{\sigma}{L^2} \right)^{\frac{1}{2}}(\epsilon \delta u_a) \\
u^\nu \hat{\nabla}_{\nu}u_{t} &= 0\\
u^\nu \hat{\nabla}_{\nu}u_{a} &= \left( 1-\frac{\sigma}{L^2} \right)^{-\frac{1}{2}}(\epsilon \partial_t \delta u_a)-\left( 1-\frac{\sigma}{L^2} \right)^{-1}\frac{\sigma}{L^2}(\epsilon \bar\nabla_a \delta r) \\
\hat{\nabla}_t {\mathcal K} &=  {\mathcal O}(\epsilon) \\
\hat{\nabla}_a {\mathcal K} &=  \left( 1-\frac{\sigma}{L^2} \right)^{-\frac{3}{2}}(\epsilon \partial^2_t \bar\nabla_a \delta r)-\left( 1-\frac{\sigma}{L^2} \right)^{-\frac{3}{2}}\frac{\sigma}{L^2}(\epsilon \bar\nabla_a \delta r) \\
&+ \left( 1-\frac{\sigma}{L^2} \right)^{-\frac{1}{2}}(-\epsilon \bar\nabla_a \bar\nabla^2 \delta r) - (D-2) \left( 1-\frac{\sigma}{L^2} \right)^{-\frac{1}{2}}(\epsilon \bar\nabla_a \delta r) \\
\hat{\nabla}^2u_t &= {\mathcal O}(\epsilon) \\
\hat{\nabla}^2u_a &= -\left( 1-\frac{\sigma}{L^2} \right)^{-1}(\epsilon \partial_t^2 \delta u_a) + \left( 1-\frac{\sigma}{L^2} \right)^{-\frac{3}{2}}\frac{\sigma}{L^2} (\epsilon \partial_t \bar\nabla_a \delta r) \\
&+ \epsilon \bar\nabla^2 \delta u_a + \left( 1-\frac{\sigma}{L^2} \right)^{-\frac{1}{2}} (\epsilon \partial_t \bar\nabla_a \delta r)
\end{split}
\end{equation}
Using equations \eqref{eq:list} the linearized vector membrane equation in the angular directions evaluates to
\begin{equation}\label{lineq}
\begin{split}
E_a^{total} &\equiv \left( \frac{\sigma}{L^2} \right)\left( 1-\frac{\sigma}{L^2} \right)^{-\frac{1}{2}}(\epsilon \delta u_a) + \left( 1-\frac{\sigma}{L^2} \right)^{-\frac{1}{2}}\epsilon \frac{\bar\nabla^2\delta u_a}{D-2} + \left( 1-\frac{\sigma}{L^2} \right)^{-1} \epsilon \frac{\bar\nabla_a\bar\nabla^2\delta r}{D-2} \\
&+ \left( 1-\frac{\sigma}{L^2} \right)^{-1} (\epsilon \bar\nabla_a \delta r) + \left( 1-\frac{\sigma}{L^2} \right)^{-1} (-\epsilon \partial_t\bar\nabla_a \delta r) + \left( 1-\frac{\sigma}{L^2} \right)^{\frac{1}{2}} (\epsilon \delta u_a) \\ &- \left( 1-\frac{\sigma}{L^2} \right)^{-\frac{1}{2}} (\epsilon \partial_t\delta u_a) +\left( 1-\frac{\sigma}{L^2} \right)^{-1}\left(\frac{\sigma}{L^2}\right) (\epsilon \bar\nabla_a\delta r)
\end{split}
\end{equation}
and,
\begin{eqnarray}
\nonumber \triangle E^{tot}_b &=& \triangle E_t \mathcal{P}^t_b + \triangle E_a \mathcal{P}^a_b\\
\nonumber &=& Q^2_0\left[ \left( 1-\frac{\sigma}{L^2}\right) ^{-\frac{1}{2}}\epsilon \delta u_b \cancelto{\mathcal{O}(\epsilon)}{\left(  \frac{\widehat{\nabla}_{\mu}\mathcal{K}}{\mathcal{K}}- u.\widehat{\nabla} u_{t}\right)} \right]+Q^2_0\left(  \frac{\widehat{\nabla}_{a}\mathcal{K}}{\mathcal{K}}- u.\widehat{\nabla} u_{a}\right) \delta^a_b\\
\nonumber &=& \frac{Q^2_0}{\left( 1-\frac{\sigma}{L^2}\right)(D-2)-\frac{\sigma}{L^2}}\left[\left( 1-\frac{\sigma}{L^2}\right)^{-1}(\epsilon \partial^2_t \overline{\nabla}^b \delta r)-\left( 1-\frac{\sigma}{L^2}\right)^{-1}\frac{\sigma}{L^2}(\epsilon \overline{\nabla}^b \delta r) \right. \\
\nonumber && \left. - (\epsilon \overline{\nabla}^b \overline{\nabla}^2 \delta r)-(D-2)(\epsilon \overline{\nabla}^b \delta r)\right]-Q^2_0 \left[ \left( 1-\frac{\sigma}{L^2}\right)^{-\frac{1}{2}}(\epsilon \partial_t \delta u^b)-\left( 1-\frac{\sigma}{L^2}\right)\right] 
\end{eqnarray}
Let $\mathcal{A}=\frac{Q^2_0}{\left( 1-\frac{\sigma}{L^2}\right)(D-2)-\frac{\sigma}{L^2}}$, then,
\begin{eqnarray}
\nonumber E^{tot}_a&=& \frac{\sigma}{L^2}\left( 1-\frac{\sigma}{L^2}\right)^{-\frac{1}{2}}(\epsilon \delta u_a)+\left( 1-\frac{\sigma}{L^2}\right)^{-\frac{1}{2}}\frac{\epsilon \overline{\nabla}^2\delta u_a}{D-2}+\left( 1-\frac{\sigma}{L^2}\right)^{-1}\epsilon \frac{\overline{\nabla}_a \overline{\nabla}^2 \delta r}{D-2}\\
\nonumber &&+\left( 1-\frac{\sigma}{L^2}\right)^{-1}(\epsilon \overline{\nabla}_a \delta r)+\left( 1-\frac{\sigma}{L^2}\right)^{-1}(-\epsilon \partial_t \overline{\nabla}_a \delta r)+ \left( 1-\frac{\sigma}{L^2}\right)^{\frac{1}{2}}\epsilon \delta u_a\\
\nonumber && -\left( 1-\frac{\sigma}{L^2}\right)^{-\frac{1}{2}}(\epsilon \partial_t \delta u_a)+\left( 1-\frac{\sigma}{L^2}\right)^{-1}\frac{\sigma}{L^2}(\epsilon \overline{\nabla}_a \delta r)- \mathcal{A}(\epsilon \overline{\nabla}_a\overline{\nabla}^2\delta r)\\
\nn \label{qnm6} &&-\mathcal{A}(D-2)(\epsilon \overline{\nabla}_a \delta r)-Q^2_0 \left( 1-\frac{\sigma}{L^2}\right)^{-\frac{1}{2}} (\epsilon \partial_t \delta u_a)+ Q^2_0 \left( 1-\frac{\sigma}{L^2}\right)^{-1}\frac{\sigma}{L^2}(\epsilon \overline{\nabla}_a \delta r)\\
\end{eqnarray}

This is a vector equation in sphere coordinates. If we take divergence, $\overline{\nabla}^a$ with respect to $\Omega_{ab}$ then we get,
\begin{eqnarray}
\nonumber \overline{\nabla}^a E^{tot}_a&=& \frac{\sigma}{L^2}\left( 1-\frac{\sigma}{L^2}\right)^{-\frac{1}{2}} (\epsilon \overline{\nabla}^a \delta u_a)+\left( 1-\frac{\sigma}{L^2}\right)^{-\frac{1}{2}}\epsilon \frac{\overline{\nabla}^a\overline{\nabla}^2\delta u_a}{D-2}+\epsilon \left( 1-\frac{\sigma}{L^2}\right)^{-1}\overline{\nabla}^2\overline{\nabla}^2\delta r\\
\nonumber &&+\left( 1-\frac{\sigma}{L^2}\right)^{-1}\epsilon \overline{\nabla}^2 \delta r-\left( 1+\frac{\sigma}{L^2}\right)^{-1}\epsilon \overline{\nabla}^2\delta r- \left( 1-\frac{\sigma}{L^2}\right)^{-1}\epsilon \partial_t \overline{\nabla}^a \delta u_a\\
\nonumber &&- \left( 1-\frac{\sigma}{L^2}\right)^{-\frac{1}{2}}\epsilon \partial_t \overline{\nabla}^a \delta u_a +\left( 1-\frac{\sigma}{L^2}\right)^{-1} \frac{\sigma}{L^2} \epsilon \overline{\nabla}^2 \delta r -\mathcal{A}\epsilon \overline{\nabla}^2\overline{\nabla}^2\delta r\\
\nonumber &&-\mathcal{A}(D-2)\epsilon \overline{\nabla}^2 \delta r-Q^2_0 \left( 1-\frac{\sigma}{L^2}\right)^{-\frac{1}{2}}\partial_t \overline{\nabla}^a\delta u_a+Q^2_0 \left( 1-\frac{\sigma}{L^2}\right)^{-1}\frac{\sigma}{L^2} \epsilon \overline{\nabla}^2 \delta r\\
\end{eqnarray}

Now we will consider the scalar membrane equation,
\begin{eqnarray}
\nonumber \widehat{\nabla}.u&=&\widehat{\nabla}^t(u_0+\epsilon \delta u_t)+\epsilon\widehat{\nabla}^a \delta u_a\\
&=&\epsilon \left(\overline{\nabla}^a\delta u_a+(D-2)\epsilon \left( 1-\frac{\sigma}{L^2}\right)^{-\frac{1}{2}}\partial_t\delta r\right)+\mathcal{O}(1)
\end{eqnarray}
For convenience we can decompose $ \delta u_a$ as $ \delta u_a = v_a + \overline{\nabla}_a \Phi$, where $\overline{\nabla}.v=0$.
Then, $\widehat{\nabla}.u=0$ implies,
\begin{equation}
\overline{\nabla}^2\Phi=-(D-2)\left( 1-\frac{\sigma}{L^2}\right)^{-\frac{1}{2}}\partial_t \delta r
\end{equation}
Now decomposing the perturbations in spherical harmonic modes,
\begin{eqnarray}\label{har_decompos}
\nonumber \delta r&=& \varSigma_l \delta r_{l} Y_l e^{-\iota \omega^s_l t}\\
v^a &=& \varSigma_l v_l Y^a_{l} e^{-\iota \omega^v_l t}\\
\nonumber Q &=& Q_0 + \varSigma_l \delta Q_l Y_{l} e^{-\iota \omega^q_l t}
\end{eqnarray}

we get the equation in leading order in $\mtO(D)$ to solve the scalar QNM frequencies. 
\begin{equation}
-(1+Q^2_0)\omega^{s^2}+2 i (1-l) \omega^s + l(l-1)(1-Q^2_0)-\frac{\sigma}{L^2}l(1+Q^2_0)=0
\end{equation}
\begin{equation}
\Rightarrow \omega^s_{\pm} = - i(l-1)\pm \sqrt{\left\lbrace \frac{1}{(1+Q^2_0)^2}-\left( \frac{Q^2_0}{1+Q^2_0}\right)^2 l  \right\rbrace (l-1) - l \frac{\sigma}{L^2}}
\end{equation}

which matches with the  results of \cite{Tanabe:2016opw}. Their results are obtained from an entirely different approach and the matching shows the consistency of our analysis.\\

Now we calculate the vector QNM frequencies. In \eqref{qnm6} we have already solved for $\delta r$ and $\Phi$. So considering the divergence less vector part  we get,
\begin{eqnarray}
\nonumber &&\frac{\sigma}{L^2}\left( 1-\frac{\sigma}{L^2}\right)^{-\frac{1}{2}}\epsilon v_a + \left( 1-\frac{\sigma}{L^2}\right)^{-\frac{1}{2}} \epsilon \frac{\overline{\nabla}^2 v_a}{D-2}+\left( 1-\frac{\sigma}{L^2}\right)^{\frac{1}{2}} \epsilon v_a\\
&&-\left( 1-\frac{\sigma}{L^2}\right)^{-\frac{1}{2}}\epsilon \partial_t v_a -Q^2_0 \left( 1-\frac{\sigma}{L^2}\right)^{-\frac{1}{2}} \epsilon \partial_t v_a=0
\end{eqnarray}
Then using \eqref{har_decompos} we get,
\begin{equation}\label{AdS_vect_qnm}
\omega^v_l=-i \left( \frac{l-1}{1+Q^2_0}\right) 
\end{equation}
In the limit $Q\to 0$ the result matches with  \cite{Emparan:2014jca,SB}. The vector fequencies for the charged case are calculated in \cite{Tanabe:2016opw}. In that paper,
$$\hat{\omega}_v=\hat{m}-i \left( \frac{l_v-1}{1+Q^2_0}\right)$$
$\hat{m}$ is zero for non rotating black hole. Hence it matches with our result.
\footnote{In \cite{Tanabe:2016opw} there are two vector modes for Reissner-Nordstorm blackhole, $$\hat{\omega}_v=-i \left( \frac{l}{1+Q^2_0}\right),\: \hat{\omega}_0=-i \left( \frac{l-2}{1+Q^2_0}\right)$$ where $l$ the scalar mode frequency can be written in terms of the vector mode frequency $l_v$ as $l=l_v\pm1$. So, the vector frequency in terms of $l_v$ is $\hat{\omega}_v=-i \left( \frac{l_v-1}{1+Q^2_0}\right)$. Readers can go through \cite{Emparan:2014jca,Tanabe:2016opw} for more details on this.}

For Charged field quasi-normal mode case, the quantities that we need up to relevant order are,
\begin{eqnarray}
\nonumber \widehat{\nabla}^2 Q &=& \epsilon \overline{\nabla}^2 \delta Q+\mathcal{O}(1)\\
\nonumber u.\widehat{\nabla} Q &=& -\epsilon \left( 1-\frac{\sigma}{L^2}\right)^{-\frac{1}{2}} \partial_t \delta Q+\mathcal{O}\left( \frac{1}{D}\right)\\
\mathcal{K} &=& D\left( 1-\frac{\sigma}{L^2}\right)^{\frac{1}{2}}+\mathcal{O}(1)\\
\nonumber u.K.u &=&  \frac{\partial_t \overline{\nabla}^2\delta r}{D-2} + \partial_t\delta r +  \left( 1-\frac{\sigma}{L^2}\right)^{-\frac{3}{2}}\left( \partial^2_t\delta r+\frac{\sigma}{L^2}\frac{\overline{\nabla}^2\delta r}{D-2} \right)+\mathcal{O}\left( \frac{ \epsilon}{D} \right)  \\
\nonumber {u.\widehat{\nabla}\mathcal{K}}&=& \mathcal{O}(1)
\end{eqnarray}

Using these quantities in the charge membrane equation and comparing the coefficients of $  e^{-i \omega^q_l t} $, we obtain,
\begin{eqnarray}
-l \delta q &=& \partial_t \delta q\\
\Rightarrow \omega^q_l &=& -i l
\end{eqnarray}
The charge frequency also matches with \cite{Tanabe:2016opw} completely.
The matching of the quasi-normal modes with the pre-calculated results is a good consistency check of our analysis.

\section{ Conclusions and Future Directions}
To summarise in a nutshell, the objective of this paper is to solve the Einstein-Maxwell equations in presence of cosmological constant using the large D perturbative expansion up to the first sub-leading order. The inverse of the space-time dimension serves as the perturbation parameter. We have solved for the first subleading correction of the metric and the gauge field. We do not need to choose any specific coordinate system and it holds for both asymptotic AdS and dS geometry (or more precisely to any smooth geometry that satisfies pure Einstein equation in presence of cosmological constant).

 The membrane-gravity duality suggests that there is a one to one correspondence with the gravity solutions and the dynamical membrane which is defined by its shape and a velocity field on it. In our case, as we are dealing with a U(1) gauge field coupled to gravity, the dual membrane also has a charge density field on it. The membrane dynamics is captured entirely in terms of the scalar, vector and charge membrane equations. We have obtained these dual membrane equations and have seen that they emerge nicely from the  Einstein-Maxwell system.
 
It turns out that the charge membrane equation contains a curvature dependent term even in the first subleading correction. It is one of the most interesting results in this paper. It implies that unlike the uncharged case, the naive covariantization of the flat space membrane equations will not work for charged membrane. In the last part of our paper, we have shown a consistency check of our membrane equations by computing the light quasi-normal modes of few  known solutions of the Einstein-Maxwell equations both in Poincare patch and global AdS. Our QNM results match with the already calculated QNM frequencies, done from a purely gravitational approach in the literature.

 A very natural extension of this work would be to extend the calculation to the next order in $\left(1\over D\right)$ expansion. From our experience of the uncharged case, it seems that it is the order where we expect the entropy production. \\
It would also be interesting  to classify the stationary black hole / brane type geometries using the time independent solutions of our membrane equation which is done in \cite{Mandlik:2018wnw}. The obvious next step would be to explore the near-stationary geometries (which are dual charged fluid dynamics in our case) and compare them with the large-$D$ limits of hydrodynamics coupled with a conserved charge.

\section*{Acknowledgement}
We are extremely grateful to Sayantani Bhattacharyya for suggesting us this problem and also guiding us through numerous difficulties during the course of this problem. We are also thankful to her for going through the draft of this paper several times and helping us with useful suggestions. We would also like to thank Arjun Bagchi, Binata Panda, Parthajit Biswas, Yogesh Dandekar, Daniel Grumiller, Subhajit Majumdar, Mangesh Mandlik,  Shiraz Minwalla, Anup Kumar Mondal, Arunabha Saha and Somyadip Thakur for suggestions and helpful discussions. We would like to acknowledge the hospitality of NISER- Bhubaneswar and SINP-Kolkata during the progress of this work. We would also like to acknowledge our gratitude to the people of India for their steady and generous support to research in the basic sciences.\\

\vspace{10mm}

\appendix

\section{Proof of $\xbar{\n}_M F^{MN}= {\n}_M F^{MN}$} \label{appndx:delmfmn}
\paragraph{Metric:} For simplicity we write $G^{[0]}_{MN}$ as $g_{\sss{MN}}$ and $\bar{G}_{MN}$ as $\eta_{MN}$ throughout the appendices. (Please note that $\eta_{MN}$ is not necessarily a flat metric. It is a solution of the Pure gravity equation in \eqref{fineom}.)
\be{}
g_{MN}=G^{[0]}_{MN}=\eta_{MN}+f\: O_M\:O_N
\ee
\paragraph{Inverse metric}:
\be{}
g^{MN}=\eta^{MN}-f\: O^M\:O^N,\quad \text{such that}\: \quad g^{MN}\:g_{MN^\prime}=\delta^{N^\prime}_N
\ee
\paragraph{Field strength:}
\bes
\begin{eqnarray}
F_{MN}&=&\n_MA_N-\n_NA_M= \n_M(\tilde{f}O_N)-\n_N(\tilde{f}O_M)\\
F_{AB}\:\eta^{AM}\:\eta^{BN}&=& \n^M(\tilde{f}O^N)-\n^N(\tilde{f}O^M)
\end{eqnarray}
\ees
Raising the indices wrt full metric $g^{AM}$,
\bes
\bea{gaugered}
\nonumber F_{AB}\:g^{AM}\:g^{BN}&=&(\eta^{AM}-fO^AO^M)(\eta^{BN}-fO^BO^N)
\lb\n_M(\tilde{f}O_N)-\n_N(\tilde{f}O_M)\rb\quad \\
\nonumber&=&\lb\n^M(\tilde{f}O^N)-\n^N(\tilde{f}O^M)\rb+f\tilde{f}\lb
O^N(O\cdot\n)O^M-O^M(O\cdot\n)O^N\rb\\
\eea
\ees

We are using a subsidiary condition here,
\be{subO}
P^N_{M}(O\cdot \n)O_N=0
\ee
We can show from \eqref{subO}
\begin{equation}
    (O.\nabla) O^M = [n^N(O.\nabla)O_N] O^M=-[O^MO_N(O.\nabla)n^N] 
\end{equation}
Hence, from \eqref{gaugered},

\bea{}
\nonumber f\tilde{f}\lb
O^N(O\cdot\n)O^M-O^M(O\cdot\n)O^N\rb&=&
f\tilde{f}\lb -O^N O^M O_N(O\cdot\nabla)n^N+O^M O^N O_M(O\cdot\nabla)n^M\rb\\
&=&0
\eea

Hence from \eqref{gaugered}

\be{}
\mathbf{F_{AB}\:g^{AM}\:g^{BN}=F_{AB}\:\eta^{AM}\:\eta^{BN}}
\ee
\paragraph{Determinant of $g_{AB}$:}
\bes
\begin{eqnarray}
\nonumber g_{AB}&=&\eta_{AB}+fO_A O_B\\
     \nonumber &=& \eta_{AC}[\delta^C_B+fO^C O_B]\\
     \nonumber \det(g)&=&\det(\eta)\cdot \det[\mathbb{I}+f(O\cdot O)]\\
    &=&\det(\eta)
\end{eqnarray}
\ees

Hence,

\bea{}
\nonumber \xbar{\n}_M(F_{AB}\:g^{AM}\:g^{BN})
&=&\frac{1}{\sqrt{-g}}\p_M[\sqrt{-g}F_{AB}\:g^{AM}\:g^{BN}]\\
\nn &=&\frac{1}{\sqrt{-\eta}}\p_M[\sqrt{-\eta}F_{AB}\:\eta^{AM}\:\eta^{BN}]\\
&=& \n_M (F_{AB}\:\eta^{AM}\:\eta^{BN})
\eea

\section{Computation of the equation of motion} \label{appndx:eom}

\paragraph{Christoffel Connections:}
\bes
\bea{}
\nonumber\ga^A_{BC}&=&\frac{1}{2}g^{A A^\prime}\Big( \p_B\:g_{A^\prime C}+ \p_C\:g_{A^\prime B}- \p_{A^\prime}\:g_{B C} \Big)\\
\nonumber &=&\frac{1}{2} \eta^{A A^\prime}\lb \p_B\:\eta_{A^\prime C}+ \p_C \:\eta_{A^\prime B}- \p_{A^\prime} \: \e_{BC}\rb\\
\nonumber&&+ \frac{1}{2}\lb\n_B(f O^A O_C)+\n_C(f O^A O_B)-\n^A(f O_B O_C) \rb\\
&&-\frac{1}{2}(f O^A O_A^\prime)\lb \p_B(f O_{A^\prime}O_C)+\p_C(f O_{A^\prime}O_B) -\p_{A^\prime}(f O_BO_C)  \rb\\
&=&\tilde{\ga}^A_{BC}+[\ga_{\text{L}}]^A_{BC}+
[\ga_{\text{NL}}]^A_{BC}
\eea
\ees
Where, 
\bes
\bea{}
\nonumber \tilde{\ga}^A_{BC}&=&\ga\quad \text{wrt}\quad \eta\\
&=&\frac{1}{2} \eta^{A A^\prime}\lb \p_B\:\eta_{A^\prime C}+ \p_C \:\eta_{A^\prime B}- \p_{A^\prime} \: \e_{BC}\rb\\
\nonumber [\ga_{\text{L}}]^A_{BC}&=&\ga\quad \text{linear wrt}\quad G^{(0)}_{AB}\\
&=&\frac{1}{2}\lb\n_B(f O^A O_C)+\n_C(f O^A O_B)-\n^A(f O_B O_C) \rb\\
\nonumber [\ga_{\text{NL}}]^A_{BC}&=&\ga\quad \text{non-linear wrt}\quad G^{(0)}_{AB}\\
\nonumber &=&-\frac{1}{2}(f O^A O_A^\prime)\lb \p_B(f O_{A^\prime}O_C)+\p_C(f O_{A^\prime}O_B) -\p_{A^\prime}(f O_BO_C)\rb\\
&=&\frac{1}{2}fO^A(O.\nabla)(fO_B O_C)\\
\eea
\ees
Also, by using the null vector condition,
\bes
\bea{}
O^A O_A&=&0\\
\n_B(O^A O_A)&=&2 O_A(\n_BO^A )=0,\:\quad O_A(\n_BO^A )=0
\eea
\ees
we can easily see that,
\bea{}
[\ga_{\text{L}}]^A_{BA}=0,\quad [\Gamma_{\text{NL}}]^A_{BA}=0
\eea
Here all the derivatives $\n$ are wrt the background metric $\eta_{AB}$. For our convenience we are writing
\bea{}
[\ga_{\text{L}}]^A_{BC}+
[\ga_{\text{NL}}]^A_{BC}=\delta\ga^A_{BC}
\eea
\paragraph{Ricci tensor:} Now we will compute the Ricci tensor $R_{AB}$ from the Christopher connections.

\begin{eqnarray}
    R_{AB} &=& \p_c [\tilde{\ga}+\delta\ga]^C_{AB}-\p_B[\tilde{\ga}+\delta\ga]^C_{AC}\\
   \nonumber && +[\tilde{\ga}+\delta\ga]^C_{CE}[\tilde{\ga}+\delta\ga]^E_{AB} - [\tilde{\ga}+\delta\ga]^C_{BE}[\tilde{\ga}+\delta\ga]^E_{CA}\\
     &=&\tilde{R}_{AB} + [R_L]_{AB} + [R_{NL}]_{AB}
\end{eqnarray}
where,
\bes
\begin{eqnarray}
  \nonumber  \tilde{R}_{AB} &=& \text{Ricci tensor wrt}\quad \eta_{AB}\\
    &=& \partial_C \tilde{\Gamma}^C_{AB} - \partial_B \tilde{\Gamma}^C_{CA} +\tilde{\Gamma}^C_{CE}\tilde{\Gamma}^E_{AB} - \tilde{\Gamma}^C_{BE}\tilde{\Gamma}^E_{CA}\\
    \nonumber{[R_{L}]}_{AB} &=& \text{ Linear part wrt} \: G^{(0)}_{AB} \\
    &=&\nabla_C {[\Gamma_L]}^C_{AB}\\
    \nonumber{[R_{NL}]}_{AB} &=&\text{ Non-linear part wrt}\: G^{(0)}_{AB} \\
    &=& \nabla_C {[\Gamma_{NL}]}^C_{AB} - {[\Gamma_L +\Gamma_{NL}]}^C_{BE}{[\Gamma_L +\Gamma_{NL}]}^E_{CA}
    \end{eqnarray}
\ees
we will use the above relations to compute each term separately in both of the equations of motion in \eqref{fineom}.

\section{Calculation of Source} \label{appndx:source}
\subsection{Notations and Identities}

\textbf{Identities:}
  \begin{equation}
  \begin{split}
  \nabla_B f &= -\left({DN\over \psi}\right)(f - \tilde f^2)~ n_B +2(Q \tilde f -\tilde f^2)~ {\cal Q}_B \\
  \\
  \nabla_B\tilde f &=\tilde f\bigg[- \left({DN\over \psi}\right) n_B +{\cal Q}_B\bigg]\\
  \\
 (O\cdot \nabla)  f &= -\left({DN\over\psi}\right) (f-\tilde f^2) +2\left( O\cdot{\cal Q}\right)\left(Q\tilde f - \tilde f^2\right) \\
 \\
 (O\cdot \nabla) \tilde f &= -\left({DN\over\psi} - O\cdot{\cal Q}\right)\tilde f\\
 \\
 (O\cdot \nabla)^2  f &= \left({DN\over\psi}\right)^2 (f-3\tilde f^2) -4\left({DN\over\psi}\right)\left( O\cdot{\cal Q}\right)\left(Q\tilde f - 2\tilde f^2\right) -(f-\tilde f^2)(O\cdot\nabla)\left(DN\over\psi\right)+{\cal O}(1) \\
 &= -\left({DN\over\psi} \right)(O\cdot\nabla) f -2 \left({DN\over\psi} \right)^2\tilde f^2-2\left({DN\over\psi} \right) (O\cdot{\cal Q})(Q\tilde f -3\tilde f^2)\\
 &~~~ -(f-\tilde f^2)(O\cdot\nabla)\left(DN\over\psi\right)+{\cal O}(1) \\
 \\
 \nabla^2 f &= \left(1 + Q^2 - 2 Q^2 \psi^{-D}\right) \nabla^2\left(\psi^{-D}\right)-4\left({DN\over\psi}\right)(n\cdot{\cal Q})\left(Q\tilde f -\tilde f^2\right)\\
 &~~-2\left({DN\over\psi}\right)^2\tilde f^2 + 2\left(\nabla\cdot{\cal Q}\right)\left(Q\tilde f - \tilde f^2\right) + {\cal O}(1)
  \end{split}
  \end{equation}
  
Note that in the last identity, the first line vanishes because of the subsidiary conditions.\\
\textbf{Notation:}
\begin{equation}
\begin{split}
&f\equiv (1 + Q^2) \psi^{-D} - Q^2 \psi^{-2D}\\
& \tilde f \equiv Q~ \psi^{-D}\\
&z = n^A (O\cdot\nabla) O_A = \frac{(u\cdot\nabla) K}{K} - u\cdot K\cdot u\\
&{\cal Q}_C \equiv \frac{\nabla_C Q}{Q}
\end{split}
\end{equation}

\subsection{Einstein Equation} \label{appndx:einsteinsource}

\begin{equation}\label{RLAB}
\begin{split}
R^{(L)}_{AB} =&~(D-1) \lambda f O_A O_B +\underbrace{\frac{1}{2}\left(O_B\nabla_A + O_A\nabla_B\right) \bigg[f (\nabla\cdot O)+ (O\cdot\nabla) f\bigg]}_{L_1} -\underbrace{\left({\nabla^2f\over 2}\right)O_A O_B}_{L_2}\\
&~+\frac{1}{2}\bigg[f (\nabla\cdot O)+ (O\cdot\nabla) f\bigg]\left(\nabla_A O_B +\nabla_B O_A\right) -(\nabla_C f)\nabla^C(O_A O_B)
-{f\over 2}\nabla^2(O_A O_B)\\
&~+{z\over 2}\left(O_B\nabla_A + O_A\nabla_B\right) f+ {\cal O}(1)
\end{split}
\end{equation}

\begin{equation}\label{RNLAB}
\begin{split}
R^{(NL)}_{AB} =&~\underbrace{{f\over 2}\bigg[(O\cdot\nabla f)(\nabla\cdot O) + (O\cdot\nabla)^2 f\bigg]O_A O_B}_{N_1}\\
&~+zf  \bigg[ f(\nabla\cdot O) +  (O\cdot\nabla) f\bigg] O_A O_B + {\cal O}(1)
\end{split}
\end{equation}

\begin{equation}\label{FFAB}
\begin{split}
{1\over 2}F_{AC}{F^C}_B =&~\underbrace{\bigg[(O\cdot\nabla)\tilde f\bigg] \left(O_B\nabla_A +O_A\nabla_B\right) \tilde f }_{T_1}+ \underbrace{\bigg( f\left[(O\cdot\nabla) \tilde f + z\tilde f\right]^2 - [\nabla_C\tilde f][\nabla^C\tilde f]\bigg) O_AO_B}_{T_2} \\
~&+\frac{z}{2} \left(O_B\nabla_A + O_A\nabla_B\right)\tilde f^2 +{1\over 2} \left[ \left(O_B\nabla_A + O_A\nabla_B\right)O^C\right]\left(\nabla_C\tilde f^2\right)\\
&-{1\over 2}\left(\nabla_C\tilde f^2\right)\nabla^C(O_A O_B)+ {\cal O}(1)
\end{split}
\end{equation}

\begin{equation}\label{Fsquar}
\begin{split}
F^2\equiv F_{AB}F^{AB} =-4\left[(O\cdot\nabla)\tilde f\right]^2 + {\cal O}(D)=-4~K^2\tilde f^2 + {\cal O}(D)
\end{split}
\end{equation}
Here the highlighted term are the ones that could contribute at order ${\cal O}(D^2)$.
 
Final form of the equation.
\begin{equation}
R^{(L)}_{AB} + R^{(NL)}_{AB} +{1\over 2} F_{AC}{F^C}_B  +\left({1\over 4D}\right)\left(F_{A'B'}F^{A'B'} \right)G_{AB}-(D-1)\lambda G_{AB}=0
\end{equation}

\subsubsection{Calculation at ${\cal O}(D^2)$}
\begin{equation}\label{step1}
\begin{split}
T_1\equiv&\bigg[(O\cdot\nabla)\tilde f\bigg] \left(O_B\nabla_A +O_A\nabla_B\right) \tilde f \\
=~&{1\over 2}\left[-{DN\over\psi} + O\cdot{\cal Q}\right] \left(O_B\nabla_A +O_A\nabla_B\right) \tilde f^2\\
=~&-{1\over 2}\left(O_B\nabla_A +O_A\nabla_B\right) \left[\left({DN\over \psi}\right)\tilde f^2\right] + {\tilde f ^2\over 2}\left(O_B\nabla_A +O_A\nabla_B\right)\left(DN\over\psi\right) \\
&+\left({O\cdot{\cal Q}\over 2}\right) \left(O_B\nabla_A +O_A\nabla_B\right) \tilde f^2
\end{split}
\end{equation}

\begin{equation}\label{step2}
\begin{split}
L_1\equiv&\frac{1}{2}\left(O_B\nabla_A + O_A\nabla_B\right) \bigg[f (\nabla\cdot O)+ (O\cdot\nabla) f\bigg]\\
=~&\frac{1}{2}\left(O_B\nabla_A + O_A\nabla_B\right) \bigg[f\left(\nabla\cdot O - {DN\over\psi}\right) +\left({DN\over \psi}\right)\tilde f^2 +2(O\cdot{\cal Q})(Q\tilde f -\tilde f^2)\bigg]  \\
=~&\frac{1}{2}\left(O_B\nabla_A + O_A\nabla_B\right) \bigg[f\left(\nabla\cdot O - {DN\over\psi}\right) +\left({DN\over \psi}\right)\tilde f^2\bigg] \\
&+\left( O\cdot{\cal Q}\right)\left(O_B\nabla_A + O_A\nabla_B\right) (Q\tilde f -\tilde f^2) + {\cal O}(1)
\end{split}
\end{equation}
Adding \eqref{step1} and \eqref{step2}
\begin{equation}\label{1plus2}
\begin{split}
L_1 + T_1 
\equiv&~\frac{1}{2}\left(O_B\nabla_A + O_A\nabla_B\right)\bigg[f\left(\nabla\cdot O - {DN\over\psi}\right)\bigg]\\
&~-\left(\frac{DN}{\psi}\right)\left( Q\tilde f-\tilde f^2\right)\left( O\cdot{\cal Q}\right)\left(O_Bn_A + O_An_B\right)\\
&~ + {\tilde f^2\over 2}(O_B\nabla_A + O_A\nabla_B)\left({DN\over\psi}\right)
\end{split}
\end{equation}

\begin{equation}\label{step3}
\begin{split}
N_1\equiv ~&{f\over 2}\bigg[(O\cdot\nabla f)(\nabla\cdot O) + (O\cdot\nabla)^2 f\bigg]O_A O_B\\
=~&{f\over 2}\bigg[(O\cdot\nabla f)\left(\nabla\cdot O -\frac{DN}{\psi}\right)-2 \left({DN\over\psi} \right)^2\tilde f^2\\
&~~~~~-2\left({DN\over\psi} \right) (O\cdot{\cal Q})(Q\tilde f -3\tilde f^2) -(f-\tilde f^2)(O\cdot\nabla)\left(DN\over\psi\right)\bigg] O_A O_B
\end{split}
\end{equation}

\begin{equation}\label{step4}
\begin{split}
T_2\equiv ~&\bigg( f\left[(O\cdot\nabla) \tilde f + z\tilde f\right]^2 - [\nabla_C\tilde f][\nabla^C\tilde f]\bigg) O_AO_B\\
=~&\bigg[\left({DN\over\psi} \right) ^2\left(f\tilde f^2 -\tilde f^2\right) -2\left({DN\over\psi} \right) \left(O\cdot{\cal Q} + z\right)\tilde f ^2f \bigg] O_A O_B+ {\cal O}(1)
\end{split}
\end{equation}

\begin{equation}\label{step5}
\begin{split}
L_2\equiv ~&-\left({\nabla^2f\over 2}\right)O_A O_B\\
=~&-{1\over2}\bigg[\left(1 + Q^2 - 2 Q^2 \psi^{-D}\right) \nabla^2\left(\psi^{-D}\right)-4\left({DN\over\psi}\right)(n\cdot{\cal Q})\left(Q\tilde f -\tilde f^2\right)\\
 &~~~~~~~~~-2\left({DN\over\psi}\right)^2\tilde f^2 + 2\left(\nabla\cdot{\cal Q}\right)\left(Q\tilde f - \tilde f^2\right) \bigg] O_A O_B+ {\cal O}(1)
\end{split}
\end{equation}

Adding \eqref{step3} , \eqref{step4} and \eqref{1plus2} with $L_2$ we find
\begin{equation}\label{L2PN1PT2}
\begin{split}
&L_2 + N_1 +T_2\\
\equiv~&{f\over 2}\left[(O\cdot\nabla)f\right]\left[\nabla\cdot O- {DN\over\psi}\right]O_A O_B - {1\over 2}\left(1 + Q^2 - 2 Q^2 \psi^{-D}\right) \nabla^2\left(\psi^{-D}\right)O_A O_B\\
~&-{\cal X}~ O_A O_B\\
&\text{where}\\
&{\cal X} \equiv~(Q\tilde f -\tilde f^2)\bigg[(\nabla\cdot {\cal Q})+(O\cdot{\cal Q})\left({DN\over\psi}\right)f\bigg]+{f\over 2}(f-\tilde f^2)(O\cdot\nabla)\left(DN\over\psi\right)+ 2 z \left({DN\over\psi}\right)\tilde f^2f\\
\end{split}
\end{equation}
Here we have used the subsidiary condition $n\cdot{\cal Q} =0$. It is ok to use the subsidiary condition for order ${\cal O}(D)$ terms as $n\cdot{\cal Q}$ does not appear at order ${\cal O}(D^2)$.\\
In ${\cal E}^{(1)}_{AB}$, the form of order ${\cal O}(D^2)$ piece now,
\begin{equation}\label{orderdsq}
\begin{split}
{\cal E}^{(1)}_{AB} = ~&\frac{1}{2}\left(O_B\nabla_A + O_A\nabla_B\right)\bigg[f\left(\nabla\cdot O - {DN\over\psi}\right)\bigg] +{f\over 2}\left[(O\cdot\nabla)f\right]\left[\nabla\cdot O- {DN\over\psi}\right]O_A O_B\\
~& + {\cal O}(D)
\end{split}
\end{equation}

\subsubsection{Calculation of  ${\cal O}(D)$ piece}
We shall write $S_{AB}$ at order ${\cal O}(1)$ as 
\begin{equation}\label{splitsource}
\begin{split}
S_{AB} \equiv&{\mathfrak S}_0 ~n_A n_B+{\mathfrak S}_1 ~O_A O_B + {\mathfrak S}_2 ~(n_A O_B + n_B O_A) + {\mathfrak S}_3~ P_{AB}\\
&+({\mathfrak V}^{(1)}_A O_B +{\mathfrak V}^{(1)}_B O_A) + ({\mathfrak V}^{(2)}_A n_B +{\mathfrak V}^{(2)}_B n_A)  + {\mathfrak T}_{AB}\\
\\
&n^A {\mathfrak V}^{(1)}_A = O^A {\mathfrak V}^{(1)}_A=n^A {\mathfrak V}^{(2)}_A=O^A {\mathfrak V}^{(2)}_A=0\\
&n^A {\mathfrak T}_{AB} = O^A {\mathfrak T}_{AB} =0,~~~P_{AB} = \bar G_{AB} - n_A n_B + u_A u_B\\
\end{split}
\end{equation}
To compute the different component we have to do the following projections.
\begin{equation}\label{excomp}
\begin{split}
&{\mathfrak S}_0 = O^A O^BS_{AB} ,~~~{\mathfrak S}_1 = u^Au^B S_{AB} ,~~~{\mathfrak S}_2 = u^A O^BS_{AB} ,~~~{\mathfrak S}_3=\left(1\over D\right)P^{AB}S_{AB}\\
&{\mathfrak V}^{(1)}_A = u^BP^C_A S_{CB} ,~~~{\mathfrak V}^{(2)}_A = O^BP^C_A S_{CB},~~~{\mathfrak T}_{AB} =P^C_A  P^{C'}_BS_{CC'}
\end{split}
\end{equation}
Now we shall decompose and simplify different terms into scalar vector and  tensor structures that would be requited for ${\cal O}(D)$  computation.
\begin{equation}\label{simpst1}
\begin{split}
&f (\nabla\cdot O) + (O\cdot\nabla) f = \left({DN\over \psi}\right) \tilde f^2 + {\cal O}(D)=K\tilde f^2 + {\cal O}(D)\\
\end{split}
\end{equation}

\begin{equation}\label{simplist2}
\begin{split}
L_3\equiv&{1\over 2}\left[f (\nabla\cdot O) + (O\cdot\nabla) f \right]\left(\nabla_A O_B + \nabla_B O_A\right)\\
 =~&K \tilde f^2\bigg[(u.K.u)~O_A O_B +{ z\over2} ~(n_A O_B + n_B O_A) +{1\over 2}\left(O_A P^C_B + O_B P^C_A\right)\left((u\cdot\nabla)O_C+u^{C'}K_{CC'}\right)\\
 &~~~~~~+{1\over 2} P^C_A P^{C'}_B \left(\nabla_{C'} O_C +\nabla_C O_{C'}\right)\bigg]
\end{split}
\end{equation}

\begin{equation}\label{simplist3}
\begin{split}
L_4\equiv& -(\nabla_C f)\nabla^C(O_A O_B)\\
=~&2 K (f-\tilde f^2) \left(\frac{(u\cdot\nabla)K}{K} \right) O_A O_B  + K\left(f -\tilde f^2\right)\left(O_A P^C_B + O_B P^C_A\right) (u\cdot\nabla) O_C+ {\cal O}(1)\\
\end{split}
\end{equation}

\begin{equation}\label{simplist4}
\begin{split}
L_5\equiv& -{f\over 2}\nabla^2(O_A O_B)\\
=~&-2f \left[(u\cdot\nabla) K\right]O_A O_B  +{f\over 2}\left(K^2\over {D}\right)\left(n_A O_B + n_B O_A\right)-{f\over 2}\left(O_A P^C_B + O_B P^C_A\right) \nabla^2O_C + {\cal O}(1)\\
\end{split}
\end{equation}

\begin{equation}\label{simplist5}
\begin{split}
L_6\equiv& {z\over 2}\left(O_B\nabla_A + O_A\nabla_B\right) f
=-K \left({z\over 2}\right)\left(f-\tilde f^2\right)\left(n_B O_A + n_A O_B\right)+ {\cal O}(1)\\
\end{split}
\end{equation}

\begin{equation}\label{simplist6}
\begin{split}
N_2\equiv zf  \bigg[ f(\nabla\cdot O) +  (O\cdot\nabla) f\bigg] O_A O_B = zK f\tilde f^2~ O_A O_B+ {\cal O}(1)\\
\end{split}
\end{equation}

\begin{equation}\label{simplist7}
\begin{split}
T_3\equiv\frac{z}{2} \left(O_B\nabla_A + O_A\nabla_B\right)\tilde f^2=-z K\tilde f^2 (n_A O_B+ n_B O_A)+ {\cal O}(1)\\
\end{split}
\end{equation}

\begin{equation}\label{simplist8}
\begin{split}
T_4\equiv&~{1\over 2} \left[ \left(O_B\nabla_A + O_A\nabla_B\right)O^C\right]\left(\nabla_C\tilde f^2\right)\\
=&~-K\tilde f^2\bigg[ 2(u\cdot K\cdot u)O_A O_B + z (n_A O_B +n_B O_A)+ \left(O_A P^C_B + O_B P^C_A\right)\left(u^{C'}K_{CC'}\right)\bigg]\\
\end{split}
\end{equation}

\begin{equation}\label{simplist9}
\begin{split}
T_5\equiv&-{1\over 2}\left(\nabla_C\tilde f^2\right)\nabla^C(O_A O_B)=2\tilde f^2\left[(u\cdot\nabla  K) ~O_A O_B +{K\over 2}\left(O_A P^C_B + O_B P^C_A\right)(u\cdot\nabla)O_C\right]
\end{split}
\end{equation}

\begin{equation}\label{tracepiece}
\begin{split}
T_6\equiv\left({F^2\over 4D}\right)G_{AB} = \left(K^2\over D\right)\tilde f^2\bigg[(1-f)O_A O_B -( n_B O_A+  n_AO_B) - P_{AB}\bigg]
\end{split}
\end{equation}

In \eqref{simplist3}, \eqref{simplist4}, \eqref{simplist9} and \eqref{typeAS2} we have used the identities.
\begin{equation}\label{iden}
\begin{split}
(n\cdot\nabla) O_A &= \left(\frac{(u\cdot\nabla)K}{K} \right)  O_A + P^C_A (u\cdot\nabla)O_C + {\cal O}\left(\frac{1}{D}\right)\\
\nabla^2O_A &= 2 (u\cdot\nabla)K ~O_A - \left({K^2\over D}\right) n_A + P^C_A\nabla^2O_C + {\cal O}(1)\\
(O_A \n_B +O_B \n_A)(\n \cd u) &= (n \cd \n)(\n \cd u)( n_A O_B+n_B O_A)+{\cal O}(1)
\end{split}
\end{equation}

Now , the way we are organizing our computation,  the $ {\cal O}(D)$ pieces in the source, will have two different types of contribution: the first one coming from  those terms in \eqref{RLAB}, \eqref{RNLAB} and \eqref{FFAB} which naively looks like ${\cal O}(D^2)$, but an order ${\cal O}(D)$ piece is also hidden in it, once we cancel the leading piece. This is true only for ${\mathfrak S}_1$  and ${\mathfrak S}_2$. We shall refer to such terms as `Type A' contribution. The rest of the order ${\cal O}(D)$ will be called `Type-B' contribution.

\begin{equation}
\begin{split}
&\text {Type -A contribution to ${\mathfrak S}_1$} \\
=~&{f\over 2}\left[(O\cdot\nabla)f\right]\left[\nabla\cdot O- {DN\over\psi}\right]\\
&\text{Total contribution to ${\mathfrak S}_1$  from \eqref{1plus2} and \eqref{L2PN1PT2} }\\
=~&{f\over 2}\left[(O\cdot\nabla)f\right]\left[\nabla\cdot O- {DN\over\psi}\right] - {\cal X}+\tilde f^2(u\cdot\nabla K)\\
=~&{f\over 2}(f -\tilde f^2) \bigg[(u\cdot\nabla)K + K(\nabla\cdot u) \bigg] - (Q \tf -\tilde f^2)\bigg[\nabla\cdot {\cal Q} - f K (u\cdot{\cal Q})\bigg] - 2 z K \tilde f^2 f \\
&~~+\tilde f^2(u\cdot\nabla K)
\end{split}
\end{equation}
Here I have used all subsidiary conditions.
\begin{equation}\label{simpsub}
\begin{split}
&{DN\over\psi} = K + \frac{(n\cdot\nabla )K}{K} + {\cal O}\left(1\over D\right),~~n\cdot{\cal Q} =0\\
\end{split}
\end{equation}

\begin{equation}\label{typeAS2}
\begin{split}
&\text {Type -A contribution to ${\mathfrak S}_2$} \\
=~&-{1\over 2}\left(DN\over \psi\right)\left(f - \tilde f^2\right)\left(\nabla\cdot O  - {DN\over\psi}\right)-\frac{f}{2}(n \cd \n)(\n \cd u) \\
&\text {Total contribution to ${\mathfrak S}_2$ from \eqref{1plus2}}\\
=~&-{1\over 2}\left(DN\over \psi\right)\left(f - \tilde f^2\right)\left(\nabla\cdot O  - {DN\over\psi}\right) +{\tilde f^2\over 2}(O\cdot\nabla) K\\
&~~~- \left(DN\over \psi\right)(Q\tilde f - \tilde f^2)\left(O\cdot{\cal Q}\right)-\frac{f}{2}(n \cd \n)(\n \cd u)\\
=~&\bigg[{1\over 2}\left(f - \tilde f^2\right)\left(n\cdot\nabla K + K\nabla\cdot u\right)+K\left(u\cdot{\cal Q}\right)(Q\tilde f - \tilde f^2)+ {\tilde f^2\over 2}(O\cdot\nabla)K-\frac{f}{2}(n \cd \n)(\n \cd u) \bigg]\\
=~&\bigg[{f\over 2}\left(n\cdot\nabla K + K\nabla\cdot u -(n \cd \n)(\n \cd u)\right)-{\tilde f^2\over 2}\left(u\cdot\nabla K + K\nabla\cdot u\right)+K\left(u\cdot{\cal Q}\right)(Q\tilde f - \tilde f^2)\bigg]
\end{split}
\end{equation}

\begin{equation}
\begin{split}
&\text {Type -A contribution to ${\mathfrak V}^{(1)}_A$} 
={\tilde f^2\over 2}\left(P^C_A\nabla_C K\right)
\end{split}
\end{equation}

\subsection{The final form of the source  $S_{AB}$}
We shall write the answer in terms of $\hat\nabla\cdot u \equiv \Pi^{AB}\nabla_A u_B$.
We shall use the following identities to simplify the answer
\begin{equation}\label{newiden}
\begin{split}
&\hat\nabla\cdot u = \nabla\cdot u +{(u\cdot\nabla)K\over K} +{\cal O}(1)\\
&P^C_B\nabla^2O_C = P^C_B\left[\nabla_C K -\hat\nabla^2u_C + K (u^DK_{DC} - (u\cdot\nabla) u_C)\right] + {\cal O}(1)\\
&(n\cdot\nabla) K =\left(\hat \nabla^2 K\over K\right) - \left(K^2\over D\right) - \bar R_{nn} + {\cal O}(1)\\
&(n\cdot\nabla)(\nabla\cdot u) =- \bar R_{u O} +\left(\hat \nabla^2 K\over K\right)  - 2(u\cdot\nabla) K + K(u\cdot K\cdot u)+ {\cal O}(1)\\
\end{split}
\end{equation}

Here $\hat\nabla^2 u_C$ and $\hat\nabla^2 K$ denote the following
$$\hat\nabla^2 u_C \equiv \Pi_A ^B\nabla_B\left(\Pi_C^K\Pi^{AK'}\nabla_{K'}u_K\right),~~~~\hat \nabla^2  K \equiv \Pi^{AA'}\nabla_A(\Pi^B_{B'}\nabla_B K)$$

\begin{equation}\label{sourceec}
\begin{split}
{\mathfrak S}_1=~&-\left(Q \tf - \tilde f^2\right) \left[\nabla\cdot{\mathcal Q} - f K (u\cdot{\cal Q})\right] + {f\over 2}\left(f -\tilde f^2\right) K(\hat\nabla\cdot u)- K (u\cdot K\cdot u)\tilde f^2 (1-f)\\
&+\tf^2\left(1-f\right)(u\cdot\nabla K) +\left(K^2\over D\right)\tilde f^2(1-f)\\
\\
{\mathfrak S}_2=~&
K \left(u\cdot {\cal Q}\right) \left(Q \tilde f -\tilde f^2\right)+\tilde f^2\left[ K (u\cdot K\cdot u) - (u\cdot \nabla) K -{K\over 2}(\hat\nabla\cdot u)\right]\\
&+{f\over 2}\left[K(\hat\nabla\cdot u)+ (n\cdot\nabla) K + {K^2\over D} -2(u\cdot\nabla)K+ K (u\cdot K\cdot u)-(n \cd \n)(\n \cd u)\right] -\left(K^2\over D\right)\tilde f^2\\
=~&K \left(u\cdot {\cal Q}\right) \left(Q \tilde f -\tilde f^2\right)+\tilde f^2\left[ K (u\cdot K\cdot u) - (u\cdot \nabla) K -{K\over 2}(\hat\nabla\cdot u)\right]\\
&+{f\over 2}\left[K(\hat\nabla\cdot u)\right] -\left(K^2\over D\right)\tilde f^2\\
\\
{\mathfrak S}_3 =~&0\\
\\
{\mathfrak V}_A^{(1)}=~&P^C_A\bigg[{K\over 2}f \left(u^{C'}K_{C'C} - (u\cdot\nabla )u_C\right) +{f\over 2} \hat\nabla^2u_C-{K\over 2}\tilde f^2  (u\cdot\nabla )u_C -\left({f-\tilde f^2}\over 2\right)\nabla_C K\bigg]\\
\\
{\mathfrak V}_A^{(2)}=~&0\\
\\
{\mathfrak T}_{AB}=~&\tilde f^2\bigg[\left({K\over 2}\right)P^C_A P^{C'}_B \left(\nabla_C O_{C'} +\nabla_{C'} O_C\right)-\left({K^2\over D}\right)P_{AB}\bigg]
\end{split}
\end{equation}

\subsection{Source from the Maxwell Equation $S_N$}\label{appndx:maxwellsource}

 We have already shown in appendix (A) that we can do the raising of $\n_M F^{MN}$ entirely wrt $\eta^{AB}$ and the covariant derivative is also wrt the background $\eta^{AB}$.
\bes
\bea{}
\nn \n_M F^{MN} &=&\n_M (\n^M A^N - \n^N A^M)\\
 &=&\n^2 A^N-\n_M \n^N A^M\\
\n^2 A^N&=&(\n^2 \tf)O^N+2(\n^M \tf)(\n_M O^N)+\tf \n^2 O^N\\
\n_M \n^N A^M&=&(O\cd\n)(\n^n \tf)+(\n^n \tf)(\n \cd O)+(\n_M \tf)(\n^N O^M)+\tf \n_M\n^N O^M\quad \quad
\eea 
\ees

Form of the source 
\begin{equation}\label{maxsource}
\begin{split}
S^N &=\underbrace{(\nabla^2 \tilde f)O^N}_{M_1}+\underbrace{(-O\cdot\nabla)(\nabla^N \tilde f)}_{M_2}+\underbrace{(-\nabla^N \tilde f)(\nabla \cdot O)}_{M_3}+\underbrace{\tilde f \nabla^2 O^N}_{M_4}\\
&+\underbrace{2(\nabla^M \tilde f)(\nabla_M O^N)}_{M_5}+\underbrace{(-\nabla_M \tilde f)(\nabla^N O^M)}_{M_6}+\underbrace{(-\tilde f \nabla_M\nabla^N O^M)}_{M_7}
\end{split}
\end{equation}


\subsubsection{More Identities and notation}
\begin{equation}\label{imax}
\begin{split}
&\nabla^2\tilde f = \left(\nabla\cdot{\cal Q} \right)\tilde f + Q~\nabla^2\psi^{-D} +{\cal O}(1)\\
&(n\cdot\nabla) n^N = \left(u\cdot\nabla K\over  K\right) (O^N-n^N) + P^{NC}\left({\nabla_C K\over K} \right)\\
&(n\cdot\nabla) (\nabla\cdot u) ={\hat\nabla^2 K\over K} - 2(u\cdot\nabla) K + K (u\cdot K\cdot u) -\bar R_{uO}\\
&\nabla^N (\nabla\cdot O)  = n^N \bigg[2(u\cdot\nabla)K - K(u\cdot K\cdot u)-\left(K^2\over D\right)\bigg] + O^N \left[(u\cdot\nabla)K\right] + P^{NC}\nabla_C K\\
\\
&\nabla_M\nabla^N O^M = n^N \left[2(u\cdot\nabla)K - K(u\cdot K\cdot u)-\left(K^2\over D\right)\right] + O^N \left[(u\cdot\nabla) K+ \bar R_{uO}\right] + P^{NC}\nabla_C K\\
\\
&P^C_B\nabla^2O_C = P^C_B\left[\nabla_C K -\hat\nabla^2u_C + K (u^DK_{DC} - (u\cdot\nabla) u_C)\right] + {\cal O}(1)\\
&(n\cdot\nabla) (\nabla\cdot O)=~\bar R_{OO}  - \left(K^2\over D\right) + 2(u\cdot\nabla) K - K(u\cdot K\cdot u)+ {\cal O}(1)\\
&~~~~~~~~~~~~~~~~~~= - \left(K^2\over D\right) + 2(u\cdot\nabla) K - K(u\cdot K\cdot u)+ {\cal O}(1)\\
\end{split}
\end{equation}
Here $\bar R_{Ou} \equiv  O\cdot\bar R\cdot u$

\subsubsection{Different terms simplified}
\begin{equation}\label{M1}
\begin{split}
M_1\equiv~&(\nabla^2 \tilde f) O^N
=Q~\nabla^2\left(\psi^{-D}\right) O^N +\tilde f ~(\nabla\cdot{\cal Q})O^N
\end{split}
\end{equation}

\begin{equation}\label{M2}
\begin{split}
M_2\equiv~(-O\cdot\nabla)(\nabla^N \tilde f)
=-\tilde f\left(DN\over\psi\right)^2 n^N 
+ \tilde f K\bigg[&\left({(O\cdot\nabla)K\over K}-(u\cdot{\cal Q}) -z\right)  n^N+z~ O^N\\
&+P^{NC}\left({\nabla_C K\over K} -u^AK_{AC}\right) + {\cal Q}^N\bigg]
\end{split}
\end{equation}

\begin{equation}\label{M3}
\begin{split}
M_3\equiv~&(-\nabla^N\tilde f)(\nabla\cdot O)=\tilde f\left(DN\over\psi\right)(\nabla\cdot O)~ n^N-\tilde f ~(\nabla\cdot O)~{\cal Q}^N\\
\end{split}
\end{equation}
We shall first simplify the sum of the first three terms. These are the terms which can naively contribute at order ${\cal O}(D^2)$ and once they are canceled, will contribute `type-A' terms to the final source. 

\begin{equation}\label{M1M2M3}
\begin{split}
M_1 + M_2+ M_3 =~&\tilde f  \left(DN\over\psi\right)\left[\nabla\cdot O- \left(DN\over\psi\right)\right]~n^N + Q~\nabla^2(\psi^{-D})~ O^N
+\tilde f\left(K - \nabla\cdot O\right){\cal Q}^N\\
&+ \tilde f K\bigg[\left({(O\cdot\nabla)K\over K}-(u\cdot{\cal Q}) -z\right)  n^N+\left({\nabla\cdot{\cal Q}\over K} +z\right)~ O^N\\
&~~~~~~~~~~~+P^{NC}\left({\nabla_C K\over K} -u^AK_{AC}\right)\bigg]\\
\\
=~&- \tilde f ~\bigg[K~\nabla\cdot u +2(u\cdot\nabla)K+K(u\cdot{\cal Q}) -K(u\cdot K\cdot u)\bigg] n^N\\
&+\tilde f~\left[\nabla\cdot{\cal Q} +(u\cdot\nabla K) - K(u\cdot K\cdot u)\right]~ O^N\\
&+\tilde f~P^{NC}\left(\nabla_C K-K~u^AK_{AC}\right)\\
\end{split}
\end{equation}

\begin{equation}\label{M4}
\begin{split}
M_4\equiv~&\tilde f \nabla^2 O^N
=\tilde f\bigg[-\left({K^2\over D}\right) n^N + 2\left[ (u\cdot\nabla) K \right] O^N + P^{NC}\nabla^2 O_C\bigg]
\end{split}
\end{equation}

\begin{equation}\label{M5}
\begin{split}
M_5\equiv~&2(\nabla^M \tilde f)(\nabla_M O^N)
=-2\tilde f \left(DN\over\psi\right)(n\cdot\nabla) O^N\\
=~&-2\tilde f (u\cdot\nabla) K~O^N - 2\tilde f K ~P^{NC}\left[u^A K_{AC}  - (u\cdot\nabla) u_C\right]
\end{split}
\end{equation}

\begin{equation}\label{M6}
\begin{split}
M_6\equiv~&(-\nabla_M \tilde f)(\nabla^N O^M)= \tilde f\left(DN\over\psi\right)n_M (\nabla^N O^M)\\
=~&\tilde f\bigg[\bigg([u\cdot\nabla] K - K[ u\cdot K\cdot u]\bigg)~ n^N + K(u\cdot K\cdot u) O^N + K~P^{NC} u^A K_{AC}\bigg]
\end{split}
\end{equation}

\begin{equation}\label{M7}
\begin{split}
M_7\equiv~&-\tilde f~ \nabla_M\nabla^N O^M= -\tilde f\bigg[\nabla^N(\nabla\cdot O) +\bar G^{PN} [\nabla_M,\nabla_P]O^M\bigg]\\
=&~-\tilde f\bigg(\left[2(u\cdot\nabla)K - K(u\cdot K\cdot u)-\left(K^2\over D\right)\right]n^N+ O^N \left[(u\cdot\nabla)K+\bar R_{uO}\right] + P^{NC}\left[\nabla_C K+\bar  R_{CA}O^A\right]\bigg)
\end{split}
\end{equation}

The final form of the source $S^N$
\begin{equation}\label{finsource}
\begin{split}
S^N=~&-\tilde f\left[K( \hat\nabla\cdot u + u\cdot{\cal Q}) +(u\cdot\nabla) K  - K(u\cdot K\cdot u) -( \bar R_{uu} +\bar R_{nn})\right] n^N\\
&+\tilde f\left(\nabla\cdot{\cal Q} - \bar R_{uO}\right)O^N\\
&+\tilde f~ P^{NC}\left[\nabla_C K -\hat\nabla^2 u_C -  K u^A(\nabla_A O_C) -\bar R_{CA}O^A \right]
\end{split}
\end{equation}


\section{Calculation of Homogeneous piece}\label{appndx:homo}
\begin{equation}\label{decomph}
\begin{split}
H_{AB}&=H_{AB}^{\text{scalar}}+H_{AB}^{\text{vector}}+H_{AB}^{\text{trace}}+H_{AB}^{\text{tensor}}+H_{AB}^{\text{Gauge scalar}}+H_{AB}^{\text{Gauge vector}}\\
H^M&=H^{M}_{\text{scalar}}+H^{M}_{\text{vector}}
+H^{M}_{\text{trace}}+H^{M}_{\text{tensor}}+H^{M}_{\text{Gauge scalar}}+H^{M}_{\text{Gauge vector}}
\end{split}
\end{equation}

\subsection{Calculation of homogeneous part   ${H}_{AB}$}
We are writing the first sub-leading metric correction $G_{AB}^{(1)}$ as $h_{AB}$ for simplicity.
\bes
\bea{}
  \nn  h_{AB} &=& \underbrace{\sum_{i} S^{(i)}_1(\zeta) \mathcal{S}^{(i)} O_A O_B}_\text{Scalar Sector} + \underbrace{\sum_{i} \frac{S_2^{(i)}(\zeta)}{D}\mathcal{P}_{AB}}_\text{Trace} \\
    &&+ \underbrace{\sum_{i} \mathcal{V}^{(i)}(\zeta)(V^{(i)}_A O_B + V^{(i)}_B O_A)}_\text{Vector Sector} + \underbrace{\sum_{i} \mathcal{T}^{(i)}(\zeta)t^{(i)}_{AB}}_\text{trace-less tensor sector}
\eea
\ees
Here, 
\bes
\bea{}
\zeta&=&D(\psi-1)\\
\mathcal{P}_{AB} &=& \eta_{AB} -n_A n_B + u_A u_B = \eta_{AB} -(n_A O_B + n_B O_A) + O_A O_B\quad \quad \:\\
O\cd V^{(i)}&=&n\cd V^{(i)} = 0\\
O_A t^{(i){AB}} &=& n_A t^{(i){AB}} = 0\\
\eta_{AB}t^{(i){AB}}&=&0
\eea
\ees
 $\mathcal{S}^{(i)}$, $V^{(i)}_A$ and $t^{(i)}_{AB}$ are the `slow' data given on the membrane. Any derivative of the these data are $\mathcal{O}(1)$ except the divergences, {\it i.e.} $\nabla.V^{(i)}, \nabla_A t^{(i)AB}$.\\
 
\textbf{The metric with the first sub-leading correction:}
\bes
\begin{eqnarray}
   \nonumber \mbox{\textbf{Notation: \:\:}}g_{AB} &=& \eta_{AB} + f O_A O_B + \frac{1}{D}h_{AB}\\
   &=&g^{(0)}_{AB}+\frac{1}{D}h_{AB}\\
    \nonumber \mbox{\textbf{And the inverse metric: \:\:}} g^{AB} &=& \eta^{AB} - f O^A O^B - \frac{1}{D}h^{AB} + \mathcal{O}\left(\frac{1}{D^2}\right)\\
     &=&g^{(0)        AB}-\frac{1}{D}h^{AB}
     \eea\ees
   \textbf{The christoffel connections are calculated wrt $g_{AB}$:}  
     \bea{}
 \nonumber \Gamma^A_{BC} &=& \Gamma^{(0) A}_{BC}\mbox{  (calculated with respect to initial ansatz,}\:\:{g^{(0)}_{AB}})\\
    \nonumber               & & + \frac{f\prime}{2D} n_D h^{AD} O_B O_C + \frac{1}{2D}[-\nabla^A h_{BC} +\nabla_B h^A_C+\nabla_C h^A_B ]\\
                       \nn    & & \frac{f}{2D} O^A (O.\nabla)h_{BC}+ \mathcal{O}\left(\frac{1}{D^2}\right)\\
                            &=& \Gamma^{(0)A}_{BC} + \delta\Gamma^{A}_{BC} + \mathcal{O}\left(\frac{1}{D^2}\right)
\end{eqnarray}

{\bf Ricci Tensor:}\\

Now Ricci tensor w.r.t. full metric $g_{AB}$,
\begin{equation}
    R_{AB} = \partial_C \Gamma^C_{AB}-\partial_B \Gamma^C_{CA} + \Gamma^C_{CD}\Gamma^D_{AB} -\Gamma^C_{AD}\Gamma^D_{CB}
\end{equation}
can be written in the form,
\begin{eqnarray}
    \nonumber R_{AB} &=& R^{(0)}_{AB} \mbox{   (with respect to only ansatz metric}\\
    \nonumber        & & + \underbrace{\nabla_C \delta\Gamma^C_{AB}}_\text{\bf Term1} \underbrace{- \nabla_B \delta\Gamma^C_{CA}}_\text{\bf Term2}  \\ \nonumber & & \underbrace{-\delta\Gamma^C_{AD}[\Gamma_{L}+\Gamma_{NL}]^D_{CB}-[\Gamma_{L}+\Gamma_{NL}]^C_{AD}\delta\Gamma^D_{CB}}_\text{\bf Term3}\\ & &
    \underbrace{+\delta\Gamma^C_{CE}[\Gamma_{L}+\Gamma_{NL}]^E_{AB}}_\text{\bf Term4}
\end{eqnarray}

\begin{eqnarray}
    \nonumber \mbox{\bf Term1} &=& \nabla_C \delta\Gamma^C_{AB}\\
    \nonumber       &=& \left[\frac{DN^2}{2\psi^2}(f-3\tilde{f}^2)(n_c h^{CD} n_D) - \frac{N}{2\psi}(f-\tilde{f}^2)n_D( \nabla_C h^{CD})\right]O_A O_B + \frac{N}{2} \nabla_C(f O^C) \dot{h}_{AB} \\
    \nonumber       & & + \frac{DN^2}{2}f \ddot{h}_{AB} - \frac{DN^2}{2}\left(\ddot{h}_AB+\frac{\dot{h}_AB}{\psi}\right) \\ & & +\frac{N}{2} \left\{ n_B \nabla_C \dot{h}^C_A + n_A \nabla_C \dot{h}^C_B \right\}\\
    \nonumber       & &\\
    \nonumber \mbox{\bf Term2} &=& -\nabla_B \delta\Gamma^C_{AC}\\
                  &=& -\frac{1}{2D}\nabla_A \nabla_B h^C_C = -\frac{DN^2}{2}\ddot{h}^C_C n_A n_B\\
    \nonumber     & &\\
    \nonumber \mbox{\bf Term3} &=& -\delta\Gamma^C_{AD}[\Gamma_{L}+\Gamma_{NL}]^D_{CB}-[\Gamma_{L}+\Gamma_{NL}]^C_{AD}\delta\Gamma^D_{CB}\\
    \nonumber                  &=& \frac{DN^2}{2\psi}(f-\tilde{f}^2)n^M [O_A \dot{h}_{MB} + O_B \dot{h}_{MA}] \\ 
    \nonumber                  & &\\
    \nonumber \mbox{\bf Term4} &=& +\delta\Gamma^C_{CE} [\Gamma_{L} + \Gamma_{NL} ]^E_{AB}\\
                               &=& -\frac{DN^2}{4\psi}(f-\tilde{f}^2)\dot{h}^C_C [(f-1)O_A O_B +(n_A O_B + n_B O_A)] \\ 
    \nonumber                  & &\\ \nonumber
\end{eqnarray}
where notations are
\bea{}
\n_C h_{AB}&=& DN \dot{h}_{AB}n_c +\mtO(1),\:\: \dot{h}_{AB}
=\p_{\zeta}{h}_{AB}
\eea
\paragraph{Calculation of homogeneous part   from the gauge field in $\mathcal{E}^{(1)}_{AB}$:}
Now we are  calculating the homogeneous contribution from the sub-leading order gauge field corrections $\frac{1}{D}a_M$ added to the initial ansatz $A_M^{(0)}=\sqrt{2}\tilde{f}O_M$.
The gauge field with the first sub-leading correction looks as,
\begin{equation}
    A_M = \underbrace{\sqrt{2}\tilde{f}O_M}_\text{Initial ansatz} + \underbrace{\frac{\sqrt{2}}{D}A^{(1)}_M}_\text{Correction}
\end{equation}
where, 
\bes
\bea{}
&&A^{(1)}_M = \mathcal{A}^{(s)}O_M+\mathcal{A}_M^{(v)}= \underbrace{\sum_i^{N_S} a_s^{(i)}(\zeta)\mathcal{S}^{i} O_M}_{\text{scalar}} + \underbrace{\sum_i^{N_V} a_v^{(i)}(\zeta)V_A^{i}}_{\text{gauge vector}}\\
 &&\mathcal{A}_M^{(v)} \:u^M = \mathcal{A}_M^{(v)} \:n^M = 0
\eea
\ees 

\begin{eqnarray}
   \nonumber F_{MN} &=& \nabla_M A_N - \nabla_N A_M\\
 \nn  &=& \underbrace{\sqrt{2}[(\nabla_M \tilde{f}) O_N -( \nabla_N \tilde{f}) O_M + \tilde{f} (\nabla_M O_N - \nabla_N O_M)]}_{F^{(0)}_{MN}} + \underbrace{\frac{\sqrt{2}}{D}(\nabla_M A^{(1)}_N - \nabla_N A^{(1)}_M)}_{\frac{1}{D}F^{(1)}_{MN}}\\
\eea
           We define the field strength wrt $g^{(0)}_{AB}$ as $F^{(0)}_{MN} $
        \bea{}
    F^{(0)}_{MN} &=& \sqrt{2}[\nabla_M ( \tilde{f} O_N )- \nabla_N (\tilde{f} O_M)]\\
    \mbox{\& } F^{(1)}_{MN} &=& \sqrt{2}(\nabla_M A^{(1)}_N - \nabla_N A^{(1)}_M)\\
\nn    -\frac{1}{2} F_{AC} F_{BD} g^{CD} &=& \underbrace{-\frac{1}{2} F^{(0)}_{AC} F^{(0)}_{BD} g^{CD}}_\text{Source} \underbrace{-\frac{1}{2D} [F^{(1)}_{AC} F^{(0)}_{BD}+F^{(0)}_{AC} F^{(1)}_{BD}] g^{CD}}_\text{\bf Term5} \underbrace{+ \frac{1}{2D}F^{(0)}_{AC} F^{(0)}_{BD} h^{CD}}_\text{\bf Term6}\\
\\
  \nonumber  +  \frac{1}{4D}F^2 g_{AB} &=& \frac{N\tilde{f}}{2\psi}[2n^C O.\nabla A^{(1)}_C-(n.\nabla A^{(1)}_A)O^A]\\
  &=&\mathcal{O}(1) \mbox{ (no correction in homogeneous part.)}
\end{eqnarray}
\begin{eqnarray}
    \nonumber \mbox{\bf Term5} &=& -\frac{1}{2D} [F^{(1)}_{AC} F^{(0)}_{BD}+F^{(0)}_{AC} F^{(1)}_{BD}] g^{CD}\\
    \nonumber &=& -\frac{N}{\psi}\tilde{f} [(n_B+fO_B)(O.\nabla)A^{(1)}_A+(n_A+fO_A)(O.\nabla)A^{(1)}_B\\
    \nonumber && + n^C(O_B \nabla_A + O_A \nabla_B)A^{(1)}_C - \{(n.\nabla A^{(1)}_A) O_B +(n.\nabla A^{(1)}_B) O_A \}] \\
    &&  \\
    \nonumber \mbox{\bf Term6} &=& \frac{1}{2D}F^{(0)}_{AC} F^{(0)}_{BD} h^{CD}\\
    &=& \frac{DN^2}{\psi^2}\tilde{f}^2(n_C h^{CD} n_D) O_A O_B
\end{eqnarray}

So, the total homogeneous part of the tensor equation of motion is,
\begin{equation}
    H_{AB} = {\bf Term1}+{\bf Term2}+{\bf Term3}+{\bf Term4}+{\bf Term5}+{\bf Term6}
\end{equation}
 \paragraph{Homogeneous part due to $\mathbf{S^{(i)}_1(\zeta) \mathcal{S}^{(i)} O_A O_B}$}
Let,
\begin{equation}
    h_{AB} = \sum_i S^{(i)}_1(\zeta) \mathcal{S}^{(i)} O_A O_B
\end{equation}
Then,
\begin{eqnarray}
    \nonumber \mbox{\bf Term1} &=&\left( \frac{DN^2}{2}\right) \lb \left(\ddot{S}_1+\frac{\dot{S}_1}{\psi}\right)\mathcal{S}^{(i)}(n_A O_B +n_B O_A-O_A O_B)\rb+ \frac{DN^2}{2}f\left(\ddot{S}_1-\frac{\dot{S}_1}{\psi}\right)\mathcal{S}^{(i)}O_A O_B\\
 &&+ \left(\frac{DN^2}{\psi}\right) \tf^2 \left( \dot{S}_1-S_1\right)\mathcal{S}^{(i)} O_A O_B\\
 \\
 \nonumber \mbox{\bf Term2} &=& 0 \\
 \\
    \nonumber \mbox{\bf Term3} &=& - \sum_i N f^\prime \dot{S}^{(i)}_1 \mathcal{S}^{(i)} O_A O_B \\
 &=& \sum_i\frac{DN^2}{\psi}(f-\tf^2)\dot{S}_1\mathcal{S}^{(i)} O_A O_B\\
 \\
    \nonumber \mbox{\bf Term4} &=& 0 \\
    \nonumber \mbox{\bf Term5} &=& 0 \\
    \\
    \nonumber \mbox{\bf Term6} &=&  \sum_i\frac{\tf^{\prime^2}}{D}S_1^{(i)} \mathcal{S}^{(i)} O_A O_B\\
    &=&  \sum_i\frac{DN^2}{\psi}\tf^2 S_1\mathcal{S}^{(i)} O_A O_B
\end{eqnarray}
So,
\begin{eqnarray}
    \nonumber H_{AB} &=& \frac{DN^2}{2}(f-1)\sum_i\left( \ddot{S}^{(i)}_1 +\frac{\dot{S}^{(i)}_1}{\psi}\right)\mathcal{S}^{(i)} O_A O_B\\
     && + \frac{DN^2}{2}\sum_i\left( \ddot{S}^{(i)}_1 + \frac{\dot{S}^{(i)}_1}{\psi} \right)\mathcal{S}^{(i)} (n_A O_B + n_B O_A)
\end{eqnarray}

\paragraph{Homogeneous part due to $\sum_i \mathbf{\frac{S_2^{(i)}\mathcal{S}^{(i)}(\zeta)}{D} P_{AB}}$}
Let,
\begin{equation}
    h_{AB} = \sum_i\frac{S_2^{(i)}(\zeta)}{D}\mathcal{S}^{(i)}P_{AB}
\end{equation}
Then,
\begin{eqnarray}
    \nonumber \mbox{\bf Term1} &=& \sum_i -\frac{DN^2}{2}\lb \( (1-f)\ddot{S}^{(i)}_2+\frac{\dot{S}_2^{(i)}}{\psi}\)\mathcal{S}^{(i)} P_{AB}+2 \dot{S}_2^{(i)}\mathcal{S}^{(i)}n_A n_B\rb+\frac{DN^2}{2\psi}\tf^2 \dot{S}_2^{(i)}\mathcal{S}^{(i)}P_{AB}\\
    \nonumber && \\
    \nonumber \mbox{\bf Term2} &=& -\sum_i\frac{DN^2}{2}\ddot{S}_2 ^{(i)}\mathcal{S}^{(i)}n_A n_B \\
    \nonumber \mbox{\bf Term3} &=& 0 \\
    \nonumber \mbox{\bf Term4} &=& \sum_i \frac{Nf^\prime}{4}\dot{S}_2  ^{(i)}\mathcal{S}^{(i)}[(f-1) O_A O_B + (n_A O_B + n_B O_A)] \\
  \nn  &=&-\sum_i\frac{DN^2}{4\psi}(f-\tf^2)\dot{S}_2^{(i)}\mathcal{S}^{(i)} [(f-1) O_A O_B + (n_A O_B + n_B O_A)]\\
    \nonumber \mbox{\bf Term5} &=& 0 \\
    \nonumber \mbox{\bf Term6} &=& 0
\end{eqnarray}
So,
\begin{eqnarray}
    \nonumber H_{AB} &=& \sum_i -\frac{DN^2}{4\psi}(f-\tf^2)\dot{S}_2^{(i)} [(f-1) O_A O_B + (n_A O_B + n_B O_A)] -\frac{DN^2}{2}\ddot{S}_2^{(i)} n_A n_B \\
  \nn  &&  -\frac{DN^2}{2}\lb \( (1-f)\ddot{S}^{(i)}_2+\frac{\dot{S}_2^{(i)}}{\psi}\)\mathcal{S}^{(i)} P_{AB}+2 \dot{S}_2^{(i)}\mathcal{S}^{(i)}n_A n_B\rb+\frac{DN^2}{2\psi}\tf^2 \dot{S}_2^{(i)}\mathcal{S}^{(i)}P_{AB}+\mathcal{O}(1)\\
\end{eqnarray}
\paragraph{Homogeneous part due to $\mathbf{\sum_i \mathcal{V}^{(i)}(O_A V^{(i)}_B + O_B V^{(i)}_A)}$}
Let,
\begin{equation}
    h_{AB} = \sum_i \mathcal{V}^{(i)}(O_A V^{(i)}_B + O_B V^{(i)}_A)
\end{equation}
Then,
\begin{eqnarray}
    \nonumber \mbox{\bf Term1} &=& \frac{DN^2}{2}\sum_i\left( f\ddot{\mathcal{V}}^{(i)} + Q^2 \psi^{-2D} \dot{\mathcal{V}}^{(i)} \right)(O_A V^{(i)}_B + O_B V^{(i)}_A)\\
    \nonumber  && + \frac{DN^2}{2}\sum_i\left( \ddot{\mathcal{V}}^{(i)} + \frac{\dot{\mathcal{V}}^{(i)}}{\psi} \right) (u_A V^{(i)}_B + u_B V^{(i)}_A) \\
    \nonumber && + \frac{1}{2}\sum_i(\nabla.V^{(i)})\left[ \frac{f^{\prime}}{D}\mathcal{V}^{(i)} O_A O_B + N \dot{\mathcal{V}}^{(i)}(O_A n_B + O_B n_A) \right] \\
    \nonumber && \\
    \nonumber \mbox{\bf Term2} &=& 0 \\
    \nonumber \mbox{\bf Term3} &=& - \frac{Nf^\prime}{2}\sum_i\dot{\mathcal{V}}^{(i)}(O_A V^{(i)}_B + O_B V^{(i)}_A) \\
    \nonumber \mbox{\bf Term4} &=& 0 \\
    \nonumber \mbox{\bf Term5} &=& 0 \\
    \nonumber \mbox{\bf Term6} &=& 0
\end{eqnarray}
So,
\begin{eqnarray}
    \nonumber H_{AB} &=& \frac{DN^2}{2}\sum_i\left( \ddot{\mathcal{V}}^{(i)} + \frac{\dot{\mathcal{V}}^{(i)}}{\psi}\right)
    \lb f (O_A V^{(i)}_B + O_B V^{(i)}_A)+(u_A V^{(i)}_B + u_B V^{(i)}_A)\rb\\
    \nn &&+ \frac{N}{2}\sum_i(\nabla.V^{(i)})\lb \dot{\mathcal{V}}^{(i)}(O_A n_B + O_B n_A)-(f-\tf^2)\frac{\mathcal{V}^{(i)}}{\psi}O_A O_B  + \mathcal{O}(1)\\
\end{eqnarray}
\paragraph{Homogeneous part due to $\mathbf{\sum_i\tau^{(i)}(\zeta) t^{(i)}_{AB}}$:}
Let,
\begin{equation}
    h_{AB} = \sum_i\tau^{(i)}(\zeta) t^{(i)}_{AB}
\end{equation}
Then,
\begin{eqnarray}
    \nonumber \mbox{\bf Term1} &=& -\frac{DN^2}{2}\sum_i\left[ (1-f)\ddot{\tau}^{(i)} + \frac{\dot{\tau}^{(i)}}{\psi} \right] t^{(i)}_{AB} + \frac{N}{2}\sum_i\dot{\tau}^{(i)}[n_B \nabla_C (t^{(i)C}_A)+n_A \nabla_C (t^{(i)C}_B)]\\
    \nonumber &&  + \frac{DN^2}{2} Q^2 \psi^{-2D} \sum_i\dot{\tau}^{(i)} t^{(i)}_{AB} \\
    \nonumber && \\
    \nonumber \mbox{\bf Term2} &=& 0 \\
    \nonumber \mbox{\bf Term3} &=& 0 \\
    \nonumber \mbox{\bf Term4} &=& 0 \\
    \nonumber \mbox{\bf Term5} &=& 0 \\
    \nonumber \mbox{\bf Term6} &=& 0
\end{eqnarray}
So,
\begin{eqnarray}
    \nonumber H_{AB} &=& -\frac{DN^2}{2}\sum_i\left[ (1-f)\ddot{\tau}^{(i)} + \frac{\dot{\tau}^{(i)}}{\psi} \right] t^{(i)}_{AB} + \frac{N}{2}\sum_i\dot{\tau}^{(i)}[n_B \nabla_C (t^{(i)C}_A)+n_A \nabla_C (t^{(i)C}_B)]\\
    \nonumber &&  + \frac{DN^2}{2} \tf^2 \sum_i\dot{\tau}^{(i)} t^{(i)}_{AB} \\
\end{eqnarray}
\paragraph{Homogeneous part due to $\mathbf{\sum_i^{N_S} a_s^{(i)}(\zeta)\mathcal{S}^{i} O_M}$:}
Let,
\begin{equation}
    \mathcal{A}^{(s)}_M = \sum_i^{N_S} a_s^{(i)}(\zeta)\mathcal{S}^{i} O_M
\end{equation}
Then,
\begin{eqnarray}
    \nonumber \mbox{\bf Term1} &=& 0 \\
    \nonumber \mbox{\bf Term2} &=& 0 \\
    \nonumber \mbox{\bf Term3} &=& 0 \\
    \nonumber \mbox{\bf Term4} &=& 0 \\
    \nonumber \mbox{\bf Term5} &=& \sum_i 2 \dot{a}_s^{(i)} N \mathcal{S}^{i} \tilde{f}^\prime [n_A O_B + n_B O_A + (f-1)O_A O_B]  \\
    \nonumber \mbox{\bf Term6} &=& 0
\end{eqnarray}
So,
\begin{eqnarray}
    \nonumber H_{AB} &=& -\frac{DN^2}{\psi} \sum_i 2 \dot{a}_s^{(i)}  \tilde{f} \mathcal{S}^{i}[n_A O_B + n_B O_A + (f-1)O_A O_B]
\end{eqnarray}
%
\paragraph{Homogeneous part due to $\mathbf{\sum_i^{N_V} a_v^{(i)}(\zeta)V_A^{i}}$:}
Let,
\begin{equation}
    \mathcal{A}^{(v)}_M = \sum_i^{N_V} a_v^{(i)}(\zeta)V_M^{i}
\end{equation}
Then,
\begin{eqnarray}
    \nonumber \mbox{\bf Term1} &=& 0 \\
    \nonumber \mbox{\bf Term2} &=& 0 \\
    \nonumber \mbox{\bf Term3} &=& 0 \\
    \nonumber \mbox{\bf Term4} &=& 0 \\
    \nonumber \mbox{\bf Term5} &=& -\frac{DN^2}{\psi}\sum_i \dot{a}_v^{(i)} \tilde{f} \lb u_A V_B^{(i)} +u_B V_A^{(i)} + f(O_A V_B^{(i)} +O_B V_A^{(i)} \rb   \\
    \nonumber \mbox{\bf Term6} &=& 0
\end{eqnarray}
So,
\begin{equation}
     H_{AB} = -\frac{DN^2}{\psi} \sum_i \dot{a}_v^{(i)} \tilde{f} \lb u_A V_B^{(i)} +u_B V_A^{(i)} + f(O_A V_B^{(i)} +O_B V_A^{(i)} \rb 
\end{equation}

\subsection{Maxwell Homogeneous Part}
\subsubsection{Gauge field equation homogeneous part due to metric correction}
The leading ansatz for gauge field is, $A_M=\tdf O_M$ which gives the field tensor as,

\be{Fab}
F_{AB}=\n_A(\tdf O_B)-\n_B(\tdf O_A)
\ee
Here we will calculate homogeneous part due to the metric correction,
\bea{gh}
g_{AB}&=&\eta_{AB}+f O_A O_B +\frac{1}{D}h_{AB}\\
g^{AB}&=&\eta^{AB}-f O^A O^B -\frac{1}{D}h^{AB}
\eea  
where $h^{AB}$ is raised with respect to $\eta^{AB}$.
Now field strength tensor raised with respect to $g_{AB}$ is,
\bea{FMNraised}
\nonumber g^{MA}F_{AB}g^{BN}&=&(\n^M \tdf O^N-\n^N \tdf O^M)+\tdf (\n^MO^N-\n^NO^M)+\frac{N\tdf}{\psi}n_A(h^{AM}O^N-h^{AN}O^M)+\mathcal{O}(1)\\
&=&F^{MN}+\frac{N\tdf}{\psi}n_A(h^{AM}O^N-h^{AN}O^M)+\mathcal{O}\left( \frac{1}{D}\right)
\eea
Here $F^{MN}$ is raised with respect to $\eta^{AB}$.

The equation of motion for gauge field is,
\bea{gaugeeom}
\nonumber E^N &=& \bar{\n}_M (g^{MA}F_{AB}g^{BN}) \\
\nonumber &&(\mbox{covariant derivative with respect to }g_{AB})\\
\nonumber&=& \n_M (g^{MA}F_{AB}g^{BN}) + \frac{1}{2D}\n_A h^C_C F^{AN}\\
\nonumber&=& \underbrace{\n_M F^{MN}}_\text{Source part, $S^N$}+\frac{N\tdf}{\psi} \left( (n_A\n_Mh^{AM})O^N-n_AO.\n h^{AN}-\mathcal{K}n_Ah^{AN} \right)\\
&&-\frac{DN^2\tdf}{\psi}(n.h.n O^N - n_Ah^{AN})+\frac{DN^2\tdf}{2\psi}\dot{h}u^N\\
&=& S^N+H^N
\eea
\bea{gaugehomo}
\nonumber H^N &=& \left[ \frac{N\tdf}{\psi}(n_A\n_Mh^{AM}-\mathcal{K}n_A(\dot{h}^{AM}+h^{AM})n_M)-\frac{N\mathcal{K}\tdf}{2}\dot{h}^C_C \right] O^N +\frac{N\mathcal{K}\tdf}{2}\dot{h}^C_C n^N\\
&& -\frac{DN^2\tdf}{\psi}n_A\dot{h}^{AM}P^N_M
\eea

Component of this homogeneous part in the direction normal to the membrane surface is,
\bea{gauge_membrane}
n_N H^N &=&  \mathcal{K}\tdf\left[\frac{1}{D}n_A\n_Mh^{AM}-\frac{\mathcal{K}}{D}n_A(\dot{h}^{AM}+h^{AM})n_M \right]
\eea

For metric correction of the form,
\bea{}
\nn  h_{AB} &=& \underbrace{\sum_{i} S^{(i)}_1(\zeta) \mathcal{S}^{(i)} O_A O_B}_\text{Scalar Sector} + \underbrace{\sum_{i} \frac{S_2^{(i)}(\zeta)}{D} \mathcal{S}^{(i)}\mathcal{P}_{AB}}_\text{Trace} \\
&&+ \underbrace{\sum_{i} \mathcal{V}^{(i)}(\zeta)(V^{(i)}_A O_B + V^{(i)}_B O_A)}_\text{Vector Sector} + \underbrace{\sum_{i} \mathcal{T}^{(i)}(\zeta)t^{(i)}_{AB}}_\text{trace-less tensor sector}
\eea

\bea{gauge_homo_final}
H^N_{\text {metric corr}}&=&\frac{N\mathcal{K}\tdf}{2}\sum_{i} \dot{S_2}^{(i)} u^N+\frac{N\tdf}{\psi}\sum_{i} \mathcal{V}^{(i)}(\n.V^{(i)}) O^N-\frac{DN^2\tdf}{\psi}\sum_{i} \dot{\mathcal{V}}^{(i)}V^{(i)}~~
\eea

\subsubsection{Gauge field equation homogeneous part due to gauge field correction}
Now we calculate the homogeneous part due to gauge field correction. \begin{eqnarray}
\nonumber F_{MN} &=& \nabla_M A_N-\nabla_N A_M \\
&=& F_{MN}^{(0)}+\frac{1}{D}F_{MN}^{(1)}\quad (\text{both of the terms are defined before})\\
\nonumber F^{MN}&=& F_{AB}g^{AM}g^{BN}\\
\nonumber  &=&F^{(0)MN}+\frac{1}{D}\underbrace{[F^{(1)}_{AB}g^{(0)AM}g^{(0)BN}]}_{\delta F^{MN}}\\
&=&F^{(0)MN}+\frac{1}{D}\delta F^{MN}
\end{eqnarray}
Since, for $h_{AB}=0$, $\Gamma^C_{AC}=\tilde{\Gamma^C_{AC}}$, covariant derivative with respect to total metric will become covariant derivative with respect to background. 
\be{eqdmgc}
\xbar\n_M F^{MN}=\n_M F^{MN}=\xbar\dl_M[F^{(0)MN}+\frac{1}{D}\delta F^{MN}]
\ee
Here,
$\xbar\n_M $ is the covariant derivative wrt $$
g_{AB}=\eta_{AB}+fO_A O_B+\frac{1}{D}h_{AB}$$
\\ Now we want to compute the covariant derivative of the anti symmetric tensor $F^{MN}$ wrt $g_{AB}$. Then following the previous formula, 

\bea{}
\xbar\n_M{F^{MN}}\Big|_{\text Homogeneous}&=& \frac{1}{D}\n_M\delta F^{MN}+\mathcal{O}\left( \frac{1}{D}\right)\\
&=&\underbrace{\n_M[F^{(1)}_{AB}g^{(0)AM}g^{(0)BN}]}_{T}
\eea

\bes
\bea{}
\nn T&=&\n^2 A^{N(1)}-\n_M A^{N(1)} A^{M(1)}-f^\prime (O\cd \n)A^{N(1)}- \frac{DN}{\psi}f(O\cd \n)A^{N(1)}-f(O \cd \n)^2 A^{N(1)}+f^\prime O^A \n^N A^{(1)}_A\quad\quad\quad\\
\nn&&+f\frac{DN}{\psi}O^A \n^N A^{(1)}_A+f(O\cd \n)(O^A \n^N a_A)-f^\prime O^N O^B (n\cd \n)A^{(1)}_B- fO^N \n_M(O^B\n^M A^{(1)}_B)\\
&&+f^\prime O^N n_M (O\cd \n)A^{M(1)}+F  \n_M [O^N(O\cd \n )A^{M(1)}]
\eea
\ees

\paragraph{Homogeneous part in $H^{N}$ due to $A_N^{(1)}=\sum_i^{N_S} a_s^{(i)}(\zeta)\mathcal{S}^{i} O_N$:}
\be{}
H^{N}=-DN^2\sum_i \bigg(\ddot{a}_s^{(i)}(\zeta)+\frac{\dot{a}_s^{(i)}(\zeta)}{\psi}\bigg) \mathcal{S}^{i} u^N
\ee

\paragraph{Homogeneous part in $H^{N}$ due to $A_N^{(1)}=\mathbf{\sum_i^{N_V} a_v^{(i)}(\zeta)V_N^{i}}$:}

\be{}
H^{N}=\sum_i \bigg[DN^2\bigg(\ddot{a}_v^{(i)}(\zeta)+\frac{\dot{a}_v^{(i)}(\zeta)}{\psi}\bigg)(1-f)-N f^\prime \dot{a}_v^{(i)}(\zeta)\bigg]V^{(i)N}+(fO^N-n^N)(\n \cd V^{(i)})N \dot{a}_v^{(i)}(\zeta)
\ee

So, the homogeneous part

\be{}
H^N_\text{scalar}=0
\ee

\be{}
H^N_\text{trace}=\frac{DN^2\tdf}{2\psi}\sum_{i} \dot{S_2}^{(i)} u^N \mathcal{S}^{(i)}
\ee

\be{}
H^N_\text{vector}=\frac{N\tdf}{\psi}\sum_{i} \mathcal{V}^{(i)}(\n.V^{(i)}) O^N-\frac{DN^2\tdf}{\psi}\sum_{i} \dot{\mathcal{V}}^{(i)}V^{(i)N}
\ee

\be{}
H^N_\text{tensor}=0
\ee

\be{}
H^N_\text{gauge scalar}=-\sum_i DN^2\bigg(\ddot{a}_s^{(i)}+\frac{\dot{a}_s^{(i)}}{\psi}\bigg)\mathcal{S}^{(i)}u^N
\ee

\be{}
H^N_\text{gauge vector}=\sum_i \lb DN^2\left(\ddot{a}_v^{(i)}(1-f)+\frac{\dot{a}_v^{(i)}}{\psi}(1-\tf^2)\right)\rb V^{(i)N}+N(fO^N-n^N)(\n \cd V^{(i)}) \dot{a}_V^{(i)}
\ee
\bea{gauge_homo_final}
\nn H^N&=&\frac{DN^2\tdf}{2\psi}\sum_{i} \dot{S_2}^{(i)} u^N+\frac{N\tdf}{\psi}\sum_{i} \mathcal{V}^{(i)}(\n.V^{(i)}) O^N-\frac{DN^2\tdf}{\psi}\sum_{i} \dot{\mathcal{V}}^{(i)}V^{(i)N}
-\sum_i DN^2(\ddot{a}_s^{(i)}+\frac{\dot{a}_s^{(i)}}{\psi})\mathcal{S}^{(i)}u^N\\
&&+\sum_i \lb DN^2\left(\ddot{a}_v^{(i)}(1-f)+\frac{\dot{a}_v^{(i)}}{\psi}(1-\tf^2)\right)\rb V^{(i)N}+N(fO^N-n^N)(\n \cd V^{(i)}) \dot{a}_V^{(i)}
\eea

\section{Proof of the consistency conditions}\label{constraints}
\subsection{Verification of the consistency constraint \eqref{consistency} }
Using the following identity for the divergence of the  source in the tensor sector source.
\begin{equation}\label{identen}
\begin{split}
P_C^B\nabla_A t^{AC}= K\bigg[V^{(1)}_B - \left({V^{(2)}_B + V^{(3)}_B+V^{(4)}_B\over 2} \right)\bigg] + {\cal O}(1)
\end{split}
\end{equation}
After substituting the solution  for $\cal T$ and the identity above, the consistency equation takes the following form
\begin{equation}\label{consistn}
\begin{split}
&(f-1)\dot{\cal T} \left(\nabla_C t^C_A\over 2K\right)-\left(D\over K\right){\mathfrak V}^{(1)}_A\\
 =~& \left({e^{-\zeta}\over 2}\right) \bigg[\left(- V^{(1)}_A  + V^{(2)}_A + V^{(4)}_A\right) +Q^2\left(V^{(1)}_A -V^{(3)}_A\right)\bigg]
\end{split}
\end{equation}
The RHS of \eqref{consistn} is proportional to the membrane equation and therefore is of the order of ${\cal O}\left(1\over D\right)$ both on and away from the membrane.\\

\subsection{Verification of the consistency constraints \eqref{consisscal}}
We will be showing the following constraint holds.
\begin{equation}
\begin{split}
{\cal C}_\text{guage}
\equiv~&N\sum_{i=1}^{N_V}\bigg[\tf \mathcal{V}^{(i)} -(1-f)\dot{a}_v^{(i)}\bigg]\left(\n \cd V^{(i)}\over K\right) + {\mathfrak A}^{(1)}+ {\mathfrak A}^{(2)}={\cal O}\left( \frac{1}{D}\right) 
\end{split}
\end{equation}
The third equation in \eqref{veceq} can be written as,
\begin{equation}\label{scacheck}
\begin{split}
& -N \lb e^{-\zeta} \frac{d}{d \zeta}\Big( e^\zeta \Big( \tf \mathcal{V}^{(i)} -(1-f) \dot{a}_v^{(i)} \Big)\Big)\rb V_B^{(i)} + v_\text{gauge}^{(i)} V_B^{(i)}=0\\
& \text{Integrating we get,} ~\tf \mathcal{V}^{(i)} -(1-f) \dot{a}_v^{(i)}=e^{-\zeta} \int \frac{1}{N}~e^\zeta~ v_\text{gauge}^{(i)}~ d\zeta\\
& \text{Also, we will be using the boundary condition,}~ (\tf \mathcal{V}^{(i)} -(1-f) \dot{a}_v^{(i)})=0 ~~\text{at}~~ \zeta \to 0
\end{split}
\end{equation}
Using the above result in the second equation of \eqref{consisscal} and also $v_\text{gauge}^{(i)}$ from \eqref{vecs}, we obtain
\begin{equation}\label{consisscal1}
N\sum_{i=1}^{N_V}\bigg[\tf \mathcal{V}^{(i)} -(1-f)\dot{a}_v^{(i)}\bigg]\left(\n \cd V^{(i)}\over K\right)=  \zeta ~\tf~ \frac{1}{K} \Big( \n \cd V^{(1)}-\n \cd V^{(2)}-\n \cd V^{(4)} \Big)
\end{equation}
Here we will need the following identities
\begin{equation}\label{dividen}
\begin{split}
&{\nabla\cdot V^{(1)}\over K}={\cal S}^{(4)} + {\cal O}\left(1\over D\right)\\
&{\nabla\cdot V^{(2)}\over K}=-\left(R_{uu}\over K\right) + {\cal S}^{(1)} -{\cal S}^{(2)}+ {\cal O}\left(1\over D\right)\\
&{\nabla\cdot V^{(3)}\over K}=\left(R_{uu}\over K\right) +{\cal S}^{(2)}+ {\cal O}\left(1\over D\right)\\
&{\nabla\cdot V^{(4)}\over K}={\cal S}^{(1)}+ {\cal O}\left(1\over D\right)\\
\end{split}
\end{equation}
Next, we will be doing a Taylor expansion around $\psi=1$ of the second term and third term in the second equation of \eqref{consisscal}, also using the fact $(\psi-1)=\frac{\zeta}{D}$,

\begin{equation}\label{consisscal2}
\begin{split}
\mathfrak{A}^{(1)}+\mathfrak{A}^{(2)}&=-\tf \lb \mathcal{S}^{(1)}- \mathcal{S}^{(2)}+ \mathcal{S}^{(3)}+ \mathcal{S}^{(5)}- \mathcal{S}^{(6)} -  \frac{R_{uu}}{ K} \rb_{\psi=1}\\
&- \tf  (\psi-1) \frac{\partial}{\partial \psi}\lb\mathcal{S}^{(1)}- \mathcal{S}^{(2)}+ \mathcal{S}^{(3)}+ \mathcal{S}^{(5)}- \mathcal{S}^{(6)} -  \frac{R_{uu}}{ K} \rb\\
&= -\tf \lb \mathcal{S}^{(1)}- \mathcal{S}^{(2)}+ \mathcal{S}^{(3)}+ \mathcal{S}^{(5)}- \mathcal{S}^{(6)} -  \frac{R_{uu}}{ K} \rb_{\psi=1}- \tf~ \frac{\zeta}{D}~ \frac{n \cd  \n}{N} \mathcal{S}^{(3)}+\mtO\left( \frac{1}{D^2}\right)\\
&= -\tf \lb \mathcal{S}^{(1)}- \mathcal{S}^{(2)}+ \mathcal{S}^{(3)}+ \mathcal{S}^{(5)}- \mathcal{S}^{(6)} -  \frac{R_{uu}}{ K} \rb_{\psi=1}- \tf \frac{\zeta}{K}(n \cd \n)(\n \cd u)+\mtO\left( \frac{1}{D^2}\right)\\
&= -\tf \lb \mathcal{S}^{(1)}- \mathcal{S}^{(2)}+ \mathcal{S}^{(3)}+ \mathcal{S}^{(5)}- \mathcal{S}^{(6)} -  \frac{R_{uu}}{ K} \rb_{\psi=1}\\
&~~~~~~~~~- \zeta ~\tf~ \frac{1}{K} \Big( \n \cd V^{(1)}-\n \cd V^{(2)}-\n \cd V^{(4)} \Big)+\mtO\left( \frac{1}{D^2}\right)
\end{split}
\ee
Hence, adding \eqref{consisscal1} and \eqref{consisscal2},
\be{}
\begin{split}
 {\cal C}_\text{guage}&
 \equiv~N\sum_{i=1}^{N_V}\bigg[\tf \mathcal{V}^{(i)} -(1-f)\dot{a}_v^{(i)}\bigg]\left(\n \cd V^{(i)}\over K\right) + {\mathfrak A}^{(1)}+ {\mathfrak A}^{(2)}\\
&= -\tf \lb \mathcal{S}^{(1)}- \mathcal{S}^{(2)}+ \mathcal{S}^{(3)}+\mathcal{S}^{(5)}- \mathcal{S}^{(6)} - \frac{R_{uu}}{ K} \rb+\mtO\left( \frac{1}{D^2}\right)
\end{split}
\end{equation}
As the RHS is proportional to the scalar membrane equation, ${\cal C}_\text{guage}$ is $\mtO\left( \frac{1}{D}\right)$ both on and away from the membrane.\\ 

Now we will show the consistency of the other constraint.
\begin{equation}
\begin{split}
{\cal C}_\text{metric}\equiv~&{N\over 2}\sum_{i=1}^{N_V}\bigg[(1-f)\dot{\mathcal{V}}^{(i)} - (f-\tf^2){\mathcal{V}}^{(i)}\bigg] \left(\n \cd V^{(i)}\over K\right)+ (1-f)~{\mathfrak S}_2 + {\mathfrak S}_1 ={\cal O}\left( \frac{1}{D}\right)\\
\end{split}
\end{equation}

We start with the condition in \eqref{scacheck}
$$
\tf \mathcal{V}^{(i)} -(1-f) \dot{a}_v^{(i)}=e^{-\zeta} \int \frac{1}{N}~e^\zeta~ v_\text{gauge}^{(i)}~ d\zeta $$
and plug it in the second equation of \eqref{veceq}. We also use the identity,
\begin{equation}
\begin{split}
e^{-\zeta} \frac{d}{d \zeta}\lb e^\zeta \Big( (1-f)\dot{ \mathcal{V}}^{(i)} -(f-\tf^2) \mathcal{V}^{(i)}\Big)\rb =(1-f)\left(\ddot{ \mathcal{V}}^{(i)}+\dot{ \mathcal{V}}^{(i)}\right) -2\tf^2 \mathcal{V}^{(i)}
\end{split}
\end{equation}
We can then write the ${\cal C}_{\text{metric}}$ equation as,
\begin{equation}
\begin{split}
&~~~~~~~~{\cal C}_{\text{metric}}=~\sum_iv_{\text{new}}^{(i)} \left(\n \cd V^{(i)}\over K\right)+(1-f){\mathfrak S}_2+{\mathfrak S}_1\\
&\text{where,~~}v_{\text{new}}^{(i)}=\frac{N}{2}\lb(1-f)\dot{ \mathcal{V}}^{(i)}-(f-\tf^2) \mathcal{V}^{(i)}\rb\\
&~~~~~~~~~~~~~~~~~~=e^{-\zeta}\int_{\infty}^{\zeta} \lb e^{\zeta}v_\text{metric}^{(i)}(\zeta) ~-\tf \int_{\infty}^{\zeta} e^{\zeta}v_\text{gauge}^{(i)}(\zeta)\rb d\zeta\\
&v_\text{metric}^{(i)}(\zeta)~\text{and}~ v_\text{gauge}^{(i)}(\zeta) ~\text{are defined in \eqref{vecs}.}
\end{split}
\end{equation}
Integrating $v_{new}^{(i)}$ and imposing the boundary condition $v_{new}^{(i)}\to 0$ for $\zeta\to \infty$ along with the other boundary condition  from \eqref{bc1}, $v_{new}^{(i)}\to 0$ for $\zeta\to 0$ for all $i$, 
\begin{equation}
\begin{split}
&v_{new}^{(1)}=-\frac{1}{2}R(f-\tf^2),~~v_{new}^{(3)}= \frac{1}{2}(\tf^2-Q\tf)\\
&v_{new}^{(2)}=v_{new}^{(4)}=\frac{1}{2}R(f-\tf^2)-\frac{1}{2}(\tf^2-Q\tf)
\end{split}
\end{equation}
Hence,
\begin{equation}\label{cmethomo}
\begin{split}
 \sum_i v_{new}^{(i)}  \left({\n \cd V^{(i)}\over K}\right)&=-\frac{1}{2}R(f-\tf^2)\left(\mathcal{S}^{(4)}-2\mathcal{S}^{(1)}
+\mathcal{S}^{(2)}+\frac{R_{uu}}{K}\right)\\
&+(\tf^2-Q\tf)\left({R_{uu}\over K} - \mathcal{S}^{(1)}+ \mathcal{S}^{(2)}\right)\\
\text{where,~}& \left(\mathcal{S}^{(4)}-2\mathcal{S}^{(1)}
+\mathcal{S}^{(2)}+\frac{R_{uu}}{K}\right)= \frac{(n \cd \n)(\n \cd u)}{K}
\end{split}
\end{equation}
The source terms are evaluated as,
\begin{equation}\label{cmetsou}
\begin{split}
(1-f){\mathfrak S}_2+{\mathfrak S}_1&=\frac{1}{2}(f-\tf^2)\mathcal{S}^{(3)}+(Q\tf-\tf^2)(\mathcal{S}^{(5)}-\mathcal{S}^{(6)})\\
&=\frac{1}{2}(f-\tf^2)\mathcal{S}^{(3)}\Big|_{\psi=1}+R\left(\frac{f-\tf^2}{2}\right)\frac{(n \cd \n)(\n \cd u)}{K}\\
&+(Q\tf-\tf^2)\left({R_{uu}\over K} - \mathcal{S}^{(1)}+ \mathcal{S}^{(2)}\right)+\mtO\left(\frac{1}{D^2}\right)
\end{split}
\end{equation}

We have put the membrane equation ${R_{uu}\over K} - \mathcal{S}^{(1)}+ \mathcal{S}^{(2)}-\mathcal{S}^{(5)}+\mathcal{S}^{(6)}\Big|_{\psi=1}=\mtO\left(\frac{1}{D}\right)$ in the last line.

Adding \eqref{cmethomo} with \eqref{cmetsou} we arrive at,
\begin{equation}
\begin{split}
{\cal C}_{\text{metric}}=\frac{1}{2}(f-\tf^2)(\hat{\n}\cd u)\Big|_{\psi=1}+\mtO\left(\frac{1}{D^2}\right)
\end{split}
\end{equation}
As the RHS is proportional to the scalar membrane equation, ${\cal C}_\text{metric}$ is $\mtO\left( \frac{1}{D}\right)$ both on and away from the membrane.
\section{QNM for the scaled membrane equation :}
\label{QNM for the scaled membrane equation}
The scalar membrane equation is,
\begin{eqnarray}\label{sclsca}
	\tilde{\nabla} . u &=& 0\\
	\partial_a \delta u^a + (D-2) \partial_t \delta r &=& 0
\end{eqnarray}

Following the line of \cite{SB} and from \ref{sclsca}, it is obvious that the temporal and spatial frequencies are related by a factor of $\frac{1}{\sqrt{D}}$. Without loss of generality we assume that the temporal frequencies are of $\mathcal{O}(1)$ and the spatial frequencies are related by equation \ref{sclsca}. \footnote{Scaling arguments are discussed in details in \cite{SB}}
 Now we will scale the coordinates as, $x^a \rightarrow y^a = \sqrt{D} x^a$. So the metric becomes,
 \begin{equation}
 	ds^2 = - r^2 dt^2 + \frac{dr^2}{r^2} + r^2 (\frac{dy^a dy^a}{D} + dx^i dx^i)
 \end{equation}
 
 where $a$, $b$ indices take p number of values and $i$, $j$ indices take $1$ to $D-p-2$. We can consider fluctuations only in the p number of $x^a$ directions. 
 Now we can write,
 
 \begin{eqnarray}
 	r &=& 1 + \frac{Y(t,y^b)}{D}\\
 	u &=& \left( U_0 + \frac{U_1(t,y^b)}{D} \right)dt + \frac{1}{D} U_a(t,y^b) dy^a \\
 	Q &=& Q_0 + \frac{1}{D} q(t,y^b)
 \end{eqnarray}
 
 In this new scaled coordinate we can write the membrane equation as,
 \begin{eqnarray} \label{eq1}
 	E^{tot}_\mu &=& \mathcal{P}_\mu^\nu E_\nu\\
 	E_\nu &=& \frac{\tilde{\nabla}^2 u_\nu}{\mathcal{K}} - (1-Q^2) \frac{\tilde{\nabla}_\nu \mathcal{K}}{\mathcal{K}} + u^\rho K_{\rho \nu} - (1+Q^2) u.\tilde{\nabla} u_\nu 
 \end{eqnarray}
 
 The total contribution from \eqref{eq1} is found to be, \footnotemark \footnotetext{For more details of the calculation the reader may go through \cite{SB}}
 \begin{eqnarray}
\nonumber E_a &=& \frac{1}{D}\Big[ \partial_b \partial^b U_a + \partial^c Y \partial_c U_a + (1-Q_0^2) \partial^b \partial_b \partial_a Y  + (1-Q_0^2) \partial^b \partial_b \partial_a Y - \partial_t \partial_a Y\hspace{2mm} \\
 \nonumber	&& -(1-Q_0^2)\partial^b Y \partial_a \partial_bY- U^b \partial_b \partial_a Y - (1+Q_0^2) \partial_t U_a - (1+Q_0^2) \partial_a Y - (1+Q_0^2) U^b \partial_b U_a \Big]~~\\ 
	&&~=0
	 \end{eqnarray}
The contribution from \eqref{sclsca} is,
\begin{equation}
 	\tilde{\nabla}.u = E^s = \partial_b U^b + \partial_t Y + U^b \partial_b Y = 0
 \end{equation}
The charge equation can be written as,
\begin{eqnarray}
	\nonumber E^c &=& \frac{1}{D}\left[\partial_a \partial^a q + \partial_a Y \partial^a q - \partial_t q + U^a \partial_a q \right] - \frac{Q_0}{D} \Big( \partial_t^2 Y - \partial_a Y \partial^a Y + 2 U^a\partial_a \partial_t Y~~~~  \\
	&&+ U_a U_b \partial^a \partial^b Y - \partial_t^2 \partial_a \partial^a Y  - \partial_t \partial^bY\partial_b Y - U_a \partial^a \partial_b\partial^b Y - \partial_b Y U^a \partial_a \partial^b Y \Big) =0~~~~~~~~~~
\end{eqnarray}

These are the scaled non linear version of the membrane equation. Now we have to calculate the QNM for this system by linearizing with respect to the equilibrium system. Such an equilibrium system will be,
\begin{eqnarray}
	Y &=& \epsilon \delta Y e^{- i (\omega^s t - k_a y^a )} + \mathcal{O}(\epsilon^2) \\
	U_a &=& \epsilon \delta V_a e^{- i (\omega^v t - k_a y^a )} + \epsilon \delta Y V^s_a e^{- i (\omega^s t - k_a y^a )} +\mathcal{O}(\epsilon^2)\\
	q &=& \epsilon \delta q e^{- i (\omega^q t - k_a y^a )} + \epsilon \delta Y Q_s e^{- i (\omega^s t - k_a y^a )}+\mathcal{O}(\epsilon^2)
\end{eqnarray}
where, $k^a V_a = 0$.\\
Upto the linear order in $\epsilon$, 
\begin{equation}
\begin{split}
&~E^s=\frac{\epsilon}{D}[\partial_b \delta V^b+ \partial_t \delta Y_0]=0\\ \label{eq2}
&~\partial_b \delta V^b=- \partial_t \delta Y_0 
\end{split}
\end{equation}
Then, upto the linear order in $\epsilon$,
\begin{eqnarray}
\nonumber E_a &=& \frac{\epsilon}{D}\Big[ \partial_b \partial^b \delta V_a  + (1-Q_0^2) \partial^b \partial_b \partial_a \delta Y  - (1+Q_0^2) \partial_t \delta V_a-(1+Q_0^2) \partial_a \delta Y - \partial_t \partial_a \delta Y \Big]\hspace{2mm} \\ \label{eq3}
	&&~=0\\
 \nonumber  \partial^a E_a&=&\frac{\epsilon}{D} \Big[ \partial_a \partial_b \partial^b \delta V_a  + (1-Q_0^2) \partial^b \partial_b \partial_a \partial^a \delta Y  - (1+Q_0^2) \partial_t \partial^a \delta V_a-(1+Q_0^2)\partial^a  \partial_a \delta Y  \hspace{2mm} \\ \label{eq4}
&&~- \partial_t \partial_a\partial^a  \delta Y \Big]=0\\ \label{eq5}
 E^c &=& \frac{\epsilon}{D}\Big[ \partial_a \partial^a \delta q  - \partial_t \delta q \Big] - \frac{\epsilon Q_0}{D} \Big( \partial_t^2 \delta Y- \partial_t^2 \partial_a \partial^a \delta Y   \Big) =0
	 \end{eqnarray}

Now using, \eqref{eq3},\eqref{eq4} and \eqref{eq5} along with the simplification in \eqref{eq2} we have calculated the scalar mode QNM for different modes,
\begin{eqnarray}
	\omega^s_{\pm} &=& - i \frac{k^2}{(1+Q^2_0)} \pm k \sqrt{1- k^2 \left(\frac{Q^2_0}{1+Q_0^2}\right)^2}  \\
	\omega^v &=& -i \frac{k^2}{1+Q_0^2}\\
	\omega^q &=& -i k^2
\end{eqnarray}

These results match exactly with our earlier results upto the scaling $k\rightarrow \frac{k}{\sqrt{D}}$  (which is naturally expected) and are also consistent with \cite{EmparanHydro}. This analysis shows that even after a scaled co-ordinate transformation and a special kind of scaling of $\mathcal{O}\Big( \sqrt{\frac{1}{D}}\Big)$, it is possible to reach the  charged black brane equations (upto the linear order) in \cite{EmparanHydro} from our covariant membrane equations  by a suitable field redefinition.


\bibliographystyle{JHEP}
\bibliography{ref}

\providecommand{\href}[2]{#2}\begingroup\raggedright\begin{thebibliography}{10}

\bibitem{EmparanCoupling}
R.~Emparan, R.~Suzuki and K.~Tanabe, \emph{{Decoupling and non-decoupling
  dynamics of large D black holes}},
  \href{https://doi.org/10.1007/JHEP07(2014)113}{\emph{JHEP} {\bfseries 07}
  (2014) 113} [\href{https://arxiv.org/abs/1406.1258}{{\ttfamily 1406.1258}}].

\bibitem{Emparan:2013moa}
R.~Emparan, R.~Suzuki and K.~Tanabe, \emph{{The large D limit of General
  Relativity}}, \href{https://doi.org/10.1007/JHEP06(2013)009}{\emph{JHEP}
  {\bfseries 1306} (2013) 009}
  [\href{https://arxiv.org/abs/1302.6382}{{\ttfamily 1302.6382}}].

\bibitem{Emparan:2013xia}
R.~Emparan, D.~Grumiller and K.~Tanabe, \emph{{Large-D gravity and low-D
  strings}},
  \href{https://doi.org/10.1103/PhysRevLett.110.251102}{\emph{Phys.Rev.Lett.}
  {\bfseries 110} (2013) 251102}
  [\href{https://arxiv.org/abs/1303.1995}{{\ttfamily 1303.1995}}].

\bibitem{Emparan:2013oza}
R.~Emparan and K.~Tanabe, \emph{{Holographic superconductivity in the large D
  expansion}}, \href{https://doi.org/10.1007/JHEP01(2014)145}{\emph{JHEP}
  {\bfseries 1401} (2014) 145}
  [\href{https://arxiv.org/abs/1312.1108}{{\ttfamily 1312.1108}}].

\bibitem{Emparan:2014cia}
R.~Emparan and K.~Tanabe, \emph{{Universal quasinormal modes of large D black
  holes}}, \href{https://doi.org/10.1103/PhysRevD.89.064028}{\emph{Phys.Rev.}
  {\bfseries D89} (2014) 064028}
  [\href{https://arxiv.org/abs/1401.1957}{{\ttfamily 1401.1957}}].

\bibitem{Emparan:2014jca}
R.~Emparan, R.~Suzuki and K.~Tanabe, \emph{{Instability of rotating black
  holes: large D analysis}},
  \href{https://doi.org/10.1007/JHEP06(2014)106}{\emph{JHEP} {\bfseries 1406}
  (2014) 106} [\href{https://arxiv.org/abs/1402.6215}{{\ttfamily 1402.6215}}].

\bibitem{Prester:2013gxa}
P.~D. Prester, \emph{{Small black holes in the large D limit}},
  \href{https://doi.org/10.1007/JHEP06(2013)070}{\emph{JHEP} {\bfseries 06}
  (2013) 070} [\href{https://arxiv.org/abs/1304.7288}{{\ttfamily 1304.7288}}].

\bibitem{membrane}
S.~Bhattacharyya, A.~De, S.~Minwalla, R.~Mohan and A.~Saha, \emph{{A membrane
  paradigm at large D}},
  \href{https://doi.org/10.1007/JHEP04(2016)076}{\emph{JHEP} {\bfseries 04}
  (2016) 076} [\href{https://arxiv.org/abs/1504.06613}{{\ttfamily
  1504.06613}}].

\bibitem{Chmembrane}
S.~Bhattacharyya, M.~Mandlik, S.~Minwalla and S.~Thakur, \emph{{A Charged
  Membrane Paradigm at Large D}},
  \href{https://doi.org/10.1007/JHEP04(2016)128}{\emph{JHEP} {\bfseries 04}
  (2016) 128} [\href{https://arxiv.org/abs/1511.03432}{{\ttfamily
  1511.03432}}].

\bibitem{yogesh1}
Y.~Dandekar, A.~De, S.~Mazumdar, S.~Minwalla and A.~Saha, \emph{{The large D
  black hole Membrane Paradigm at first subleading order}},
  \href{https://doi.org/10.1007/JHEP12(2016)113}{\emph{JHEP} {\bfseries 12}
  (2016) 113} [\href{https://arxiv.org/abs/1607.06475}{{\ttfamily
  1607.06475}}].

\bibitem{Emparan:2015hwa}
R.~Emparan, T.~Shiromizu, R.~Suzuki, K.~Tanabe and T.~Tanaka, \emph{{Effective
  theory of Black Holes in the 1/D expansion}},
  \href{https://doi.org/10.1007/JHEP06(2015)159}{\emph{JHEP} {\bfseries 06}
  (2015) 159} [\href{https://arxiv.org/abs/1504.06489}{{\ttfamily
  1504.06489}}].

\bibitem{Suzuki:2015iha}
R.~Suzuki and K.~Tanabe, \emph{{Stationary black holes: Large $D$ analysis}},
  \href{https://doi.org/10.1007/JHEP09(2015)193}{\emph{JHEP} {\bfseries 09}
  (2015) 193} [\href{https://arxiv.org/abs/1505.01282}{{\ttfamily
  1505.01282}}].

\bibitem{Emparan:2015gva}
R.~Emparan, R.~Suzuki and K.~Tanabe, \emph{{Evolution and endpoint of the black
  string instability: Large D solution}},
  \href{https://doi.org/10.1103/PhysRevLett.115.091102}{\emph{Phys. Rev. Lett.}
  {\bfseries 115} (2015) 091102}
  [\href{https://arxiv.org/abs/1506.06772}{{\ttfamily 1506.06772}}].

\bibitem{Tanabe:2015hda}
K.~Tanabe, \emph{{Black rings at large D}},
  \href{https://doi.org/10.1007/JHEP02(2016)151}{\emph{JHEP} {\bfseries 02}
  (2016) 151} [\href{https://arxiv.org/abs/1510.02200}{{\ttfamily
  1510.02200}}].

\bibitem{Tanabe:2015isb}
K.~Tanabe, \emph{{Instability of the de Sitter Reissner--Nordstrom black hole
  in the $1/D$ expansion}},
  \href{https://doi.org/10.1088/0264-9381/33/12/125016}{\emph{Class. Quant.
  Grav.} {\bfseries 33} (2016) 125016}
  [\href{https://arxiv.org/abs/1511.06059}{{\ttfamily 1511.06059}}].

\bibitem{Suzuki:2015axa}
R.~Suzuki and K.~Tanabe, \emph{{Non-uniform black strings and the critical
  dimension in the $1/D$ expansion}},
  \href{https://doi.org/10.1007/JHEP10(2015)107}{\emph{JHEP} {\bfseries 10}
  (2015) 107} [\href{https://arxiv.org/abs/1506.01890}{{\ttfamily
  1506.01890}}].

\bibitem{Emparan:2015rva}
R.~Emparan, R.~Suzuki and K.~Tanabe, \emph{{Quasinormal modes of (Anti-)de
  Sitter black holes in the 1/D expansion}},
  \href{https://doi.org/10.1007/JHEP04(2015)085}{\emph{JHEP} {\bfseries 04}
  (2015) 085} [\href{https://arxiv.org/abs/1502.02820}{{\ttfamily
  1502.02820}}].

\bibitem{EmparanHydro}
R.~Emparan, K.~Izumi, R.~Luna, R.~Suzuki and K.~Tanabe, \emph{{Hydro-elastic
  Complementarity in Black Branes at large D}},
  \href{https://doi.org/10.1007/JHEP06(2016)117}{\emph{JHEP} {\bfseries 06}
  (2016) 117} [\href{https://arxiv.org/abs/1602.05752}{{\ttfamily
  1602.05752}}].

\bibitem{Romero-Bermudez:2015bma}
A.~M. Garc{\'\i}a-Garc{\'\i}a and A.~Romero-Berm{\'u}dez, \emph{{Conductivity
  and entanglement entropy of high dimensional holographic superconductors}},
  \href{https://arxiv.org/abs/1502.03616}{{\ttfamily 1502.03616}}.

\bibitem{Tanabe:2016opw}
K.~Tanabe, \emph{{Charged rotating black holes at large D}},
  \href{https://arxiv.org/abs/1605.08854}{{\ttfamily 1605.08854}}.

\bibitem{Sadhu:2016ynd}
A.~Sadhu and V.~Suneeta, \emph{{Nonspherically symmetric black string
  perturbations in the large dimension limit}},
  \href{https://doi.org/10.1103/PhysRevD.93.124002}{\emph{Phys. Rev.}
  {\bfseries D93} (2016) 124002}
  [\href{https://arxiv.org/abs/1604.00595}{{\ttfamily 1604.00595}}].

\bibitem{Herzog:2016hob}
C.~P. Herzog, M.~Spillane and A.~Yarom, \emph{{The holographic dual of a
  Riemann problem in a large number of dimensions}},
  \href{https://arxiv.org/abs/1605.01404}{{\ttfamily 1605.01404}}.

\bibitem{Rozali:2016yhw}
M.~Rozali and A.~Vincart-Emard, \emph{{On Brane Instabilities in the Large $D$
  Limit}}, \href{https://doi.org/10.1007/JHEP08(2016)166}{\emph{JHEP}
  {\bfseries 08} (2016) 166}
  [\href{https://arxiv.org/abs/1607.01747}{{\ttfamily 1607.01747}}].

\bibitem{Chen:2015fuf}
B.~Chen, Z.-Y. Fan, P.~Li and W.~Ye, \emph{{Quasinormal modes of Gauss-Bonnet
  black holes at large D}},
  \href{https://doi.org/10.1007/JHEP01(2016)085}{\emph{JHEP} {\bfseries 01}
  (2016) 085} [\href{https://arxiv.org/abs/1511.08706}{{\ttfamily
  1511.08706}}].

\bibitem{Chen:2016fuy}
B.~Chen and P.-C. Li, \emph{{Instability of Charged Gauss-Bonnet Black Hole in
  de Sitter Spacetime at Large $D$}},
  \href{https://arxiv.org/abs/1607.04713}{{\ttfamily 1607.04713}}.

\bibitem{Chen:2017hwm}
B.~Chen and P.-C. Li, \emph{{Static Gauss-Bonnet Black Holes at Large $D$}},
  \href{https://doi.org/10.1007/JHEP05(2017)025}{\emph{JHEP} {\bfseries 05}
  (2017) 025} [\href{https://arxiv.org/abs/1703.06381}{{\ttfamily
  1703.06381}}].

\bibitem{Chen:2017wpf}
B.~Chen, P.-C. Li and Z.-z. Wang, \emph{{Charged Black Rings at large D}},
  \href{https://doi.org/10.1007/JHEP04(2017)167}{\emph{JHEP} {\bfseries 04}
  (2017) 167} [\href{https://arxiv.org/abs/1702.00886}{{\ttfamily
  1702.00886}}].

\bibitem{Konoplya:2013sba}
R.~A. Konoplya and A.~Zhidenko, \emph{{Instability of D-dimensional extremally
  charged Reissner-Nordstrom(-de Sitter) black holes: Extrapolation to
  arbitrary D}}, \href{https://doi.org/10.1103/PhysRevD.89.024011}{\emph{Phys.
  Rev.} {\bfseries D89} (2014) 024011}
  [\href{https://arxiv.org/abs/1309.7667}{{\ttfamily 1309.7667}}].

\bibitem{Guo:2015swu}
E.-D. Guo, M.~Li and J.-R. Sun, \emph{{CFT dual of charged AdS black hole in
  the large dimension limit}},
  \href{https://doi.org/10.1142/S0218271816500851}{\emph{Int. J. Mod. Phys.}
  {\bfseries D25} (2016) 1650085}
  [\href{https://arxiv.org/abs/1512.08349}{{\ttfamily 1512.08349}}].

\bibitem{Chen:2017rxa}
B.~Chen, P.-C. Li and C.-Y. Zhang, \emph{{Einstein-Gauss-Bonnet Black Strings
  at Large $D$}}, \href{https://doi.org/10.1007/JHEP10(2017)123}{\emph{JHEP}
  {\bfseries 10} (2017) 123}
  [\href{https://arxiv.org/abs/1707.09766}{{\ttfamily 1707.09766}}].

\bibitem{Herzog:2017qwp}
C.~P. Herzog and Y.~Kim, \emph{{The Large Dimension Limit of a Small Black Hole
  Instability in Anti-de Sitter Space}},
  \href{https://doi.org/10.1007/JHEP02(2018)167}{\emph{JHEP} {\bfseries 02}
  (2018) 167} [\href{https://arxiv.org/abs/1711.04865}{{\ttfamily
  1711.04865}}].

\bibitem{radiation}
S.~Bhattacharyya, A.~K. Mandal, M.~Mandlik, U.~Mehta, S.~Minwalla, U.~Sharma
  et~al., \emph{{Currents and Radiation from the large $D$ Black Hole
  Membrane}}, \href{https://doi.org/10.1007/JHEP05(2017)098}{\emph{JHEP}
  {\bfseries 05} (2017) 098}
  [\href{https://arxiv.org/abs/1611.09310}{{\ttfamily 1611.09310}}].

\bibitem{yogesh2}
Y.~Dandekar, S.~Mazumdar, S.~Minwalla and A.~Saha, \emph{{Unstable `black
  branes' from scaled membranes at large $D$}},
  \href{https://doi.org/10.1007/JHEP12(2016)140}{\emph{JHEP} {\bfseries 12}
  (2016) 140} [\href{https://arxiv.org/abs/1609.02912}{{\ttfamily
  1609.02912}}].

\bibitem{Dandekar:2017aiv}
Y.~Dandekar, S.~Kundu, S.~Mazumdar, S.~Minwalla, A.~Mishra and A.~Saha,
  \emph{{An Action for and Hydrodynamics from the improved Large D membrane}},
  \href{https://arxiv.org/abs/1712.09400}{{\ttfamily 1712.09400}}.

\bibitem{SB}
S.~Bhattacharyya, P.~Biswas, B.~Chakrabarty, Y.~Dandekar and A.~Dinda,
  \emph{{The large D black hole dynamics in AdS/dS backgrounds}},
  \href{https://arxiv.org/abs/1704.06076}{{\ttfamily 1704.06076}}.

\bibitem{SB2}
S.~Bhattacharyya, P.~Biswas and Y.~Dandekar, \emph{{Black holes in presence of
  cosmological constant: Second order in $\frac{1}{D}$}},
  \href{https://arxiv.org/abs/1805.00284}{{\ttfamily 1805.00284}}.

\bibitem{Saha:2018elg}
A.~Saha, \emph{{The large D Membrane Paradigm For Einstein-Gauss-Bonnet
  Gravity}},  \href{https://arxiv.org/abs/1806.05201}{{\ttfamily 1806.05201}}.

\bibitem{Chen:2018nbh}
B.~Chen, P.-C. Li, Y.~Tian and C.-Y. Zhang, \emph{{Holographic Turbulence in
  Einstein-Gauss-Bonnet Gravity at Large $D$}},
  \href{https://arxiv.org/abs/1804.05182}{{\ttfamily 1804.05182}}.

\bibitem{arbitrarydim}
S.~Bhattacharyya, R.~Loganayagam, I.~Mandal, S.~Minwalla and A.~Sharma,
  \emph{{Conformal Nonlinear Fluid Dynamics from Gravity in Arbitrary
  Dimensions}},
  \href{https://doi.org/10.1088/1126-6708/2008/12/116}{\emph{JHEP} {\bfseries
  0812} (2008) 116} [\href{https://arxiv.org/abs/0809.4272}{{\ttfamily
  0809.4272}}].

\bibitem{WaldBook}
R.~M. Wald, \emph{{General Relativity}}. University of Chicago Press, 1984,
  \href{https://doi.org/10.7208/chicago/9780226870373.001.0001}{10.7208/chicago/9780226870373.001.0001}.

\bibitem{Mandlik:2018wnw}
M.~Mandlik and S.~Thakur, \emph{{Stationary Solutions from the Large D Membrane
  Paradigm}},  \href{https://arxiv.org/abs/1806.04637}{{\ttfamily 1806.04637}}.

\end{thebibliography}\endgroup

\end{document}